\documentclass[prb,aps,amssymb,twocolumn,nofootinbib,superscriptaddress]{revtex4-2}
\usepackage{amsmath}
\usepackage{amsthm}
\usepackage{listings}
\usepackage{braket}
\usepackage[dvips]{graphicx}
\usepackage{hyperref}
\usepackage{psfrag}
\usepackage{comment}
\usepackage{bm}
\usepackage{color}
\usepackage{lineno}

\newcommand{\beq}{\begin{equation}}
	\newcommand{\eneq}{\end{equation}}

\begin{document}
	\title{Effects of first- and second-order topological phases on equilibrium crystal  shapes}
	
\author{Yutaro Tanaka}
\author{Shuichi Murakami}
\affiliation{Department of Physics, Tokyo Institute of Technology, 2-12-1 Ookayama, Meguro-ku, Tokyo 152-8551, Japan}

\begin{abstract}
We study equilibrium crystal shapes of a topological insulator (TI), a  topological crystalline insulator (TCI) protected by mirror symmetry, and a second-order topological insulator (SOTI) protected by inversion symmetry.  By adding magnetic fields to the three-dimensional TI, we can realize the mirror-symmetry-protected TCI and the inversion-symmetry-protected SOTI. They each have topological boundary states in different positions:~The TCI has gapless states on the surfaces that are invariant under the symmetry operation, and the SOTI has gapless states at the intersections between certain surfaces.  
In this paper, we discuss how these boundary states affect the surface energies and the equilibrium crystal shapes in terms of the calculations of the simple tight-binding model by using the Wulff construction. By comparing the changes in the shapes of the TI with those of the trivial insulator through the process of applying the magnetic fields, we show that the presence or absence of the topological boundary states affects the emergence of the specific facets in a different way from the trivial insulator. 
\end{abstract}
	
\maketitle
	
\section{introduction}
Crystal shapes have a large variety since they reflect microscopic physics of the respective materials. Theoretically, the equilibrium crystal shape is the shape that minimizes the surface free energy. 
The equilibrium crystal shape is determined by the surface free energy via the Wulff  construction \cite{wulffconst, wulffconst2, wulffconst3, PhysRev.82.87}, and  one can apply this method to predict the shapes of nanoparticles  \cite{Marks_1994, Xia_2009, barmparis2015nanoparticle}.  It is helpful to study the equilibrium crystal shapes because the shapes of the nanoparticles play important roles in controlling their properties  \cite{sun2002shape, Bratlie:2007vf, yang2008anatase, lovette2008crystal,  B711490G, ringe2011wulff, auyeung2014dna,  yang2014titania, liu2014titanium, tran2016surface, anderson2017predicting, wang2019crystal, https://doi.org/10.1002/adma.202008373}.
In particular, nanoparticles of topological insulators (TIs) exhibit unique phenomena originating from the nontrivial topology \cite{PhysRevB.82.155314, PhysRevB.86.235119, doi:10.7566/JPSJ.82.074712, lin2015using,  Siroki:2016vz, PhysRevMaterials.1.024201, PhysRevB.100.205417, PhysRevB.101.165410, PhysRevA.102.013720, https://doi.org/10.1002/adma.202008373, Castro-Enriquez:2022ut}. 
However, it is not well understood how the topological boundary states of the TIs affect their equilibrium shapes.

An interplay between symmetry and topology has led us to a wide range of topological phases of matter, including TIs \cite{RevModPhys.82.3045, RevModPhys.83.1057},  topological crystalline insulators (TCIs)  \cite{PhysRevLett.106.106802, hsieh2012topological, slager2013space, PhysRevLett.110.156403, PhysRevB.90.085304, PhysRevB.90.165114,  PhysRevB.93.195413, wang2016hourglass, lu2016symmetry, PhysRevB.93.195138, PhysRevB.96.064102, wieder2018wallpaper, PhysRevB.100.165202, kim2020glide} and second-order topological insulators (SOTIs) \cite{PhysRevLett.108.126807,PhysRevLett.110.046404,benalcazar2017quantized, PhysRevB.96.245115, PhysRevLett.119.246401, PhysRevLett.119.246402, schindler2018higher, fang2017rotation}. 
The three-dimensional (3D) TIs  and the 3D TCIs have gapless states on their surfaces \cite{RevModPhys.82.3045, RevModPhys.83.1057, PhysRevLett.106.106802, hsieh2012topological, PhysRevB.90.085304, PhysRevB.90.165114},  which are classified as the first-order topological phase. 
The surface states of the TCI appear only on the surfaces that are invariant under the symmetry considered  \cite{hsieh2012topological, PhysRevLett.110.156403, PhysRevB.90.085304, PhysRevB.90.165114,  PhysRevB.93.195413, wang2016hourglass}. On the other hand, the SOTIs are classified as the second-order topological phase \cite{PhysRevB.97.205135, PhysRevB.97.241405, schindler2018higherbismuth, PhysRevB.98.205147, serra2018observationnature555, peterson2018quantizedNature7695, imhof2018topolectricalnatphys, PhysRevLett.123.016806, wang2018higher, sheng2019two,  PhysRevB.98.035147, okugawa2019second, ghosh2019engineering, agarwala2019higher, chen2019higher, PhysRevLett.125.017001, hirayama2020higher, chen2020universal, Arai:2021vc, Nagasato:2021tg, PhysRevResearch.3.L032029, PhysRevResearch.3.033177, PhysRevLett.127.176601,  PhysRevB.103.115118,  PhysRevB.104.245427,  PhysRevB.105.045113, PhysRevB.105.045126, PhysRevB.105.115119, PhysRevB.105.195149, PhysRevB.106.165401,  PhysRevB.106.165422,  Mu:2022tierg, Mu:2022ta} and do not exhibit the surface states  but exhibit gapless states at the intersections between the surfaces in the 3D system, which are called hinge states \cite{PhysRevLett.108.126807,PhysRevLett.110.046404, PhysRevB.96.245115, schindler2018higher}. 
Thus we expect that such topological surface and hinge states affect equilibrium crystal shapes.

In this paper,  we study the equilibrium shapes of a TI, a  TCI protected by mirror symmetry, and a SOTI protected by inversion symmetry. 
By adding a magnetic field to the 3D TI, we can realize the mirror-symmetry-protected TCI or the inversion-symmetry-protected SOTI, as discussed in Secs.~\ref{Sec:3_mirror_TCI} and \ref{Sec:5_inversion_SOTI}.   
Thus, by focusing on the changes in the surface energies and the equilibrium crystal shape by adding the magnetic fields, we study how the surface states of the TCI protected by mirror symmetry \cite{hsieh2012topological, PhysRevLett.110.156403, Tanaka:2012wj, PhysRevB.87.235317, PhysRevLett.115.086802, Cao_2021} and the hinge states of the SOTI protected by inversion symmetry \cite{PhysRevB.97.205136, PhysRevB.98.205129, PhysRevLett.122.256402, PhysRevResearch.2.043274} affect the equilibrium crystal shape.  In addition, by comparing the changes in the shapes of the TI to that of the trivial insulator by adding magnetic fields, we show that the topological surface states affect the equilibrium crystal shape, and this is unique to the topological phases.  
We note that in our previous work \cite{PhysRevLett.129.046802}, equilibrium crystal shapes of TCIs protected by glide symmetry are studied. The calculation method in the present paper is partially parallel to the previous work, but we will see that  the results are quite different because the symmetries protecting the topological phases are different. 

This paper is organized as follows. In Sec.~\ref{Sec2:_TI_surface}, we introduce a tight-binding model of a 3D TI and calculate the surface states. In Sec.~\ref{Sec:3_mirror_TCI}, by adding a magnetic field to the 3D TI while preserving mirror symmetry, we realize the mirror-protected TCI and calculate the surface states protected by mirror symmetry. In Sec.~\ref{Sec4_surface_energy_shape}, we calculate the surface energies and the equilibrium crystal shapes of the TI and the TCI.  In Sec.~\ref{Sec:5_inversion_SOTI}, we calculate the hinge states, the surface energies, and the equilibrium crystal shape of the SOTI that are realized by adding the magnetic field to the TI in Sec.~\ref{Sec:3_mirror_TCI}. 
A conclusion and discussion are given in Sec.~\ref{Sec6:conclusion_discussion}.

\section{Surface states of a topological insulator}\label{Sec2:_TI_surface}
In this section, as a preliminary step toward calculations of models of a TCI and a SOTI,  we introduce a tight-binding model of a TI and study its surface states. 
\subsection{Tight-binding model and symmetry}
\begin{figure}
\includegraphics[width=1.\columnwidth]{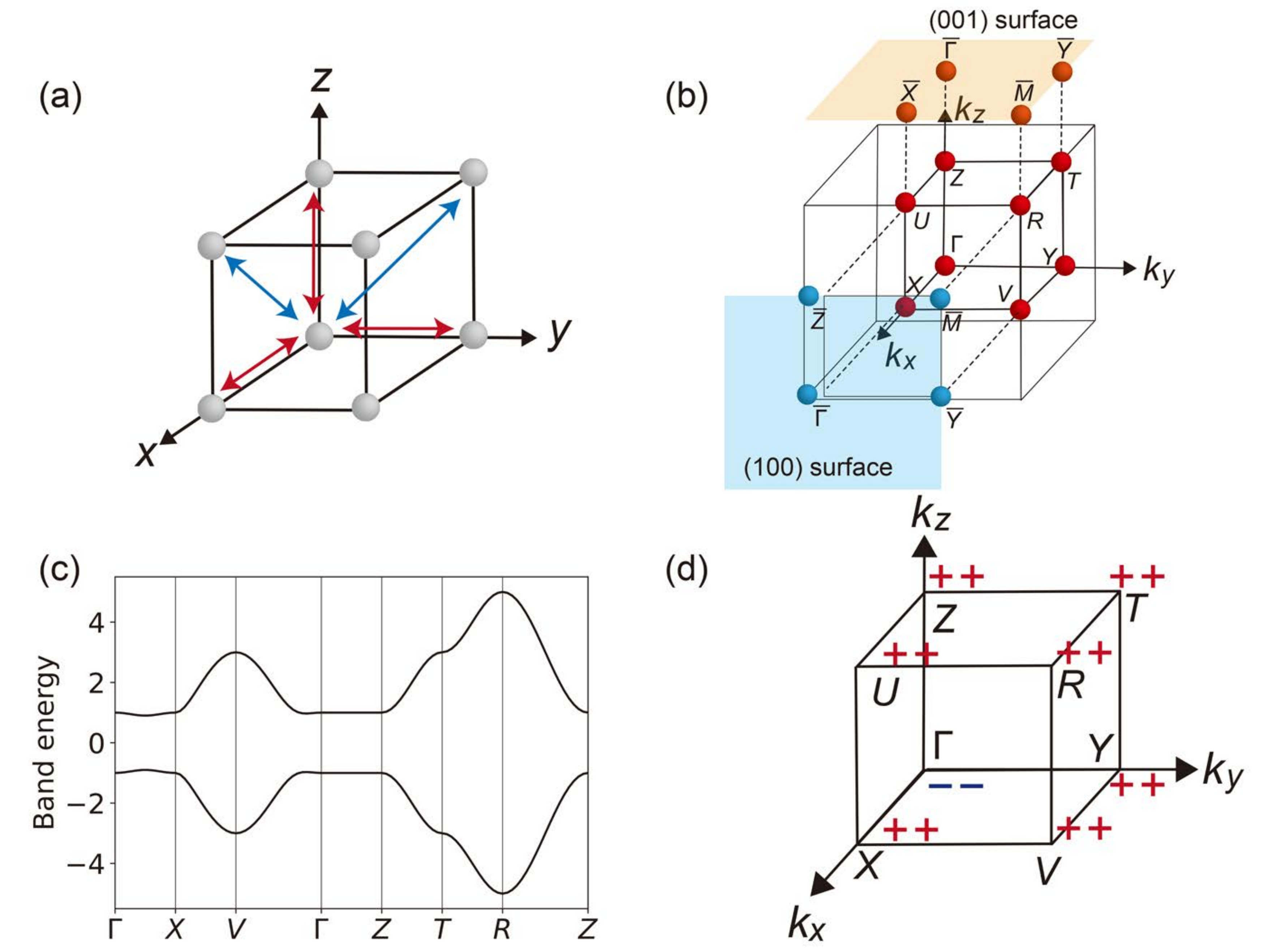}
\caption{Our tight-binding model of the 3D TI. (a) The crystal structure of our model. The red (blue) arrows indicate the nearest (next-nearest) neighbor hopping.  (b) The Brillouin zone and the (100), (010), and (001) surface Brillouin zone. (c) The bulk band structure of $\mathcal{H}_{\rm TI}(\boldsymbol{k})$ with the parameters $m=2$, $t=1$, $v=0.4$, $v'=0.5$, and $v_{z}=1$. (d) The inversion eigenvalues of occupied states at the high-symmetry points. }\label{chap8_TI_model}
\end{figure}

We start from a three-dimensional (3D) tight-binding model of a $\mathbb{Z}_{2}$ TI on a primitive tetragonal lattice with lattice vectors $\boldsymbol{a}_{1}=(a,0,0)$, $\boldsymbol{a}_{2}=(0,a, 0)$, and $\boldsymbol{a}_{3}=(0,0, a)$ with $a$ being the lattice constant:
\begin{align}
	\mathcal{H}_{\rm TI}(\boldsymbol{k})= &\Bigl( m-t \sum_{j=x,y,z}  \cos k_{j} \Bigr)\tau_{z}\otimes \sigma_0 \nonumber \\
& +(v+v'\cos k_z) \sin k_x \tau_x \otimes \sigma_x \nonumber \\
&+(v+v'\cos k_z) \sin k_y \tau_x \otimes \sigma_y\nonumber\\
&+v_{z}\sin k_z \tau_x \otimes  \sigma_{z},
 \end{align}
 where $\sigma_{i}$ and $\tau_{i}$ ($i=x,y,z$) are Pauli matrices, and $\sigma_0$ and $\tau_0$ are identity matrices. We set the lattice constant to be $a=1$. 
 This model has nearest neighbor hopping in the  [100], [010], and [001] directions and next-nearest neighbor hopping in the [101], [10$\bar{1}$] , [011], and [01$\bar{1}$] directions [Fig.~\ref{chap8_TI_model}(a)].  
 This model has time-reversal ($\mathcal{T}$) symmetry,  inversion $(\mathcal{I})$ symmetry, and fourfold rotation $C_{4z}$ symmetry with the rotation axis along the $z$ direction: 
\begin{gather}
\mathcal{T} \mathcal{H}_{\rm TI}(\boldsymbol{k}) \mathcal{T}^{-1} = \mathcal{H}_{\rm TI}(-\boldsymbol{k}), \nonumber \\
\mathcal{I} \mathcal{H}_{\rm TI}(\boldsymbol{k}) \mathcal{I}^{-1} = \mathcal{H}_{\rm TI}(-\boldsymbol{k}), \nonumber \\
C_{4z} \mathcal{H}_{\rm TI}(\boldsymbol{k}) C_{4z}^{-1}=\mathcal{H}_{\rm TI}(C_{4z}\boldsymbol{k}), 
\end{gather}
where $\mathcal{T}=-i \tau_0 \otimes \sigma_y K$ with $K$ being the complex conjugation, $\mathcal{I}=\tau_z \otimes \sigma_0$,  $C_{4z}=\tau_0 \otimes (1-i\sigma_z)/ \sqrt{2}$, and $C_{4z}\boldsymbol{k}=(-k_y, k_x, k_z)$. 
Furthermore, our model $\mathcal{H}_{\rm TI}(\boldsymbol{k})$ has mirror $M_{z}(=\mathcal{I}C_{2z})$ symmetry with respect to the $x$-$y$ mirror plane: $M_{z}\mathcal{H}_{\rm TI}(k_x, k_y, k_z) M_{z}^{-1}=\mathcal{H}_{\rm TI}(k_x, k_y, -k_z)$ with $M_{z}=-i \tau_{z}\otimes \sigma_{z}$. 
In this paper, to simplify our discussion, we consider band structures on the ($hkl$) surfaces up to maximum absolute values of the Miller index $h_{\rm max}=k_{\rm max}=l_{\rm max}=1$.

Figure~\ref{chap8_TI_model}(b) shows the Brillouin zone and the (100), (010), and (001) surface Brillouin zones. Figure~\ref{chap8_TI_model}(b) also shows high-symmetry points $K_{\mathcal{I}} =\{ \Gamma={\pi}(0,0,0),\ X={\pi}(1, 0, 0),\  Y={\pi}(0,1, 0),\  V={\pi}(1,1, 0),\  Z={\pi}(0,0,1),\  U={\pi}(1,0,1),\  T={\pi}(0,1,1),\  R={\pi}(1, 1, 1) \}$, which are invariant under $\mathcal{I}$. 
 Figure ~\ref{chap8_TI_model}(c) shows the bulk band structure of $\mathcal{H}_{\rm TI}(\boldsymbol{k})$, where the Fermi energy is set to be zero. 
Figure~\ref{chap8_TI_model}(d) shows the inversion eigenvalues of the occupied states of $\mathcal{H}_{\rm TI}(\boldsymbol{k})$ at the high-symmetry points $K_{\mathcal{I}}$. Thus, the strong $\mathbb{Z}_2$ topological invariant $\nu$ is $\nu \equiv 1$ mod 2 \cite{PhysRevB.76.045302}, and therefore our model is a 3D TI. 

\begin{figure}
\centering
\includegraphics[width=1.\columnwidth]{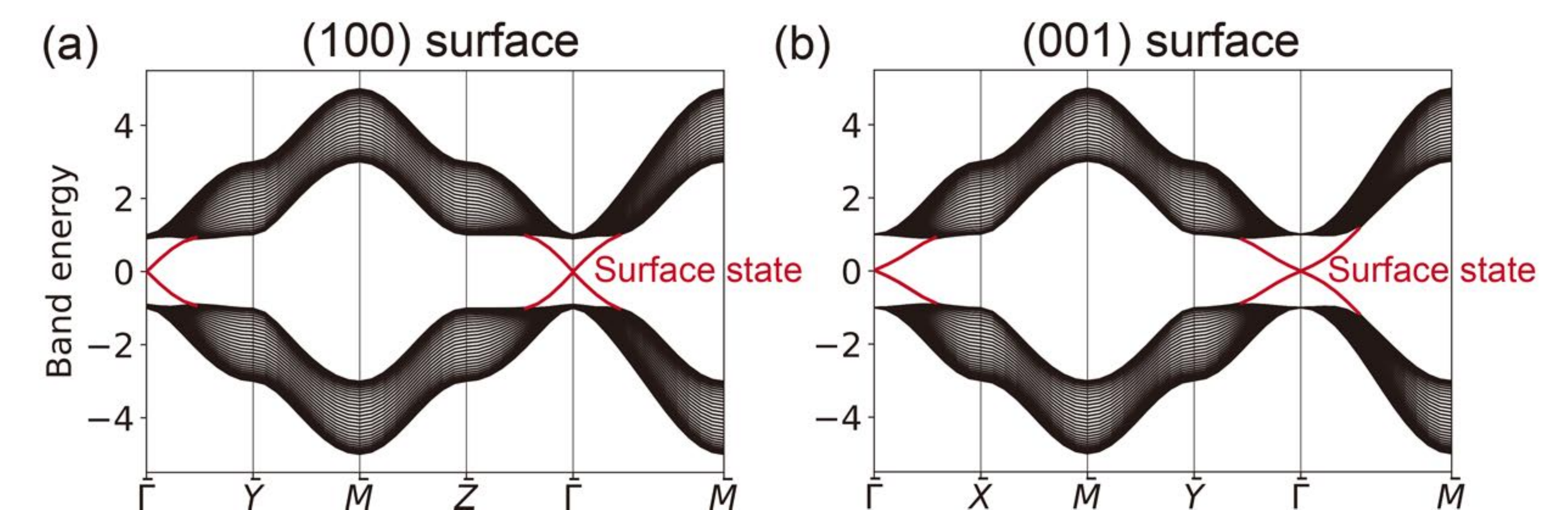}
\caption{Band structures of $\mathcal{H}_{\rm TI}(\boldsymbol{k})$ in the slab geometries with (a) the (100)  surfaces, (b) (001) surfaces. We choose the parameters $m=2$, $t=1$, $v=0.4$, $v'=0.5$, and $v_{z}=1$. The slab thickness is $30$.}\label{chap8_TI_surface}
\end{figure}

We calculate band structures of $\mathcal{H}_{\rm TI}(\boldsymbol{k})$ with periodic boundary conditions (PBCs) in two directions and with finite size in the remaining direction.
We refer to such a geometry as the slab geometry. For example, the slab geometry with the (100) surfaces indicates PBCs in the $x$ and $y$ directions and finite size in the $z$ direction.  Figure~\ref{chap8_TI_surface}(a) shows that band structure with this slab geometry, and we find that the gapless surface states appear. We also calculate the band structure in the (001) slab [Fig.~\ref{chap8_TI_surface}(b)], which indicates that the gapless surface states appear similarly to the (100) surface. 
By $\mathcal{I}$, $M_{z}$, and $C_{4z}$ symmetries, it is sufficient to calculate the (100), (001), (101), and (111) surfaces.

\subsection{Lattice vectors and reciprocal lattice vectors for slab geometry}
\begin{figure}
\begin{center}
\includegraphics[width=1.\columnwidth]{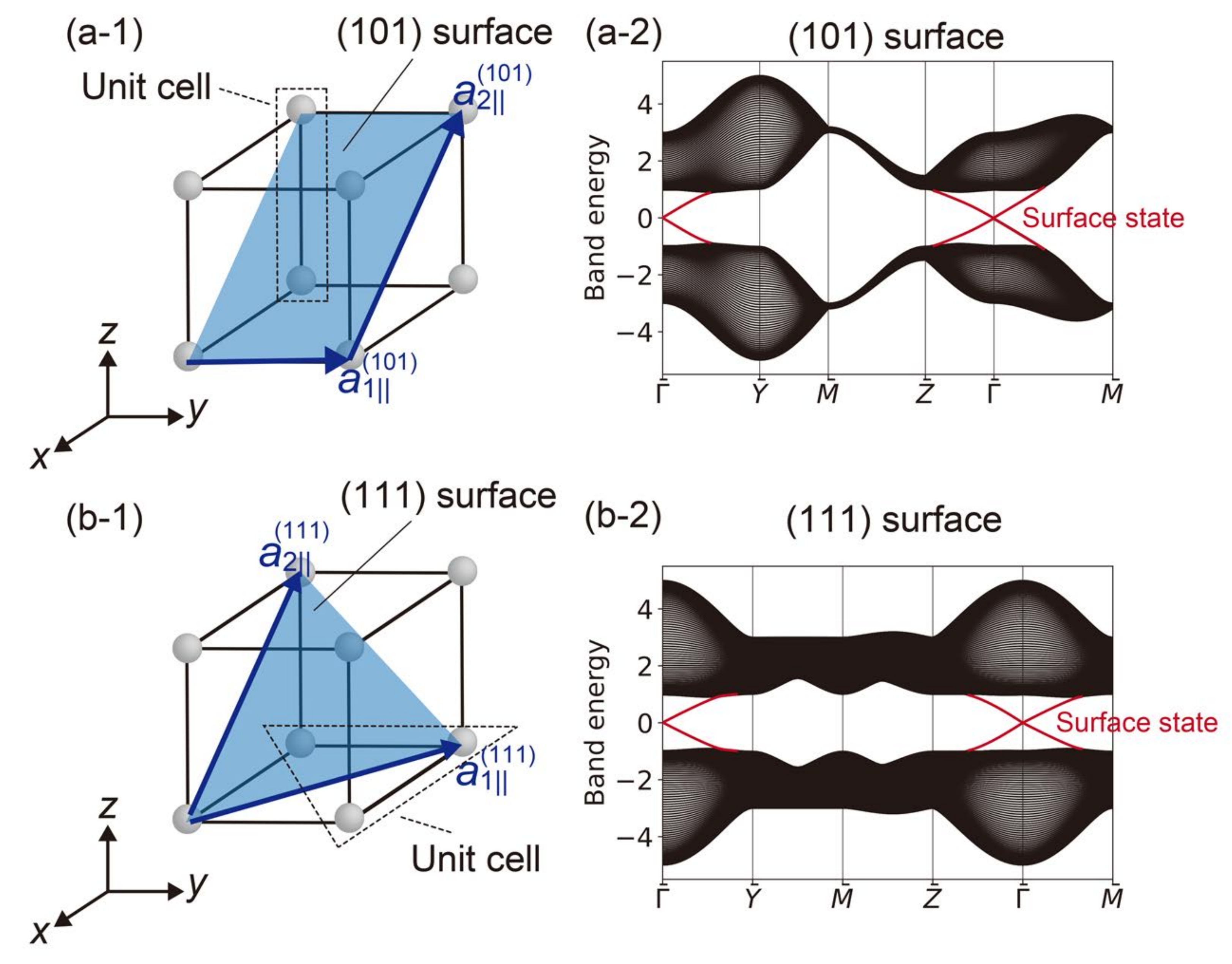}
\caption{Lattice vectors and band structures of $\mathcal{H}_{\rm TI}(\boldsymbol{k})$ in the slab geometries with (a) the (101) surface and (b) the (111) surface. (a-1), (b-1) The lattice vectors.   The high-symmetry points are given by $\bar{\Gamma}=(k_{1\parallel}, k_{2 \parallel})=(0,0)$, $\bar{Y}=(1/2, 0)$, $\bar{Z}=(0, 1/2)$, and  $\bar{M}=(1/2, 1/2)$. We choose the parameters $m=2$, $t=1$, $v=0.4$, $v'=0.5$, and $v_{z}=1$. The slab thickness is 30 in the unit of the vectors (1,0,1) in (a-2) and (1,1,1) in (b-2). }\label{chap8_TI_101_111_surface}
\end{center}
\end{figure}

In the slab geometry with ($hkl$) surfaces,  we choose lattice vectors  to be 
\begin{gather}\label{chap6_def_lat_vec}
	 \boldsymbol{a}^{(hkl)}_{1\parallel}=(-k, h,0), \nonumber  \\
	  \boldsymbol{a}^{(hkl)}_{2\parallel}=(-l,0,h),
\end{gather}
along the ($hkl$) surfaces ($h\neq 0$).
Although these vectors form a nonprimitive unit cell in general, we choose these  vectors for convenience. 
Let us consider a slab geometry with (101) surfaces. The lattice vectors for the (101) slab are given by  $\boldsymbol{a}_{1 \parallel}^{(101)}=(0,1,0)$ and   $\boldsymbol{a}_{2 \parallel}^{(101)}=(-1,0,1)$,   
as shown in Fig.~\ref{chap8_TI_101_111_surface}(a-1). Then the reciprocal lattice vectors are given by 
$\boldsymbol{b}_{1 \parallel}^{(101)}={2\pi}\boldsymbol{a}^{(101)}_{1 \parallel}$ and $\boldsymbol{b}_{2 \parallel}^{(101)}={\pi}\boldsymbol{a}^{(101)}_{2 \parallel}$.
The $k$ vector on the (101) surface is written as $\boldsymbol{k}=k_{1 \parallel} \boldsymbol{b}_{1 \parallel}^{(101)} +k_{2 \parallel} \boldsymbol{b}_{2 \parallel}^{(101)}$, and the high-symmetry points are given by $(k_{1 \parallel}, k_{2 \parallel})=(0,0)$, $(1/2,0)$, $(0, 1/2)$, and $(1/2, 1/2)$.  Figure~\ref{chap8_TI_101_111_surface}(a-2) shows the band structure in the slab geometry with the (101) surfaces, which indicates the emergence of the gapless surface states.  In addition, we consider a slab geometry with (111) surfaces. The lattice vectors are given by  $\boldsymbol{a}_{1 \parallel}^{(111)}=(-1,1,0)$ and  $\boldsymbol{a}_{2 \parallel}^{(111)}=(-1,0,1)$, as shown in Fig.~\ref{chap8_TI_101_111_surface}(b-1), and the reciprocal lattice vectors are given by 
$\boldsymbol{b}_{1 \parallel}^{(111)}={2\pi}(2\boldsymbol{a}^{(111)}_{1_\parallel}-\boldsymbol{a}^{(111)}_{2 \parallel})/3$  and $\boldsymbol{b}_{2 \parallel}^{(111)}={2\pi}(-\boldsymbol{a}^{(111)}_{1 \parallel}+2\boldsymbol{a}^{(111)}_{2 \parallel})/{3}$. We calculate the band structure  in the slab geometry with the (111) surfaces  [Fig.~\ref{chap8_TI_101_111_surface}(b-2)]. 
Because the surface states are protected by time-reversal symmetry, the gapless states appear on the (100), (001), (101), and (111) surfaces. 

\section{Surface states of an $M_z$-protected topological crystalline insulator}\label{Sec:3_mirror_TCI}
Next, we study a TCI protected by mirror $M_{z}$ symmetry.  The TCI phase can be easily realized by adding a Zeeman term to the Hamiltonian $H_{\rm TI}(\boldsymbol{k})$:
\begin{align}\label{eq:mtci_tight_bind_model}
	H_{\rm MTCI}(\boldsymbol{k})=H_{\rm TI}(\boldsymbol{k})+B_{z}\sigma_{z}. 
\end{align}
The Zeeman term $B_{z}\sigma_z$ breaks $\mathcal{T}$ symmetry and preserves $M_{z}$, $\mathcal{I}$, and $
C_{4z}$ symmetries.  When time-reversal symmetry is broken, the $\mathbb{Z}_{2}$ topological invariant $\nu$ is not well defined, and therefore  we introduce other topological invariants \cite{bradlyn2017topological, PhysRevX.7.041069, po2017symmetry, watanabe2018structure} as discussed below. 

Let $\Gamma_j$ denote one of the eight inversion invariant momenta $K_{\mathcal{I}}$. 
The eight $\Gamma_{j}$ can be indexed by three integers $n_{l}=0,1$ defined mod 2,
$
\Gamma_{j=(n_{1},n_{2},n_{3})}=\frac{1}{2}(n_{1}\boldsymbol{b}_{1}+n_{2}\boldsymbol{b}_{2}+n_{3}\boldsymbol{b}_{3}),
$
where $\boldsymbol{b}_{1}$, $\boldsymbol{b}_{2}$, and $\boldsymbol{b}_{3}$  are primitive reciprocal lattice vectors. Three $\mathbb{Z}_{2}$ weak topological indices \cite{watanabe2018structure, PhysRevB.98.115150} are defined as 
\begin{equation}\label{weakz2index}
\nu_{a}:= \sum_{\Gamma_{j}\land n_{a}=1}n_{-}(\Gamma_{j})\ \ ({\rm mod}\ 2)\ (a=1,2,3),
\end{equation}
where $n_{-}(\Gamma_{i})$ is the number of occupied states with odd parity at the inversion-invariant momenta $\Gamma_{j}$, and the summation is taken over the inversion-symmetric momenta on the plane $n_{a}=1$. 
The $\mathbb{Z}_{4}$ strong topological index \cite{watanabe2018structure, PhysRevB.98.115150} is defined as 
\begin{align}\label{z4index}
\mu_{1}:=
&\frac{1}{2}\sum_{\Gamma_{j}}\Bigl(n_{+}(\Gamma_{j})-n_{-}(\Gamma_{j})\Bigr) \ \ ({\rm mod}\ 4),
\end{align}
where $n_{+}(\Gamma_{j})$ is the number of occupied states with even parity at the inversion-symmetric momentum $\Gamma_{j}$.
For systems with inversion symmetry, topological phases are characterized by the symmetry indicator $X_{\rm BS}=(\nu_{1}, \nu_{2}, \nu_{3}, \mu_{1}$) with $\nu_{a}=0, 1$ and $\mu_{1}=0, 1, 2, 3$. 

Henceforth, we assume that the magnetic field $B_z$ is so small that the gap is not inverted by $B_z$. Then, the inversion eigenvalues of occupied states at $K_{\mathcal{I}}$ are the same as $\mathcal{H}_{\rm TI}(\boldsymbol{k})$ [Fig.~\ref{chap8_TI_model}(d)], which results in $\nu_x=\nu_y=\nu_z=0$ and $\mu_{1}=2$.  
According to Refs.~\cite{PhysRevB.98.115150, PhysRevResearch.2.043274, elcoro2021magnetic, peng2021topological}, when the inversion eigenvalues of occupied states satisfy these conditions in the presence of $M_z$ symmetry, 
the mirror Chern number is nontrivial, and the system is in the mirror-symmetry-protected TCI phase. 
The mirror Chern numbers in the $k_{z}=0$ and $k_{z}=\pi$ sectors are defined as 
\begin{equation}\label{donemirrorcherndefine}
C^{0}_{m}:= \frac{1}{2}\bigl( C^{0}_{+}-C^{0}_{-}\bigr),\ \ C^{\pi}_{m}:= \frac{1}{2}\bigl( C^{\pi}_{+}-C^{\pi}_{-}\bigr),
\end{equation}
respectively, where $C^{0}_{\pm}$ and $C^{\pi}_{\pm}$ represent the Chern numbers in the mirror sectors with mirror eigenvalues $\pm i$ in the $k_{z}=0$ and $k_{z}=\pi$ sectors respectively. 
The $\mathbb{Z}_{4}$ symmetry-based indicator $\mu_{1}$ in an insulator is related with the mirror Chern numbers via 
\begin{equation}\label{eq:mirrorchern_and_mu1}
\mu_{1}\equiv 2(C^{0}_{m}+C^{\pi}_{m})\ \  ({\rm mod\ 4}),
\end{equation}
where $\mu_{1}=0,2$ \cite{PhysRevResearch.2.043274}. The values $\mu_{1}=0,2$ mean that the bulk is insulating, while the values $\mu_{1}=1,3$ correspond to the Weyl semimetal phase \cite{PhysRevB.98.115150}. Equation~(\ref{eq:mirrorchern_and_mu1}) means that when  $\mu_{1}=2$, one of the two mirror Chern numbers $C^{0}_{m}$ and $C^{\pi}_{m}$, is an odd number, and the other is an even number, which means the emergence of topological surface states on mirror-symmetric surfaces, such as (100) and (010) surfaces. 
In our model, the mirror Chern number in the $k_z=0$ plane is $C^{0}_{m}=1$. 

\begin{figure}
\centering
\includegraphics[width=1.\columnwidth]{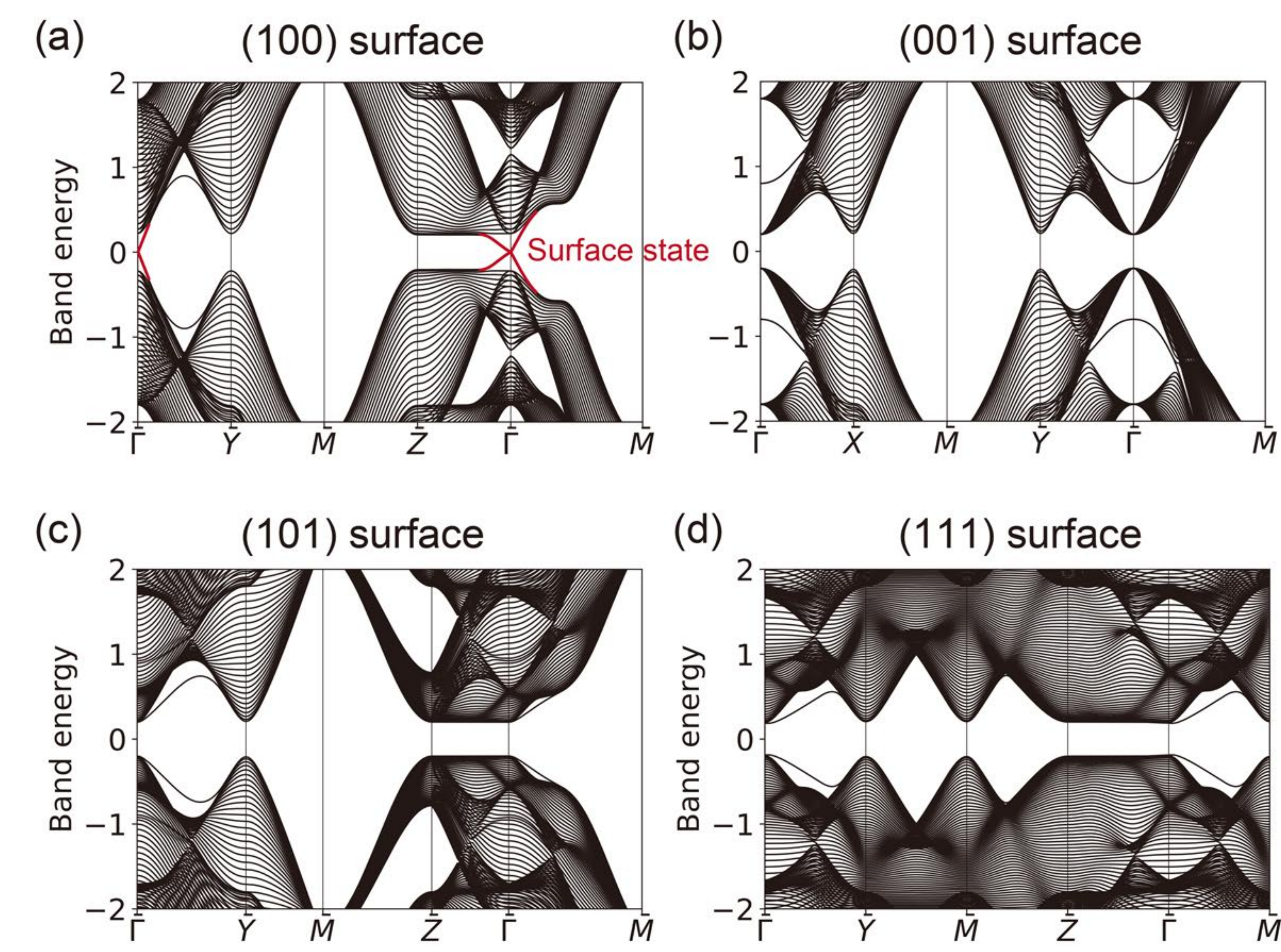}
\caption{Band structures of $\mathcal{H}_{\rm MTCI}(\boldsymbol{k})$ in the slab geometries with (a) the (100)  surfaces, (b) the (001) surfaces, (c) the (101) surfaces, and  (d) the (111) surfaces. The high-symmetry points are the same as in the case of $\mathcal{H}_{\rm TI}(\boldsymbol{k})$. The parameters are $m=2$, $t=1$, $v=0.4$,  $v'=0.5$, $v_{z}=1$, and $B_{z}=0.8$. }\label{chap8_MTI_surface}
\end{figure}

To confirm that the mirror-symmetry-protected TCI phase is realized in our model, we calculate band structures of $\mathcal{H}_{\rm MTCI}(\boldsymbol{k})$ in the slab geometries with the (100), (001), (101), and (111) surfaces [Fig.~\ref{chap8_MTI_surface}]. Figure~\ref{chap8_MTI_surface}(a) shows that the (100) surface has the gapless states because this surface is invariant under $M_{z}$.
Figures~\ref{chap8_MTI_surface}(b), \ref{chap8_MTI_surface}(c), and \ref{chap8_MTI_surface}(d) show that gapless surface states do not appear on the (001), (101), and (111) surfaces. This is because these surfaces are not invariant under $M_{z}$.  In this way, the presence or absence of the gapless surface states in the mirror-symmetry-protected TCI  depends on the surface orientation, unlike the TI protected by $\mathcal{T}$ symmetry. 

\section{Surface energy and equilibrium crystal shape of a topological insulator and a topological crystalline insulator}\label{Sec4_surface_energy_shape}

In the previous work \cite{PhysRevLett.129.046802}, we discussed equilibrium shapes of the TCIs protected by glide $G_y$ symmetry \cite{PhysRevB.90.085304, PhysRevB.91.155120, PhysRevB.91.161105}, where $G_y$ is a mirror reflection with respect to the $xz$ plane followed by translation by a half of the lattice vector along the $z$ direction, and we studied effects of their surface states on the equilibrium shape. 
As discussed in Ref.~\cite{PhysRevLett.129.046802}, the glide-symmetry-protected TCI has exotic surface states that depend on the parity of $\beta$ in the Miller index ($\alpha0\beta$). On the other hand, topological phases protected by symmorphic symmetries, such as a mirror-symmetry-protected TCI  and a SOTI protected by $\mathcal{I}$ symmetry, do not have such characteristic dependence of the boundary states on the Miller index. 
The  TCI protected by $M_z$ symmetry has gapless states on the (100) surfaces and the (010) surfaces. In contrast, when the surface is not invariant under the mirror operation, $e.g.$, the (101), (102), and (001) surfaces,  the surface does not have topologically protected gapless states.
In addition, the SOTI has the hinge states instead of the surface states,  and therefore the hinge states may lead to the effects on the equilibrium shape in a different way from the TCI.
Thus, one can expect that the mirror-symmetry-protected TCI phase and the SOTI phase lead to different contributions from the glide-symmetry-protected TCI \cite{PhysRevLett.129.046802}. 
In this section, we study the equilibrium crystal shape of the TCI protected by mirror symmetry, and in the next section we study that of the SOTI. 

\subsection{Surface energy and Wulff construction}\label{chap6_Wulff_surface_ene}
Next, we study surface energies of our model to obtain the equilibrium crystal shape.  We can calculate the surface energies of an ($hkl$) surface from the band structure in the slab geometry.  The slab has finite thickness in the direction perpendicular to the $(hkl)$ surface and has PBCs along the ($hkl$) directions. 
We introduce hoppings between the top and bottom surfaces of the slab. 
When the Hamiltonian has a hopping parameter $t'$ in the bulk, we introduce the hopping $\lambda t'$ across the top and bottom surfaces with $\lambda$ being a real parameter. When we choose $\lambda=1$, this system is equal to a bulk crystal because it has the PBC in the $[hkl]$ direction. On the other hand, when we choose $\lambda=0$, this system is just a slab geometry having the ($hkl$) surfaces. 
We define a Hamiltonian $H_{\rm slab}^{(hkl)}(\lambda)$ with such geometry, and the surface energy of $H^{(hkl)}_{\rm slab}(\lambda)$  can be defined as 
 \begin{equation}\label{chap7_eq:definition_surface_energy}
E_{\rm surf}^{(hkl)}:=\sum_{n=1}^{N} \frac{ E^{(hkl)}_{{\rm slab},n}|_{\lambda=0}-E^{(hkl)}_{{\rm slab},n}|_{\lambda=1} }{2 S^{(hkl)}},
\end{equation}
where $E^{(hkl)}_{{\rm slab},n}|_{\lambda}$ is the energy from the $n$-th occupied  band of $H^{(hkl)}_{\rm slab}(\lambda)$, $N$ is the total number of occupied bands, and $S^{(hkl)}$ represents the area of the slab surface of $H^{(hkl)}_{\rm slab}(\lambda)$. 
Note that we focus only on the energy of non-interacting electrons at zero temperature and do not consider the energies due to the electron-electron interaction and the electrostatic energies of nuclei.    

\begin{figure}
\centering
\includegraphics[width=1.\columnwidth]{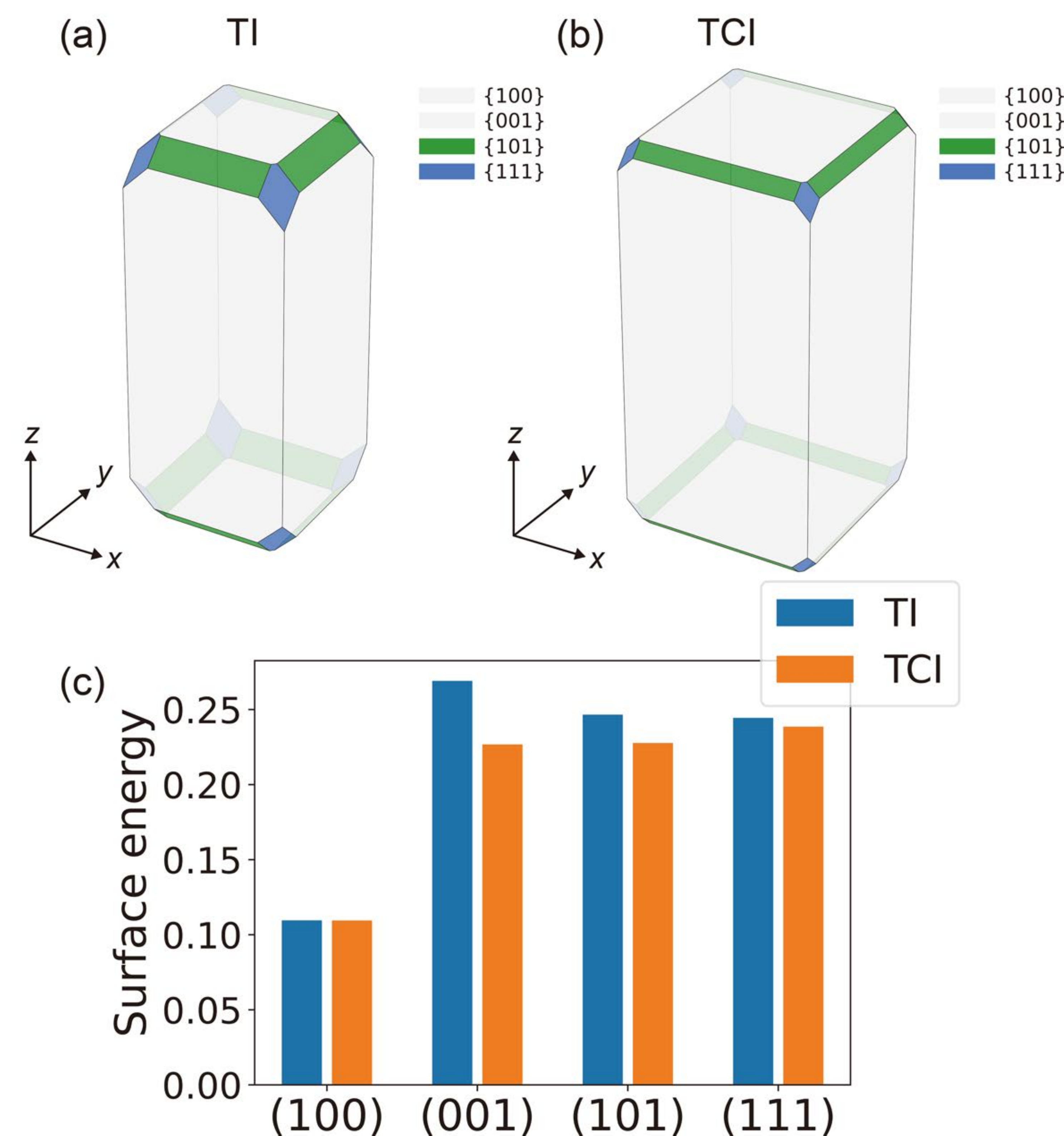}
\caption{(a), (b) Equilibrium crystal shapes obtained from the surface energies  $E_{\rm surf}^{(hkl)}$ of our models (a) $\mathcal{H}_{\rm TI}(\boldsymbol{k})$ and (b) $\mathcal{H}_{\rm MTCI}(\boldsymbol{k})$. 
(c) Surface energies of $\mathcal{H}_{\rm TI}(\boldsymbol{k})$ and $\mathcal{H}_{\rm MTCI}(\boldsymbol{k})$. The parameters are $m=2$, $t=1$, $v=0.4$, $v'=0.5$, and $v_{z}=1$. In the TI and the TCI, the  magnetic field strength is $B_z=0$ and $B_z=0.8$, respectively. }\label{chap6_shape_TI_TCI}
\end{figure}

According to the Wulff construction \cite{wulffconst}, we can obtain the equilibrium crystal shape minimizing the total surface energy by setting the distance $h_{hkl}$ between a surface with a Miller index $(hkl)$ and the crystal center to be  proportional to  $E_{\rm surf}^{(hkl)}$: 
\begin{equation}
\frac{E_{\rm surf}^{(hkl)}}{h_{hkl}} = l_{d},
\end{equation}
 where $l_{d}$ is a constant, and we set $l_{d}=1$ in the following. 
Then, the equilibrium crystal shape is given  
by the following 3D region:
 \begin{gather}
 	\mathcal{W}=\bigcap_{\boldsymbol{n}_{hkl} \in S^{2}} \Gamma_{\boldsymbol{n}_{hkl}},\label{Wulff_shape_1} \\
 	\Gamma_{\boldsymbol{n}_{hkl}}=\left\{ \boldsymbol{r} \in \mathbb{R}^{3}\ |\  \boldsymbol{r} \cdot \boldsymbol{n}_{hkl} \leq E^{(hkl)}_{\rm surf} \right\}, \label{Wulff_shape_2}
 \end{gather}
 where $\boldsymbol{n}_{hkl}$ is the outward unit normal vector to the $(hkl)$ surface, and $S^{2}$ is the unit sphere. By using the WULFFPACK  package \cite{WulffPack}, we obtain the equilibrium crystal shapes from $E_{\rm surf}^{(hkl)}$ in the following. 

\subsection{Equilibrium crystal shapes of a topological insulator and a topological crystalline insulator}  
Figures \ref{chap6_shape_TI_TCI}(a) and \ref{chap6_shape_TI_TCI}(b) show the equilibrium crystal shapes of $\mathcal{H}_{\rm TI}(\boldsymbol{k})$ and $\mathcal{H}_{\rm MTCI}(\boldsymbol{k})$, respectively.  The parameters are the same as those in the calculations of the band structure in Figs.~\ref{chap8_TI_surface}--\ref{chap8_MTI_surface}. 
Then Fig.~\ref{chap6_shape_TI_TCI}(c) shows the surface energies of the TI and the TCI; this figure indicates that the (001), (101), and (111) surface energies of the TCI are lower than those of the TI. In contrast, the (100) surface energy of the TCI is almost the same as that of the TI.  It can be seen from these surface energies that  the (001) surface of the TCI is more likely to appear than the (001) surface of the TI.  Indeed, the (001) surface of the TCI appear more extensively than that of the TI as shown in Figs.~\ref{chap6_shape_TI_TCI}(a) and \ref{chap6_shape_TI_TCI}(b). 

We can explain the difference of the crystal shapes between the TI and the TCI as follows. As we show in Fig.~\ref{chap8_MTI_surface}, the (001), (101), and (111) surfaces of the TCI do not have gapless states, while the gapless states appear on the (100) surface. The gapped surfaces make the surface energies lower,  which makes the (001), (101), and  (111)  surfaces more favorable. 

 \begin{figure}
\centering
\includegraphics[width=1.\columnwidth]{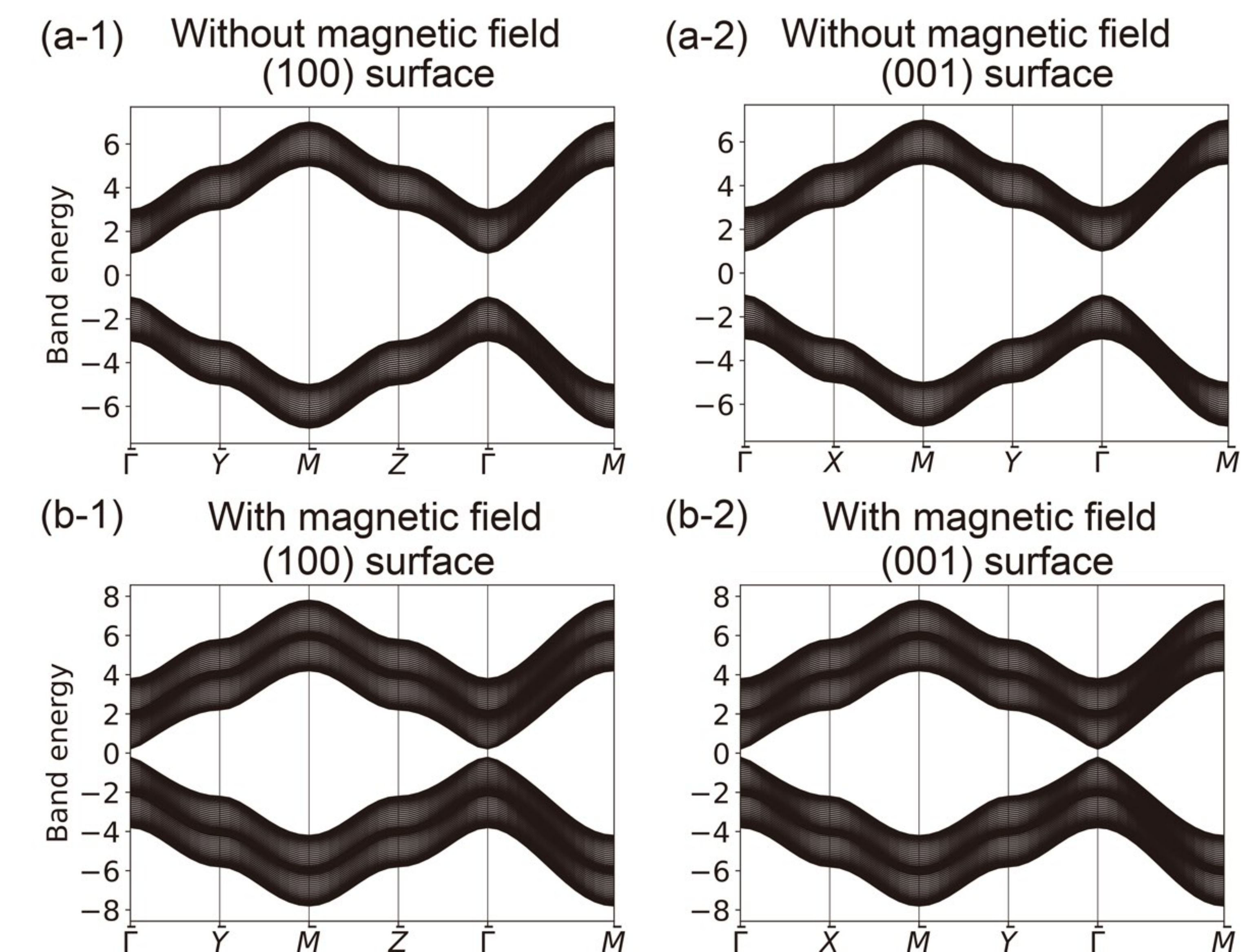}
\caption{Band structures in the slab geometries of the trivial insulators (a) without the magnetic field [$\mathcal{H}_{\rm TI}(\boldsymbol{k})$ with $m=4$] and (b) with the magnetic field [$\mathcal{H}_{\rm MTCI}(\boldsymbol{k})$ with $m=4$]. The slab geometries have the (100) surface in (a-1) and (b-1) and  the (001) surface in (a-2) and (b-2). We choose the parameter $m=4$, and the other parameters are the same as in Fig.~\ref{chap6_shape_TI_TCI}. The magnetic field strength is $B_{z}=0.8$ in (b).}\label{trivial_band}
\end{figure}

\begin{figure}
\centering
\includegraphics[width=1.\columnwidth]{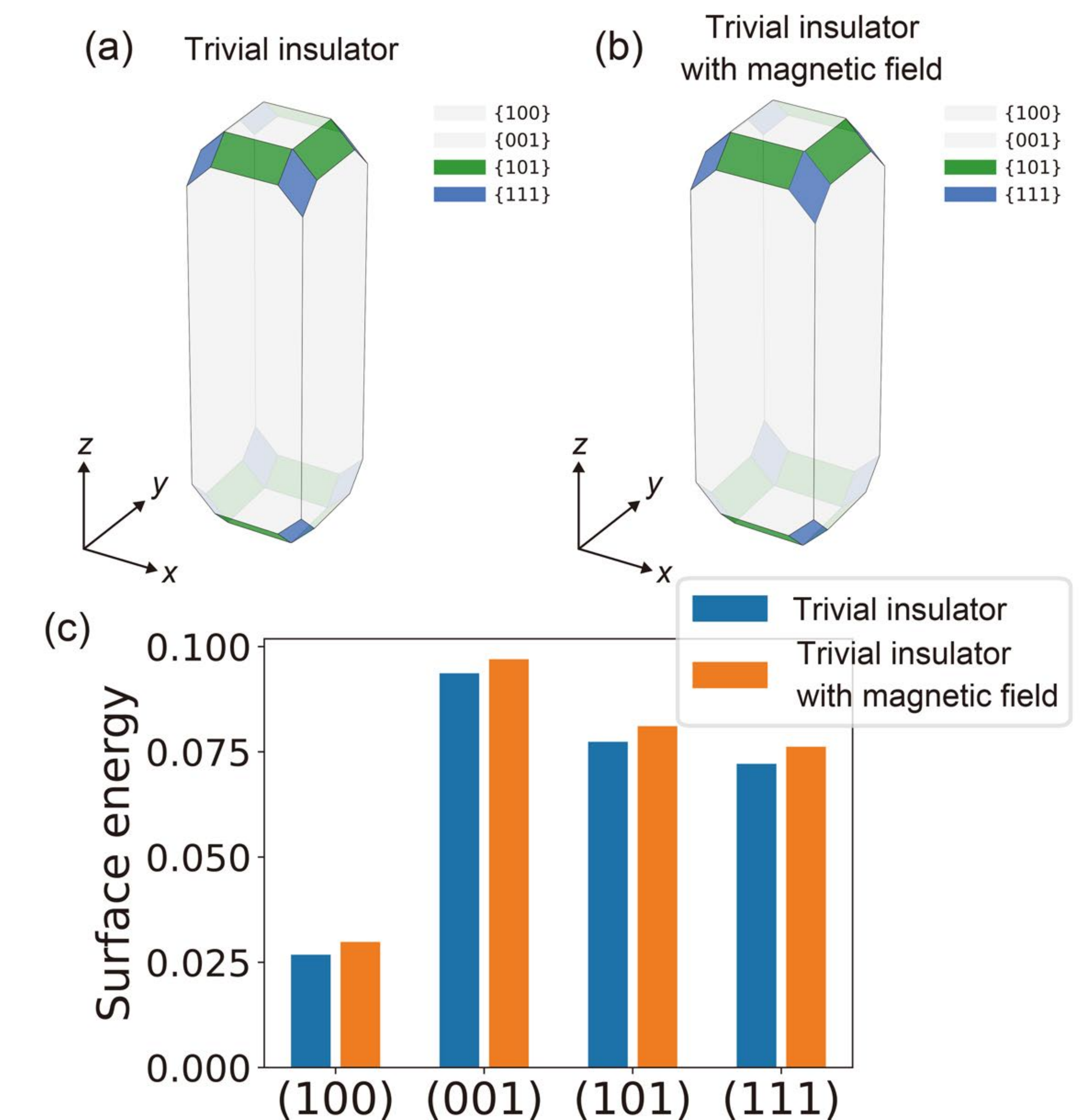}
\caption{(a), (b) Equilibrium crystal shapes of trivial insulators realized in our model (a) without the magnetic field and (b) with the magnetic field. 
(c) Surface energies of the trivial insulator with and without the magnetic field. 
The parameters of the model are the same as in Fig.~\ref{trivial_band}.
}\label{chap6_shape_trivial}
\end{figure}

To support this explanation, we also calculate the band structure and the crystal shapes of trivial insulators. 
When we choose the parameter $m=4$ in both $\mathcal{H}_{\rm TI}(\boldsymbol{k})$ and $\mathcal{H}_{\rm MTCI}(\boldsymbol{k})$ instead of $m=2$, trivial insulator phases are realized in both $\mathcal{H}_{\rm TI}(\boldsymbol{k})$ and $\mathcal{H}_{\rm MTCI}(\boldsymbol{k})$. 
This is because the change from $m=2$ to $m=4$ leads to band inversion at the $\Gamma$ point, resulting in the symmetry indicator being  $\mu_1=0$, which corresponds to the topologically trivial phase. 
Figures~\ref{trivial_band}(a) and \ref{trivial_band}(b) show the band structures of $\mathcal{H}_{\rm TI}(\boldsymbol{k})$ and $\mathcal{H}_{\rm MTCI}(\boldsymbol{k})$ with $m=4$, where the trivial insulators are realized.
From these results, we confirm that the gapless surface states do not appear in either of these cases.
Figures~\ref{chap6_shape_trivial}(a) and \ref{chap6_shape_trivial}(b) show the trivial insulators with and without the magnetic field,  and these shapes are almost the same.
This result is different  from those of the TI and the TCI in Fig.~\ref{chap6_shape_TI_TCI}.
As for the TI and the TCI, the difference of their crystal shapes originates from the magnetic field. In contrast, the magnetic field does not produce significant changes in the crystal shapes of the trivial insulator. 
 It can be seen from these results that the change in the crystal shape induced by the magnetic field in Fig.~\ref{chap6_shape_TI_TCI} is due to the topological surface state in $\mathcal{H}_{\rm TI}(\boldsymbol{k})$ and $\mathcal{H}_{\rm MTCI}(\boldsymbol{k})$ with $m=2$. 
 
Figure~\ref{chap6_shape_trivial}(c) shows the difference between the surface energies of the trivial insulators with and without the magnetic field. 
As shown in Figs.~\ref{trivial_band}(a) and \ref{trivial_band}(b), surface bands are absent in the trivial insulators, unlike the TI and the TCI. Thus, the difference in Fig.~\ref{chap6_shape_trivial}(c) does not arise from surface bands but from the changes in the bulk bands through the magnetic field. Thus, within this model, the changes in the surface energies and the crystal shapes by the magnetic field in Fig.~\ref{chap6_shape_TI_TCI} are mainly by the topological surface states because the contribution from the bulk bands is tiny.
 
\section{Surface energy and equilibrium crystal shape of a second-order topological insulator}\label{Sec:5_inversion_SOTI}
In the following, we study how the hinge states affect the crystal shape of the SOTI protected by $\mathcal{I}$ symmetry. For this purpose, to begin with, we calculate band structures of a tight-binding model of the SOTI protected by $\mathcal{I}$ symmetry
\begin{equation}
	\mathcal{H}_{\rm SOTI}(\boldsymbol{k})=\mathcal{H}_{\rm TI}(\boldsymbol{k}) + \sum_{i=x,y,z} B_{i}\sigma_{i}.
\end{equation}
This model is constructed by adding a Zeeman term to $\mathcal{H}_{\rm TI}(\boldsymbol{k})$. Unlike the model equation~(\ref{eq:mtci_tight_bind_model}) of the TCI protected by mirror symmetry, the Zeeman field is no longer along the $z$ axis.  
 The Zeeman term breaks $\mathcal{T}$, $M_{z}$, and $C_{4z}$ symmetries, but preserves $\mathcal{I}$ symmetry. Because $M_{z}$ symmetry is broken, the topological surface states protected by $M_{z}$ symmetry do not appear, unlike the mirror-symmetric TCI. 
 
 \begin{figure}
\centering
\includegraphics[width=1.\columnwidth]{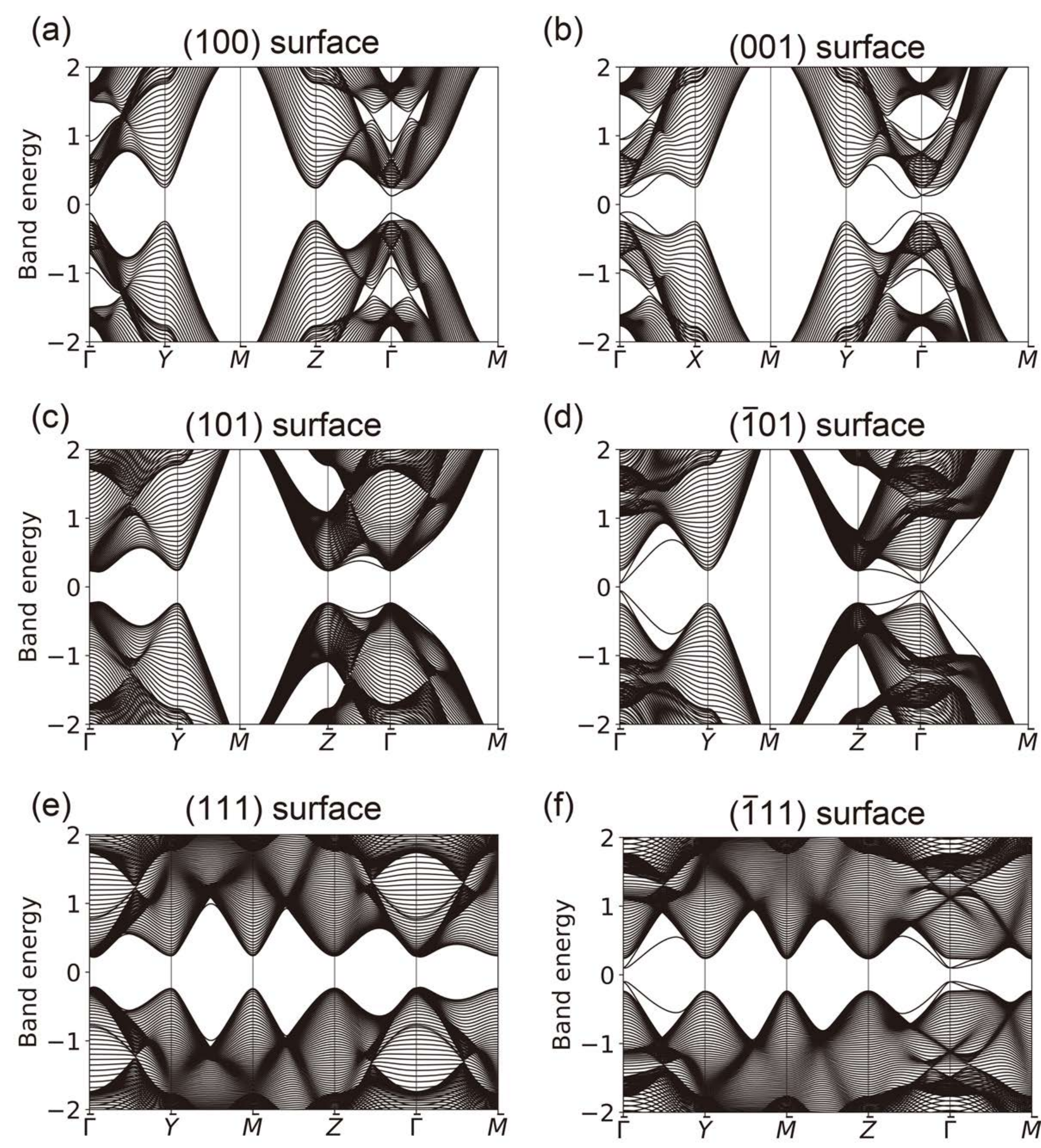}
\caption{Band structures of $\mathcal{H}_{\rm SOTI}(\boldsymbol{k})$ in the slab geometries with (a) the (100), (b) (001), (c) (101), (d) $(\bar{1}0{1})$, (e) (111), and  (f) ($\bar{1}$11) surfaces. The high-symmetry points are the same as the case of $\mathcal{H}_{\rm TI}(\boldsymbol{k})$. The parameters are $m=2$, $t=1$, $v=0.4$,  $v'=0.5$, $v_{z}=1$, $B_{x}=B_{z}=0.5$, and $B_{y}=0.3$. The thickness of the slab geometries with the ($hkl$) surface is  $30$ times the length of the vector ($h$,$k$,$l$).  }\label{chap6_SOTI_band}
\end{figure}
 
\subsection{Surfaces of a second-order topological insulator}
To see the gapped surface states of our model, we calculate band structures of $\mathcal{H}_{\rm SOTI}(\boldsymbol{k})$ in the slab geometries with various surfaces [Figs.~\ref{chap6_SOTI_band}(a)-\ref{chap6_SOTI_band}(f)]. These results show that gapless surface states do not appear. The (101) surface is not equivalent to the ($10\bar{1}$),  (011), ($0\bar{1}1$)
  or  ($0\bar{1}\bar{1}$) surfaces, and the (111) surface is not equivalent to  $(\bar{1}11)$, $(1\bar{1}1)$, or  $(\bar{1}\bar{1}1)$ because the $C_{4z}$ symmetry is broken. Therefore, we need to calculate the (100), (010), (001), (110), (1$\bar{1}$0), (101), (10$\bar{1}$), (011), (0$\bar{1}$1), (111), ($\bar{1}{1}$1), (1$\bar{1}$1), and ($\bar{1}\bar{1}$1) surfaces for our purpose. 
Figures~\ref{chap6_SOTI_band}(c) and \ref{chap6_SOTI_band}(d) show that the band structure in the slab geometry with the (101) surface is different from that with the ($\bar{1}0{1}$) surface. 
In addition, Figs.~\ref{chap6_SOTI_band}(e) and \ref{chap6_SOTI_band}(f) also show that the $(\bar{1}11)$ surface states behave differently from the (111) surface. 
These anisotropic surface states are characteristic of the SOTI and are different from those of the TI and the TCI protected by mirror symmetry. 

\begin{figure}
\centering
\includegraphics[width=1.\columnwidth]{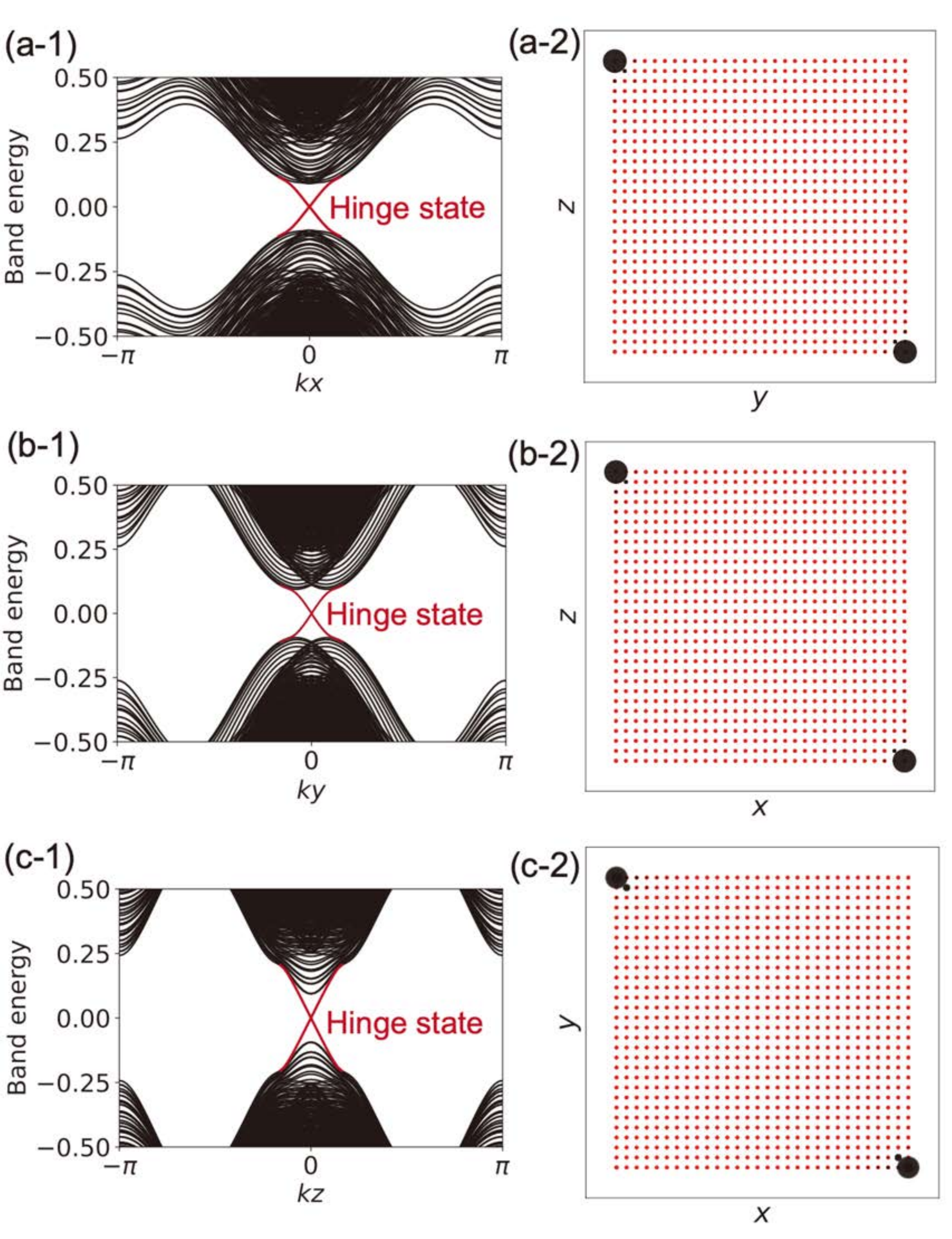}
\caption{Band structures of $\mathcal{H}_{\rm SOTI}(\boldsymbol{k})$ in the rod  geometries with the PBC in (a-1) the $x$ , (b-1) $y$, and (c-1) $z$ directions. (a-2), (b-2), (c-2) The real-space distributions of the zero-energy states, corresponding of (a-1), (b-1), and (c-1). The parameters are the same as those in Fig.~\ref{chap6_SOTI_band}. The system sizes of the rod geometries are $30\times 30$. 
}\label{chap6_hinge_band}
\end{figure}
 
 \subsection{Hinge states of a second-order topological insulator}
Next, to confirm that hinge states emerge in this model, we calculate band structures  with the PBC in one direction and with finite sizes in the other two directions. We refer to  such a geometry as a rod geometry (Fig.~\ref{chap6_hinge_band}).  Figure~\ref{chap6_hinge_band}(a-1) shows that the gapless states appear in the rod geometry with the PBC in the $x$ direction and with the finite sizes in the $y$ and $z$ directions. Figure~\ref{chap6_hinge_band}(a-2) indicates the real-space distribution of the eigenstates at $E=0$ within the $yz$ plane. From this result, we find that the hinge states appear along the $x$ direction. In addition, we obtain the band structures shown in Fig.~\ref{chap6_hinge_band}(b-1) [Fig.~\ref{chap6_hinge_band}(c-1)] and  the real-space distribution of zero-energy states shown in Fig.~\ref{chap6_hinge_band}(b-2)  [Fig.~\ref{chap6_hinge_band}(c-2)] in the rod geometry with the PBC in the $y$ ($z$) direction.
It can be seen from these results that the hinge states also appear along the $y$ and $z$ directions, at the $\mathcal{I}$-invariant positions. 

\begin{figure}
\centering
\includegraphics[width=1.\columnwidth]{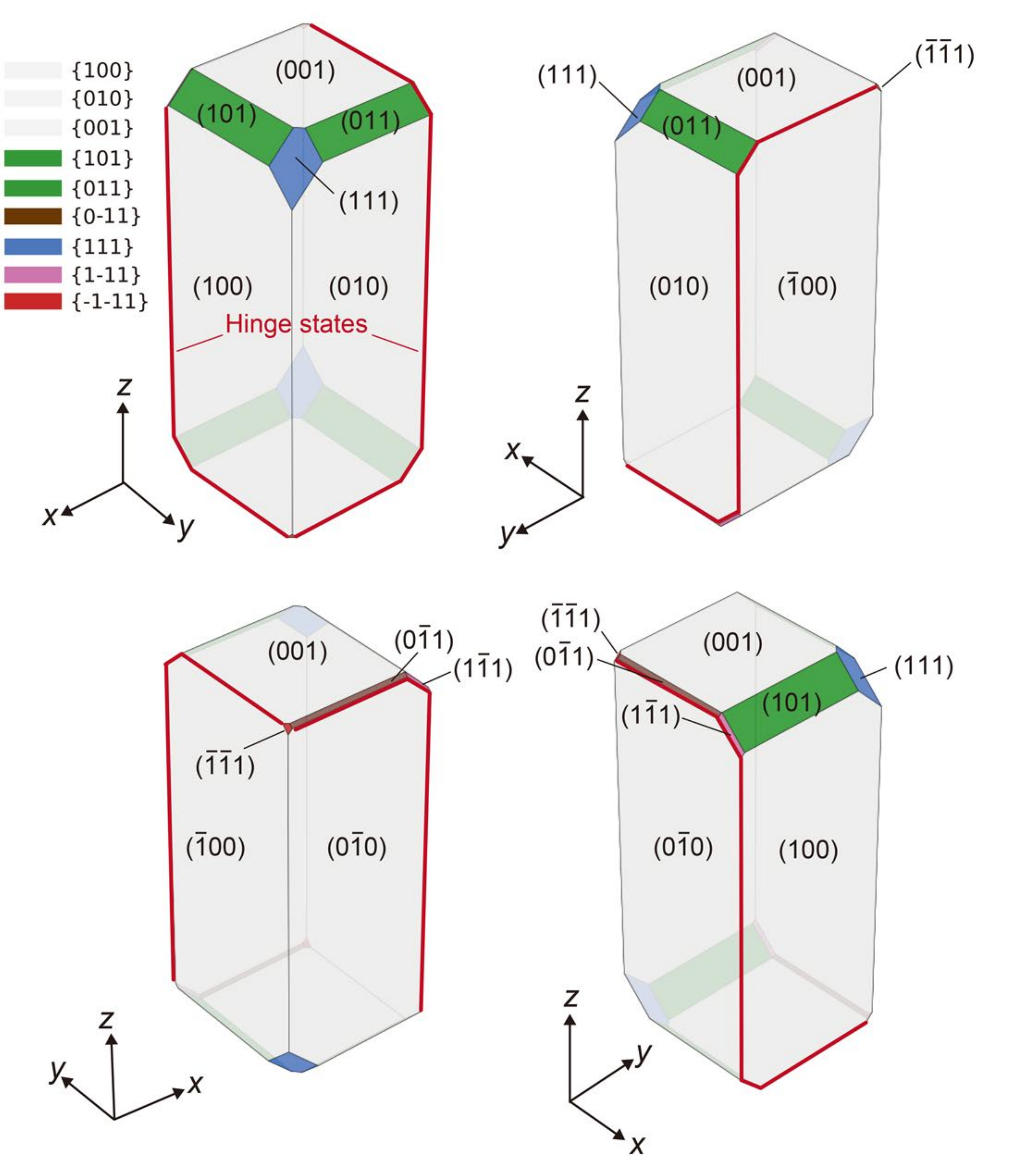}
\caption{Equilibrium crystal shape obtained from the surface energies  $E_{\rm surf}^{(hkl)}$ of our model $\mathcal{H}_{\rm SOTI}(\boldsymbol{k})$, seen from four different angles. The red lines between the surfaces indicate the positions of the hinge states. The parameters are the same as those in Fig.~\ref{chap6_SOTI_band}.}\label{chap6_hoti_shape}
\end{figure}

\begin{figure*}
\includegraphics[width=2.\columnwidth]{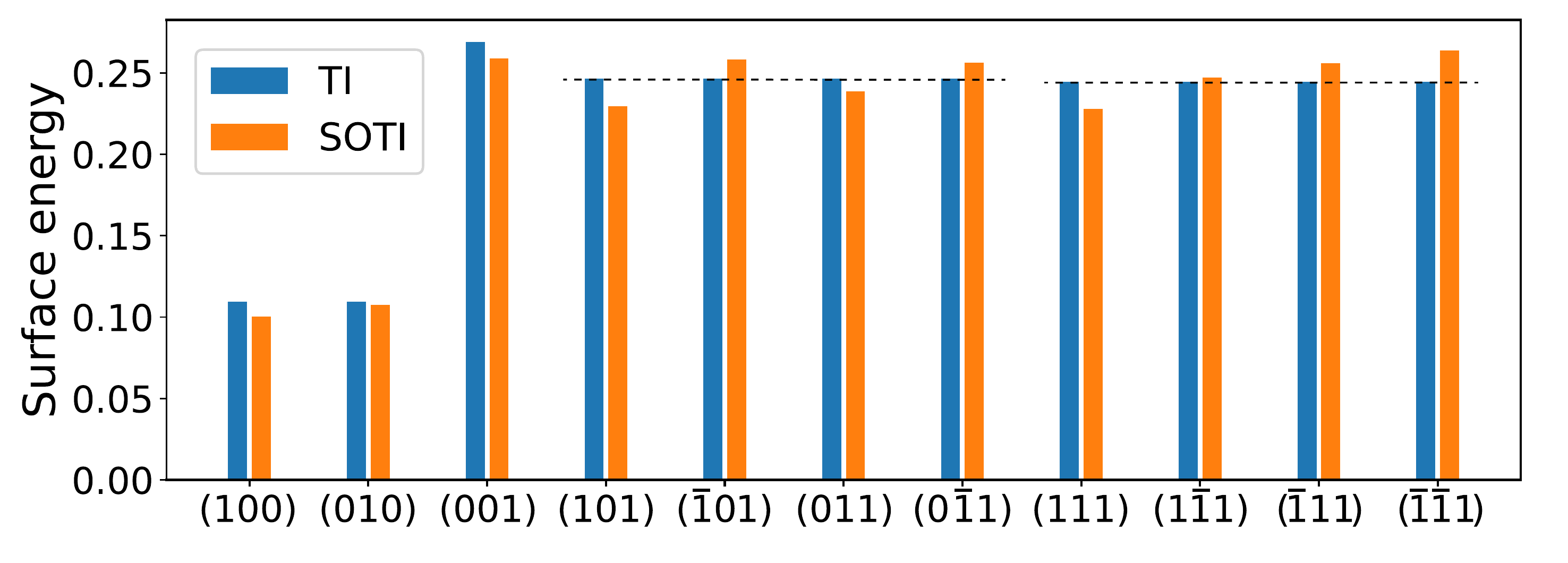}
\caption{The surface energies of $\mathcal{H}_{\rm TI}(\boldsymbol{k})$ and $\mathcal{H}_{\rm SOTI}(\boldsymbol{k})$. The parameters are $m=2$, $t=1$, $v=0.4$, $v'=0.5$, and $v_{z}=1$. The magnetic field strength in $\mathcal{H}_{\rm SOTI}(\boldsymbol{k})$ is given by $B_{x}=B_{z}=0.5$, and $B_{y}=0.3$. The thickness of the slab geometries with the ($hkl$) surface is  $30$ times the vector ($h,k,l$). }\label{chap6_comparison_surface_energy_TI_SOTI}
\end{figure*}
 
\subsection{Equilibrium crystal shape of a SOTI} 
By using a similar method to that used with the TI and the TCI, we obtain an equilibrium crystal shape of the SOTI from the surface energies $E_{\rm surf}^{(hkl)}$ defined in Eq.~(\ref{chap7_eq:definition_surface_energy}).  Figure~\ref{chap6_hoti_shape} shows the equilibrium crystal shape of the SOTI. As we described above, the (101) surface energy  is different from the ($\bar{1}$01) surface energy, unlike the TI and the TCI in the previous section. This difference in the surface energies results in the presence of the (101) surface and the absence of  the ($\bar{1}$01)  surface.  In addition, the (011) surface energy is also different from that of the (0$\bar{1}$1) surface energy, resulting in the emergence of the (011) surface being more extensive than the emergence of the (0$\bar{1}$1) surface. Such anisotropy in how the facets appear also occurs for the (111), ($\bar{1}$11) (1$\bar{1}$1), and ($\bar{1}\bar{1}$1) surfaces, as shown in Fig.~\ref{chap6_hoti_shape}.  

In order to understand such an anisotropic equilibrium shape of the SOTI with $\mathcal{I}$ symmetry, we focus on the following surface theory. 
 We introduce the surface Dirac Hamiltonian \cite{PhysRevB.97.205136, PhysRevResearch.2.043274}
\begin{equation}\label{eq:surf_dirac_Hamil}
	\mathcal{H}_{\rm Dirac}=v_s(\boldsymbol{k}_{s}\times \boldsymbol{n})\cdot \boldsymbol{\sigma} + m_{\boldsymbol{n}}\boldsymbol{n} \cdot \boldsymbol{\sigma},
\end{equation}
where $\boldsymbol{n}$ is the normal vector of the surface, $\boldsymbol{k}_s$ is the wavevector parallel to the surface, and $m_{\boldsymbol{n}}$ is the surface Dirac mass.  Here, we assume that the Dirac mass is determined uniquely by $\boldsymbol{n}$.   
Figure~\ref{chap6_hoti_shape} shows the positions of the hinge states, which appear between the surfaces with a positive Dirac mass $m_{\boldsymbol{n}}$ and the surfaces with a negative $m_{\boldsymbol{n}}$. 
For example, the (100), (010), and (001) surfaces have a positive $m_{\boldsymbol{n}}$  because these surfaces are penetrated by the outward pointing magnetic field, while the ($\bar{1}$00), (0$\bar{1}$0), and (00$\bar{1}$) surfaces have a negative $m_{\boldsymbol{n}}$. 
The presence of surfaces with a positive $m_{\boldsymbol{n}}$ and surfaces with a negative $m_{\boldsymbol{n}}$ leads to the vanishing of $m_{\boldsymbol{n}}$ between these surfaces, which corresponds to the emergence of hinge states \cite{PhysRevB.97.205136}. This behavior of $m_{\boldsymbol{n}}$ makes surface energies higher between the surfaces with a positive $m_{\boldsymbol{n}}$ and the surfaces with a negative $m_{\boldsymbol{n}}$, which makes this surface not likely to appear. Actually,  Fig.~\ref{chap6_hoti_shape} shows that such surfaces are less likely to appear. For instance, the ($\bar{1}$01) surface, which is between the (001) surface (positive $m_{\boldsymbol{n}}$) and the ($\bar{1}$00) surface (negative $m_{\boldsymbol{n}}$), does not emerge in Fig.~\ref{chap6_hoti_shape}. 
In this way, the behaviors of the Dirac mass can explain the anisotropic crystal shape of the SOTI. 

To support the above explanation of the crystal shape of the SOTI in terms of the hinge states and the Dirac mass, we show a comparison of the surface energies between the TI and the SOTI (Fig.~\ref{chap6_comparison_surface_energy_TI_SOTI}).  We find that the (101), (011), and (111) surface energies of the SOTI are lower than those of the TI. In contrast, the ($\bar{1}$01), (0$\bar{1}$1), ($\bar{1}$11), (1$\bar{1}$1), and ($\bar{1}\bar{1}$1) surfaces have higher surface energies than those of the TI. These surfaces are between the surfaces with a positive $m_{\boldsymbol{n}}$ and the surfaces with a negative $m_{\boldsymbol{n}}$. This result is consistent with our explanation of the crystal shape of the SOTI based on the behaviors of  $m_{\boldsymbol{n}}$ and the hinge states. 

In addition, we demonstrate through a more quantitative analysis that the boundary states of the SOTI lead to these anisotropic surface energies. For our purpose, we consider the ($hkl$) surface with the surface normal vector $\boldsymbol{n}=(r\sin \theta \cos \phi, r \sin \theta \sin \phi, r \cos \theta)$ and focus on the dependence of the surface energy on $\phi$ with $r$ and $\theta$ being fixed.  
 We choose $\theta=\pi/4$ to study the $(h01)$ and $(0k1)$ surfaces ($h,k=1$ or $\bar{1}$).
We also focus on the surface energies $E^{(hkl)}_{\rm surf}$ with $E^{({1}01)}_{\rm surf}$ as the reference value:
\begin{equation}\label{dif:delta_E}
	\Delta E^{(hkl)}_{({1}01)}:=E^{(hkl)}_{\rm surf}-E^{({1}01)}_{\rm surf}.
\end{equation} 
As discussed in the Appendix, $\Delta E^{(hkl)}_{({1}01)}$ can be approximated by
\begin{gather}\label{dif:delta_E_eff_aprox}
\Delta E^{(hkl)}_{({1}01)}\simeq \frac{E^{(hkl)}_{\rm Dirac}}{2S^{(hkl)}}-\frac{E^{(10{1})}_{\rm Dirac}}{2S^{(10{1})}},
\end{gather}
for the $(h01)$ and $(0k1)$ surfaces ($h,k=1$ or $\bar{1}$), 
where $E^{(hkl)}_{\rm Dirac}$ is the energy obtained from the surface Dirac Hamiltonian $\mathcal{H}_{\rm Dirac}(\boldsymbol{k})$ and given by
\begin{gather}\label{concrete_form_E_dirac}
E^{(hkl)}_{\rm Dirac}:= \frac{2 \lambda}{3v^2_s k_c^2 }\Bigl( |m_{\boldsymbol{n}}|^3-(v_s^2 k_c^2+ m^2_{\boldsymbol{n}})^{\frac{3}{2}}\Bigr), 
\end{gather}
where 
$
m_{\boldsymbol{n}}:=\boldsymbol{B}\cdot \boldsymbol{n}/ |\boldsymbol{n}|$, and  $\lambda$ and $k_c$ are real parameters. 

\begin{figure}[b]
\includegraphics[width=1.\columnwidth]{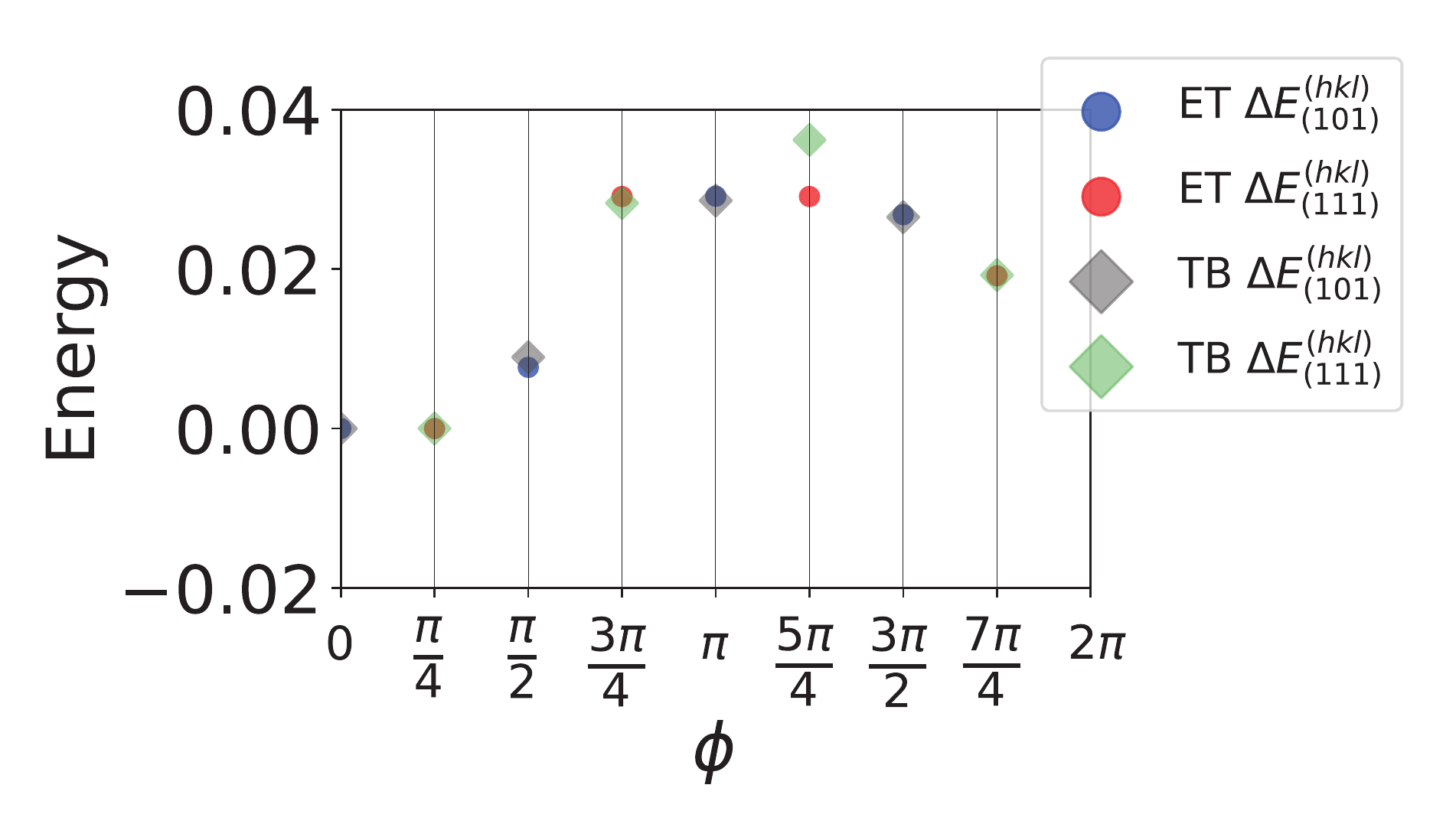}
\caption{$\Delta E^{(hkl)}_{({1}01)}$ and $\Delta E^{(hkl)}_{({1}{1}1)}$ obtained from the tight-binding model (TB) $\mathcal{H}_{\rm SOTI}(\boldsymbol{k})$ and from the effective theory (ET) [Eq.~(\ref{dif:delta_E_eff_aprox})].  The parameters in $\mathcal{H}_{\rm SOTI}(\boldsymbol{k})$ are the same as those in Fig.~\ref{chap6_SOTI_band}. The parameters in ET are $\lambda=1$, $v_s=1$. We choose $\theta=\pi/4$ and $k_c=0.41$ for $\Delta E^{(hkl)}_{({1}01)}$ and $\theta=\cos^{-1} ({1}/{\sqrt{3}})=\sin^{-1} (\sqrt{2/3})$  and   $k_c=0.47$  for $\Delta E^{(hkl)}_{({1}{1}1)}$.}\label{eff_surf_ene}
\end{figure}

Figure~\ref{eff_surf_ene} shows $\Delta E^{(hkl)}_{({1}01)}$ obtained from the direct calculations of the tight-binding model $\mathcal{H}_{\rm SOTI}(\boldsymbol{k})$. Figure~\ref{eff_surf_ene} also shows that 
Eq.~(\ref{dif:delta_E_eff_aprox}) obtained from our effective surface theory is in good agreement with the results from the tight-binding model.   
Furthermore, to analyze the $(hk1)$ surface energies ($h,k=1$ or $\bar{1}$), we introduce $\Delta E^{(hkl)}_{{1}{1}1}$ defined similarly to $\Delta E^{(hkl)}_{({1}01)}$ as  $\Delta E^{(hkl)}_{({1}{1}1)}= E^{(hkl)}_{\rm surf}-E^{({1}{1}1)}_{\rm surf}$. Figure~\ref{eff_surf_ene} also indicates $\Delta E^{(hkl)}_{({1}{1}1)}$ obtained from the tight-binding model $\mathcal{H}_{\rm SOTI}(\boldsymbol{k})$, and we find that these results are almost identical to our effective surface theory $\Delta E^{(hkl)}_{({1}{1}1)}\simeq E^{(hkl)}_{\rm Dirac}/(2S^{(hkl)})-E^{({1}{1}1)}_{\rm Dirac}/(2S^{({1}{1}1)})$. 
Thus, we conclude that the dependence of the surface energies on the angle $\phi$ can be understood in terms of the mass term in the surface Dirac Hamiltonian. 
In Fig.~\ref{eff_surf_ene}, the results are slightly different between the effective theory and the tight-binding model because the effective theory is just an approximate theory. In the tight-binding model, the strength of the coupling between the magnetic field and the spin differs in the $x$ and $z$ directions, while the coupling in the effective theory is isotropic in both of these directions.

As discussed above, the surface energies $E^{(hkl)}_{\rm surf}$ of $\mathcal{H}_{\rm SOTI}(\boldsymbol{k})$ are so different between $\phi$ and $\phi+\pi$ that they affect the equilibrium shape. This is because the Dirac mass $m_{\boldsymbol{n}}$ behaves in an anisotropic manner in the SOTI. The glide-symmetry-protected TCI discussed in the previous work \cite{PhysRevLett.129.046802} does not have such a dependence of the surface energy on $\phi$. Thus, this is a clear difference between the  $\mathcal{I}$-symmetry-protected SOTI and the glide-symmetry-protected TCI in the effects of the boundary states on the equilibrium shapes. 

\section{Conclusion and discussion}\label{Sec6:conclusion_discussion}
In this paper, we study surface energies and equilibrium crystal shapes of a topological insulator, a topological crystalline insulator, and a second-order topological insulator. 
To begin with, we calculate surface states and surface energies of the topological insulator.  We obtain the equilibrium crystal shape of the topological insulator from the surface energies by using the Wulff construction. Next, we study surface states and surface energies of a topological crystalline insulator protected by mirror symmetry, which can be  realized by adding a magnetic field to the topological insulator without breaking mirror symmetry.  The topological surface states of the topological crystalline insulator appear only on specific surfaces which are invariant under the mirror operation. By comparing the crystal shape of the topological insulator with that of the topological crystalline insulator, we discover that the presence and the absence of the topological surface states affect their crystal shapes. In addition, we find that this effect does not occur in trivial insulators, and  this effect is unique to the topological insulator and the topological crystalline insulator. In addition to the topological crystalline insulator, we discuss surface energies of a second-order topological insulator, which can also be realized by adding a magnetic field to the topological insulator while breaking mirror symmetry and preserving inversion symmetry. We obtain an equilibrium crystal shape of the second-order topological insulator from the surface energies and unveil that the hinge states of the second-order topological insulator affect its crystal shape. The hinge states make specific surface energies higher, and the surfaces are less likely to appear than the other surfaces. 
This leads to a more anisotropic crystal shape of the second-order topological insulator than the topological insulator and the topological crystalline insulator. 

In this paper, we discussed the surface energies and the equilibrium crystal shapes when we applied  magnetic fields to a topological insulator. In our model, the Zeeman effects are of the same order of magnitude as the  hopping amplitudes. In real materials, the coupling between spin and the magnetic fields is very small, and it might be difficult to experimentally observe the effect of the magnetic field in a topological insulator. On the other hand, magnetic dopants can open a sizable surface band gap of a topological insulator, compared with the external magnetic field \cite{chen2010massive, lee2015imaging, li2019intrinsic}. One example of such magnetic topological insulators is Fe-doped ${\rm Bi}_2{\rm Se}_3$, and a massive Dirac cone  has been observed with the size of the band gap being approximately 50 meV \cite{chen2010massive}. When the surface band gap induced by magnetic dopants  becomes non-negligible in comparison to the hopping amplitudes, the magnetic topological insulator is suitable for experiments to confirm our theory. 
Lastly, we comment on the stability of the edges of graphene that our results might suggest. 
Graphene has edge states at the zigzag edges, while the edge states are absent on the armchair edges \cite{doi:10.1143/JPSJ.65.1920}. 
According to Refs.~\cite{PhysRevB.81.125445, doi:10.1073/pnas.1207519109}, the armchair edges are generally stabler than the zigzag edges. 
Thus, while various contributions other than the edge states affect the edge stability, the absence of the edge states might partially contribute to the stability of the armchair edges of graphene.

\begin{acknowledgments}
This work was supported by Japan Society for the Promotion of Science (JSPS) KAKENHI Grants No.~JP21J22264, No.~JP22K18687, and No.~JP22H00108, and by MEXT Initiative to Establish Next-Generation Novel Integrated Circuits Centers (X-NICS) Grant No. JPJ011438.
\end{acknowledgments}

\appendix

\section{Effective surface theory in term of Dirac Hamiltonian}\label{app:est}
In this appendix, we show that  $\Delta E^{(hkl)}_{({1}01)}$ defined by Eq.~(\ref{dif:delta_E}) can be approximated by Eq.~(\ref{dif:delta_E_eff_aprox}).
Note that we focus on  the $(h01)$ and $(0k1)$ surfaces ($h,k=1$ or $\bar{1}$). From the definition of $E^{(hkl)}_{\rm surf}$ [Eq.~(\ref{chap7_eq:definition_surface_energy})], we get the following equation:
\begin{widetext}
\begin{align}
	\Delta E^{(hkl)}_{({1}01)}
	=&\sum_{n=1}^{N-N_{\rm surf}}\Biggl( \frac{E^{(hkl)}_{{\rm slab},n}|_{\lambda=0}-E^{(hkl)}_{{\rm slab},n}|_{\lambda=1}}{2S^{(hkl)}}-\frac{E^{({1}01)}_{{\rm slab},n}|_{\lambda=0}-E^{({1}01)}_{{\rm slab},n}|_{\lambda=1}}{2S^{({1}01)}} \Biggr)\nonumber \\
	 +& \sum_{n=N-N_{\rm surf}+1}^{N}\Biggl(
	 \frac{E^{(hkl)}_{{\rm slab},n}|_{\lambda=0}-E^{(hkl)}_{{\rm slab},n}|_{\lambda=1}}{2S^{(hkl)}}-\frac{E^{({1}01)}_{{\rm slab},n}|_{\lambda=0}-E^{({1}01)}_{{\rm slab},n}|_{\lambda=1}}{2S^{({1}01)}}
	 \Biggr),
\end{align}
\end{widetext}
where $N_{\rm surf}$ is the number of occupied bands forming the surface states, $n=1,2,\cdots N-N_{\rm surf}$ corresponds to bulk states, and $n=N-N_{\rm surf}+1,\cdots N$ corresponds to surface states. 
Here we make an approximation 
\begin{align}
	&E^{(hkl)}_{{\rm slab},n}|_{\lambda=0} \simeq E^{({1}01)}_{{\rm slab},n}|_{\lambda=0}\ \ (n=1,2,\cdots, N-N_{\rm surf}), \nonumber  \\
	&E^{(hkl)}_{{\rm slab},n}|_{\lambda=1} \simeq E^{({1}01)}_{{\rm slab},n}|_{\lambda=1}\ \ (n=1,2,\cdots, N), 
\end{align}
for the $(h01)$ and $(0k1)$ surfaces ($h,k=1$ or $\bar{1}$) because the hopping amplitudes in the $[h01]$ and $[0k1]$ directions are the same in our model, and we neglect the effect of the magnetic field. Thus, $\Delta E^{(hkl)}_{({1}01)}$ can be rewritten as a sum of contributions from surface states 
\begin{align}\label{app_approx_delta_sur}
	\Delta E^{(hkl)}_{({1}01)}
	\simeq \sum_{n=N-N_{\rm surf}+1}^{N}\Biggl( \frac{E^{(hkl)}_{{\rm slab},n}|_{\lambda=0}}{2S^{(hkl)}} - \frac{E^{({1}01)}_{{\rm slab},n}|_{\lambda=0}}{2S^{({1}01)}}
	 \Biggr),
\end{align}
for the $(h01)$ and $(0k1)$ surfaces ($h,k=1$ or $\bar{1}$). 
Also, we can expect that the right-hand side of Eq.~(\ref{app_approx_delta_sur}) is also determined mainly by the surface band structure around $\boldsymbol{k}=0$. 
Furthermore, we assume the energy bands forming the surface states around $\boldsymbol{k}=0$ can be described by the surface Dirac  Hamiltonian [Eq.~(\ref{eq:surf_dirac_Hamil})]. Thus, we get $\Delta E^{(hkl)}_{({1}01)}\simeq E^{(hkl)}_{\rm Dirac}/(2S^{(hkl)})-E^{({1}01)}_{\rm Dirac}/(2S^{({1}01)})$, where $E^{(hkl)}_{\rm Dirac}$ is defined by
\begin{equation}\label{app:eq:ed_dirac}
	E^{(hkl)}_{\rm Dirac}:=\frac{\lambda}{2A}\int^{2\pi}_{0}d\theta ' \int^{k_c}_{0}k dk E_{-}(\boldsymbol{k}),
\end{equation}
where we use the polar coordinates $(k, \theta')$ for the $k$-plane on the surface, $A$ is the area of the integral range $A=k_c^2 \pi$, and $E_{-}(\boldsymbol{k})$ is the negative eigenvalue of the Dirac Hamiltonian $\mathcal{H}_{\rm Dirac}(\boldsymbol{k})$: $E_{-}(\boldsymbol{k})=-\sqrt{v_s^2k^2+m_{\boldsymbol{n}}^2}$. 
Here $k_c$ is a cutoff wavevector for the Dirac cone. 
By performing the integral of Eq.~(\ref{app:eq:ed_dirac}), we find that $E^{(hkl)}_{\rm Dirac}$ is given by Eq.~(\ref{concrete_form_E_dirac}), and therefore we get Eq.~(\ref{dif:delta_E_eff_aprox}) in the main text. 


\begin{thebibliography}{113}%
\makeatletter
\providecommand \@ifxundefined [1]{%
 \@ifx{#1\undefined}
}%
\providecommand \@ifnum [1]{%
 \ifnum #1\expandafter \@firstoftwo
 \else \expandafter \@secondoftwo
 \fi
}%
\providecommand \@ifx [1]{%
 \ifx #1\expandafter \@firstoftwo
 \else \expandafter \@secondoftwo
 \fi
}%
\providecommand \natexlab [1]{#1}%
\providecommand \enquote  [1]{``#1''}%
\providecommand \bibnamefont  [1]{#1}%
\providecommand \bibfnamefont [1]{#1}%
\providecommand \citenamefont [1]{#1}%
\providecommand \href@noop [0]{\@secondoftwo}%
\providecommand \href [0]{\begingroup \@sanitize@url \@href}%
\providecommand \@href[1]{\@@startlink{#1}\@@href}%
\providecommand \@@href[1]{\endgroup#1\@@endlink}%
\providecommand \@sanitize@url [0]{\catcode `\\12\catcode `\$12\catcode
  `\&12\catcode `\#12\catcode `\^12\catcode `\_12\catcode `\%12\relax}%
\providecommand \@@startlink[1]{}%
\providecommand \@@endlink[0]{}%
\providecommand \url  [0]{\begingroup\@sanitize@url \@url }%
\providecommand \@url [1]{\endgroup\@href {#1}{\urlprefix }}%
\providecommand \urlprefix  [0]{URL }%
\providecommand \Eprint [0]{\href }%
\providecommand \doibase [0]{https://doi.org/}%
\providecommand \selectlanguage [0]{\@gobble}%
\providecommand \bibinfo  [0]{\@secondoftwo}%
\providecommand \bibfield  [0]{\@secondoftwo}%
\providecommand \translation [1]{[#1]}%
\providecommand \BibitemOpen [0]{}%
\providecommand \bibitemStop [0]{}%
\providecommand \bibitemNoStop [0]{.\EOS\space}%
\providecommand \EOS [0]{\spacefactor3000\relax}%
\providecommand \BibitemShut  [1]{\csname bibitem#1\endcsname}%
\let\auto@bib@innerbib\@empty
\bibitem [{\citenamefont {Wulff}(1901)}]{wulffconst}%
  \BibitemOpen
  \bibfield  {author} {\bibinfo {author} {\bibfnamefont {G.}~\bibnamefont
  {Wulff}},\ }\bibfield  {title} {\bibinfo {title} {{Zur Frage der
  Geschwindigkeit des Wachstums und der Aufl\"osung der Krystallflagen}},\
  }\href@noop {} {\bibfield  {journal} {\bibinfo  {journal} {Z. Kristallogr.}\
  }\textbf {\bibinfo {volume} {34}},\ \bibinfo {pages} {449} (\bibinfo {year}
  {1901})}\BibitemShut {NoStop}%
\bibitem [{\citenamefont {von Laue}(1943)}]{wulffconst2}%
  \BibitemOpen
  \bibfield  {author} {\bibinfo {author} {\bibfnamefont {M.}~\bibnamefont {von
  Laue}},\ }\bibfield  {title} {\bibinfo {title} {{Der Wulffsche Satz f\"ur die
  Gleichgewichtsform von Kristallen}},\ }\href@noop {} {\bibfield  {journal}
  {\bibinfo  {journal} {Z. Kristallogr.}\ }\textbf {\bibinfo {volume} {105}},\
  \bibinfo {pages} {124} (\bibinfo {year} {1943})}\BibitemShut {NoStop}%
\bibitem [{\citenamefont {Dinghas}(1944)}]{wulffconst3}%
  \BibitemOpen
  \bibfield  {author} {\bibinfo {author} {\bibfnamefont {A.}~\bibnamefont
  {Dinghas}},\ }\bibfield  {title} {\bibinfo {title} {{Uber einen geometrischen
  Satz von Wulff f\"ur die gleichgewichtsform von Kristallen}},\ }\href@noop {}
  {\bibfield  {journal} {\bibinfo  {journal} {Z. Kristallogr.}\ }\textbf
  {\bibinfo {volume} {105}},\ \bibinfo {pages} {304} (\bibinfo {year}
  {1944})}\BibitemShut {NoStop}%
\bibitem [{\citenamefont {Herring}(1951)}]{PhysRev.82.87}%
  \BibitemOpen
  \bibfield  {author} {\bibinfo {author} {\bibfnamefont {C.}~\bibnamefont
  {Herring}},\ }\bibfield  {title} {\bibinfo {title} {{Some Theorems on the
  Free Energies of Crystal Surfaces}},\ }\href
  {https://doi.org/10.1103/PhysRev.82.87} {\bibfield  {journal} {\bibinfo
  {journal} {Phys. Rev.}\ }\textbf {\bibinfo {volume} {82}},\ \bibinfo {pages}
  {87} (\bibinfo {year} {1951})}\BibitemShut {NoStop}%
\bibitem [{\citenamefont {Marks}(1994)}]{Marks_1994}%
  \BibitemOpen
  \bibfield  {author} {\bibinfo {author} {\bibfnamefont {L.~D.}\ \bibnamefont
  {Marks}},\ }\bibfield  {title} {\bibinfo {title} {{Experimental studies of
  small particle structures}},\ }\href
  {https://doi.org/10.1088/0034-4885/57/6/002} {\bibfield  {journal} {\bibinfo
  {journal} {Rep. Prog. Phys.}\ }\textbf {\bibinfo {volume} {57}},\ \bibinfo
  {pages} {603} (\bibinfo {year} {1994})}\BibitemShut {NoStop}%
\bibitem [{\citenamefont {Xia}\ \emph {et~al.}(2009)\citenamefont {Xia},
  \citenamefont {Xiong}, \citenamefont {Lim},\ and\ \citenamefont
  {Skrabalak}}]{Xia_2009}%
  \BibitemOpen
  \bibfield  {author} {\bibinfo {author} {\bibfnamefont {Y.}~\bibnamefont
  {Xia}}, \bibinfo {author} {\bibfnamefont {Y.}~\bibnamefont {Xiong}}, \bibinfo
  {author} {\bibfnamefont {B.}~\bibnamefont {Lim}},\ and\ \bibinfo {author}
  {\bibfnamefont {S.}~\bibnamefont {Skrabalak}},\ }\bibfield  {title} {\bibinfo
  {title} {{Shape-Controlled Synthesis of Metal Nanocrystals: Simple Chemistry
  Meets Complex Physics?}},\ }\href
  {https://doi.org/https://doi.org/10.1002/anie.200802248} {\bibfield
  {journal} {\bibinfo  {journal} {Angew. Chem. Int. Ed.}\ }\textbf {\bibinfo
  {volume} {48}},\ \bibinfo {pages} {60} (\bibinfo {year} {2009})}\BibitemShut
  {NoStop}%
\bibitem [{\citenamefont {Barmparis}\ \emph {et~al.}(2015)\citenamefont
  {Barmparis}, \citenamefont {Lodziana}, \citenamefont {Lopez},\ and\
  \citenamefont {Remediakis}}]{barmparis2015nanoparticle}%
  \BibitemOpen
  \bibfield  {author} {\bibinfo {author} {\bibfnamefont {G.~D.}\ \bibnamefont
  {Barmparis}}, \bibinfo {author} {\bibfnamefont {Z.}~\bibnamefont {Lodziana}},
  \bibinfo {author} {\bibfnamefont {N.}~\bibnamefont {Lopez}},\ and\ \bibinfo
  {author} {\bibfnamefont {I.~N.}\ \bibnamefont {Remediakis}},\ }\bibfield
  {title} {\bibinfo {title} {{Nanoparticle shapes by using Wulff constructions
  and first-principles calculations}},\ }\href
  {https://www.beilstein-journals.org/bjnano/articles/6/35} {\bibfield
  {journal} {\bibinfo  {journal} {Beilstein J. Nanotechnol.}\ }\textbf
  {\bibinfo {volume} {6}},\ \bibinfo {pages} {361} (\bibinfo {year}
  {2015})}\BibitemShut {NoStop}%
\bibitem [{\citenamefont {Sun}\ and\ \citenamefont {Xia}(2002)}]{sun2002shape}%
  \BibitemOpen
  \bibfield  {author} {\bibinfo {author} {\bibfnamefont {Y.}~\bibnamefont
  {Sun}}\ and\ \bibinfo {author} {\bibfnamefont {Y.}~\bibnamefont {Xia}},\
  }\bibfield  {title} {\bibinfo {title} {{Shape-controlled synthesis of gold
  and silver nanoparticles}},\ }\href
  {https://www.science.org/doi/full/10.1126/science.1077229} {\bibfield
  {journal} {\bibinfo  {journal} {Science}\ }\textbf {\bibinfo {volume}
  {298}},\ \bibinfo {pages} {2176} (\bibinfo {year} {2002})}\BibitemShut
  {NoStop}%
\bibitem [{\citenamefont {Bratlie}\ \emph {et~al.}(2007)\citenamefont
  {Bratlie}, \citenamefont {Lee}, \citenamefont {Komvopoulos}, \citenamefont
  {Yang},\ and\ \citenamefont {Somorjai}}]{Bratlie:2007vf}%
  \BibitemOpen
  \bibfield  {author} {\bibinfo {author} {\bibfnamefont {K.~M.}\ \bibnamefont
  {Bratlie}}, \bibinfo {author} {\bibfnamefont {H.}~\bibnamefont {Lee}},
  \bibinfo {author} {\bibfnamefont {K.}~\bibnamefont {Komvopoulos}}, \bibinfo
  {author} {\bibfnamefont {P.}~\bibnamefont {Yang}},\ and\ \bibinfo {author}
  {\bibfnamefont {G.~A.}\ \bibnamefont {Somorjai}},\ }\bibfield  {title}
  {\bibinfo {title} {{Platinum Nanoparticle Shape Effects on Benzene
  Hydrogenation Selectivity}},\ }\href {https://doi.org/10.1021/nl0716000}
  {\bibfield  {journal} {\bibinfo  {journal} {Nano Lett.}\ }\textbf {\bibinfo
  {volume} {7}},\ \bibinfo {pages} {3097} (\bibinfo {year} {2007})}\BibitemShut
  {NoStop}%
\bibitem [{\citenamefont {Yang}\ \emph {et~al.}(2008)\citenamefont {Yang},
  \citenamefont {Sun}, \citenamefont {Qiao}, \citenamefont {Zou}, \citenamefont
  {Liu}, \citenamefont {Smith}, \citenamefont {Cheng},\ and\ \citenamefont
  {Lu}}]{yang2008anatase}%
  \BibitemOpen
  \bibfield  {author} {\bibinfo {author} {\bibfnamefont {H.~G.}\ \bibnamefont
  {Yang}}, \bibinfo {author} {\bibfnamefont {C.~H.}\ \bibnamefont {Sun}},
  \bibinfo {author} {\bibfnamefont {S.~Z.}\ \bibnamefont {Qiao}}, \bibinfo
  {author} {\bibfnamefont {J.}~\bibnamefont {Zou}}, \bibinfo {author}
  {\bibfnamefont {G.}~\bibnamefont {Liu}}, \bibinfo {author} {\bibfnamefont
  {S.~C.}\ \bibnamefont {Smith}}, \bibinfo {author} {\bibfnamefont {H.~M.}\
  \bibnamefont {Cheng}},\ and\ \bibinfo {author} {\bibfnamefont {G.~Q.}\
  \bibnamefont {Lu}},\ }\bibfield  {title} {\bibinfo {title} {{Anatase
  ${\mathrm{TiO}}_{2}$ single crystals with a large percentage of reactive
  facets}},\ }\href {https://www.nature.com/articles/nature06964} {\bibfield
  {journal} {\bibinfo  {journal} {Nature}\ }\textbf {\bibinfo {volume} {453}},\
  \bibinfo {pages} {638} (\bibinfo {year} {2008})}\BibitemShut {NoStop}%
\bibitem [{\citenamefont {Lovette}\ \emph {et~al.}(2008)\citenamefont
  {Lovette}, \citenamefont {Browning}, \citenamefont {Griffin}, \citenamefont
  {Sizemore}, \citenamefont {Snyder},\ and\ \citenamefont
  {Doherty}}]{lovette2008crystal}%
  \BibitemOpen
  \bibfield  {author} {\bibinfo {author} {\bibfnamefont {M.~A.}\ \bibnamefont
  {Lovette}}, \bibinfo {author} {\bibfnamefont {A.~R.}\ \bibnamefont
  {Browning}}, \bibinfo {author} {\bibfnamefont {D.~W.}\ \bibnamefont
  {Griffin}}, \bibinfo {author} {\bibfnamefont {J.~P.}\ \bibnamefont
  {Sizemore}}, \bibinfo {author} {\bibfnamefont {R.~C.}\ \bibnamefont
  {Snyder}},\ and\ \bibinfo {author} {\bibfnamefont {M.~F.}\ \bibnamefont
  {Doherty}},\ }\bibfield  {title} {\bibinfo {title} {{Crystal shape
  engineering}},\ }\href {https://pubs.acs.org/doi/abs/10.1021/ie800900f}
  {\bibfield  {journal} {\bibinfo  {journal} {Ind. \& Eng. Chem. Res.}\
  }\textbf {\bibinfo {volume} {47}},\ \bibinfo {pages} {9812} (\bibinfo {year}
  {2008})}\BibitemShut {NoStop}%
\bibitem [{\citenamefont {Grzelczak}\ \emph {et~al.}(2008)\citenamefont
  {Grzelczak}, \citenamefont {P{\'e}rez-Juste}, \citenamefont {Mulvaney},\ and\
  \citenamefont {Liz-Marz{\'a}n}}]{B711490G}%
  \BibitemOpen
  \bibfield  {author} {\bibinfo {author} {\bibfnamefont {M.}~\bibnamefont
  {Grzelczak}}, \bibinfo {author} {\bibfnamefont {J.}~\bibnamefont
  {P{\'e}rez-Juste}}, \bibinfo {author} {\bibfnamefont {P.}~\bibnamefont
  {Mulvaney}},\ and\ \bibinfo {author} {\bibfnamefont {L.~M.}\ \bibnamefont
  {Liz-Marz{\'a}n}},\ }\bibfield  {title} {\bibinfo {title} {{Shape control in
  gold nanoparticle synthesis}},\ }\href {https://doi.org/10.1039/B711490G}
  {\bibfield  {journal} {\bibinfo  {journal} {Chem. Soc. Rev.}\ }\textbf
  {\bibinfo {volume} {37}},\ \bibinfo {pages} {1783} (\bibinfo {year}
  {2008})}\BibitemShut {NoStop}%
\bibitem [{\citenamefont {Ringe}\ \emph {et~al.}(2011)\citenamefont {Ringe},
  \citenamefont {Van~Duyne},\ and\ \citenamefont {Marks}}]{ringe2011wulff}%
  \BibitemOpen
  \bibfield  {author} {\bibinfo {author} {\bibfnamefont {E.}~\bibnamefont
  {Ringe}}, \bibinfo {author} {\bibfnamefont {R.~P.}\ \bibnamefont
  {Van~Duyne}},\ and\ \bibinfo {author} {\bibfnamefont {L.}~\bibnamefont
  {Marks}},\ }\bibfield  {title} {\bibinfo {title} {{Wulff construction for
  alloy nanoparticles}},\ }\href {https://pubs.acs.org/doi/10.1021/nl2018146}
  {\bibfield  {journal} {\bibinfo  {journal} {Nano Lett.}\ }\textbf {\bibinfo
  {volume} {11}},\ \bibinfo {pages} {3399} (\bibinfo {year}
  {2011})}\BibitemShut {NoStop}%
\bibitem [{\citenamefont {Auyeung}\ \emph {et~al.}(2014)\citenamefont
  {Auyeung}, \citenamefont {Li}, \citenamefont {Senesi}, \citenamefont
  {Schmucker}, \citenamefont {Pals}, \citenamefont {de~La~Cruz},\ and\
  \citenamefont {Mirkin}}]{auyeung2014dna}%
  \BibitemOpen
  \bibfield  {author} {\bibinfo {author} {\bibfnamefont {E.}~\bibnamefont
  {Auyeung}}, \bibinfo {author} {\bibfnamefont {T.~I.}\ \bibnamefont {Li}},
  \bibinfo {author} {\bibfnamefont {A.~J.}\ \bibnamefont {Senesi}}, \bibinfo
  {author} {\bibfnamefont {A.~L.}\ \bibnamefont {Schmucker}}, \bibinfo {author}
  {\bibfnamefont {B.~C.}\ \bibnamefont {Pals}}, \bibinfo {author}
  {\bibfnamefont {M.~O.}\ \bibnamefont {de~La~Cruz}},\ and\ \bibinfo {author}
  {\bibfnamefont {C.~A.}\ \bibnamefont {Mirkin}},\ }\bibfield  {title}
  {\bibinfo {title} {{DNA-mediated nanoparticle crystallization into Wulff
  polyhedra}},\ }\href {https://www.nature.com/articles/nature12739} {\bibfield
   {journal} {\bibinfo  {journal} {Nature}\ }\textbf {\bibinfo {volume}
  {505}},\ \bibinfo {pages} {73} (\bibinfo {year} {2014})}\BibitemShut
  {NoStop}%
\bibitem [{\citenamefont {Yang}\ \emph {et~al.}(2014)\citenamefont {Yang},
  \citenamefont {Yang}, \citenamefont {Wu}, \citenamefont {Li}, \citenamefont
  {Liu}, \citenamefont {Zhao}, \citenamefont {Yu}, \citenamefont {Gong},\ and\
  \citenamefont {Yang}}]{yang2014titania}%
  \BibitemOpen
  \bibfield  {author} {\bibinfo {author} {\bibfnamefont {S.}~\bibnamefont
  {Yang}}, \bibinfo {author} {\bibfnamefont {B.~X.}\ \bibnamefont {Yang}},
  \bibinfo {author} {\bibfnamefont {L.}~\bibnamefont {Wu}}, \bibinfo {author}
  {\bibfnamefont {Y.~H.}\ \bibnamefont {Li}}, \bibinfo {author} {\bibfnamefont
  {P.}~\bibnamefont {Liu}}, \bibinfo {author} {\bibfnamefont {H.}~\bibnamefont
  {Zhao}}, \bibinfo {author} {\bibfnamefont {Y.~Y.}\ \bibnamefont {Yu}},
  \bibinfo {author} {\bibfnamefont {X.~Q.}\ \bibnamefont {Gong}},\ and\
  \bibinfo {author} {\bibfnamefont {H.~G.}\ \bibnamefont {Yang}},\ }\bibfield
  {title} {\bibinfo {title} {{Titania single crystals with a curved surface}},\
  }\href {https://www.nature.com/articles/ncomms6355} {\bibfield  {journal}
  {\bibinfo  {journal} {Nat. Commun.}\ }\textbf {\bibinfo {volume} {5}},\
  \bibinfo {pages} {5355} (\bibinfo {year} {2014})}\BibitemShut {NoStop}%
\bibitem [{\citenamefont {Liu}\ \emph {et~al.}(2014{\natexlab{a}})\citenamefont
  {Liu}, \citenamefont {Yang}, \citenamefont {Pan}, \citenamefont {Yang},
  \citenamefont {Lu},\ and\ \citenamefont {Cheng}}]{liu2014titanium}%
  \BibitemOpen
  \bibfield  {author} {\bibinfo {author} {\bibfnamefont {G.}~\bibnamefont
  {Liu}}, \bibinfo {author} {\bibfnamefont {H.~G.}\ \bibnamefont {Yang}},
  \bibinfo {author} {\bibfnamefont {J.}~\bibnamefont {Pan}}, \bibinfo {author}
  {\bibfnamefont {Y.~Q.}\ \bibnamefont {Yang}}, \bibinfo {author}
  {\bibfnamefont {G.~Q.}\ \bibnamefont {Lu}},\ and\ \bibinfo {author}
  {\bibfnamefont {H.-M.}\ \bibnamefont {Cheng}},\ }\bibfield  {title} {\bibinfo
  {title} {{Titanium dioxide crystals with tailored facets}},\ }\href
  {https://pubs.acs.org/doi/10.1021/cr400621z} {\bibfield  {journal} {\bibinfo
  {journal} {Chem. Rev.}\ }\textbf {\bibinfo {volume} {114}},\ \bibinfo {pages}
  {9559} (\bibinfo {year} {2014}{\natexlab{a}})}\BibitemShut {NoStop}%
\bibitem [{\citenamefont {Tran}\ \emph {et~al.}(2016)\citenamefont {Tran},
  \citenamefont {Xu}, \citenamefont {Radhakrishnan}, \citenamefont {Winston},
  \citenamefont {Sun}, \citenamefont {Persson},\ and\ \citenamefont
  {Ong}}]{tran2016surface}%
  \BibitemOpen
  \bibfield  {author} {\bibinfo {author} {\bibfnamefont {R.}~\bibnamefont
  {Tran}}, \bibinfo {author} {\bibfnamefont {Z.}~\bibnamefont {Xu}}, \bibinfo
  {author} {\bibfnamefont {B.}~\bibnamefont {Radhakrishnan}}, \bibinfo {author}
  {\bibfnamefont {D.}~\bibnamefont {Winston}}, \bibinfo {author} {\bibfnamefont
  {W.}~\bibnamefont {Sun}}, \bibinfo {author} {\bibfnamefont {K.~A.}\
  \bibnamefont {Persson}},\ and\ \bibinfo {author} {\bibfnamefont {S.~P.}\
  \bibnamefont {Ong}},\ }\bibfield  {title} {\bibinfo {title} {{Surface
  energies of elemental crystals}},\ }\href
  {https://www.nature.com/articles/sdata201680} {\bibfield  {journal} {\bibinfo
   {journal} {Sci. Data}\ }\textbf {\bibinfo {volume} {3}},\ \bibinfo {pages}
  {160080} (\bibinfo {year} {2016})}\BibitemShut {NoStop}%
\bibitem [{\citenamefont {Anderson}\ \emph {et~al.}(2017)\citenamefont
  {Anderson}, \citenamefont {Gebbie-Rayet}, \citenamefont {Hill}, \citenamefont
  {Farida}, \citenamefont {Attfield}, \citenamefont {Cubillas}, \citenamefont
  {Blatov}, \citenamefont {Proserpio}, \citenamefont {Akporiaye}, \citenamefont
  {Arstad},\ and\ \citenamefont {Gale}}]{anderson2017predicting}%
  \BibitemOpen
  \bibfield  {author} {\bibinfo {author} {\bibfnamefont {M.~W.}\ \bibnamefont
  {Anderson}}, \bibinfo {author} {\bibfnamefont {J.~T.}\ \bibnamefont
  {Gebbie-Rayet}}, \bibinfo {author} {\bibfnamefont {A.~R.}\ \bibnamefont
  {Hill}}, \bibinfo {author} {\bibfnamefont {N.}~\bibnamefont {Farida}},
  \bibinfo {author} {\bibfnamefont {M.~P.}\ \bibnamefont {Attfield}}, \bibinfo
  {author} {\bibfnamefont {P.}~\bibnamefont {Cubillas}}, \bibinfo {author}
  {\bibfnamefont {V.~A.}\ \bibnamefont {Blatov}}, \bibinfo {author}
  {\bibfnamefont {D.~M.}\ \bibnamefont {Proserpio}}, \bibinfo {author}
  {\bibfnamefont {D.}~\bibnamefont {Akporiaye}}, \bibinfo {author}
  {\bibfnamefont {B.}~\bibnamefont {Arstad}},\ and\ \bibinfo {author}
  {\bibfnamefont {J.~D.}\ \bibnamefont {Gale}},\ }\bibfield  {title} {\bibinfo
  {title} {{Predicting crystal growth via a unified kinetic three-dimensional
  partition model}},\ }\href {https://www.nature.com/articles/nature21684}
  {\bibfield  {journal} {\bibinfo  {journal} {Nature}\ }\textbf {\bibinfo
  {volume} {544}},\ \bibinfo {pages} {456} (\bibinfo {year}
  {2017})}\BibitemShut {NoStop}%
\bibitem [{\citenamefont {Wang}\ \emph
  {et~al.}(2019{\natexlab{a}})\citenamefont {Wang}, \citenamefont {Liu},\ and\
  \citenamefont {Wang}}]{wang2019crystal}%
  \BibitemOpen
  \bibfield  {author} {\bibinfo {author} {\bibfnamefont {S.}~\bibnamefont
  {Wang}}, \bibinfo {author} {\bibfnamefont {G.}~\bibnamefont {Liu}},\ and\
  \bibinfo {author} {\bibfnamefont {L.}~\bibnamefont {Wang}},\ }\bibfield
  {title} {\bibinfo {title} {{Crystal facet engineering of photoelectrodes for
  photoelectrochemical water splitting}},\ }\href
  {https://pubs.acs.org/doi/10.1021/acs.chemrev.8b00584} {\bibfield  {journal}
  {\bibinfo  {journal} {Chem. Rev.}\ }\textbf {\bibinfo {volume} {119}},\
  \bibinfo {pages} {5192} (\bibinfo {year} {2019}{\natexlab{a}})}\BibitemShut
  {NoStop}%
\bibitem [{\citenamefont {Xie}\ \emph {et~al.}(2021)\citenamefont {Xie},
  \citenamefont {Zhang}, \citenamefont {Xie}, \citenamefont {Hou},
  \citenamefont {Ji}, \citenamefont {Pang}, \citenamefont {Chen}, \citenamefont
  {Titirici}, \citenamefont {Weng},\ and\ \citenamefont
  {Chai}}]{https://doi.org/10.1002/adma.202008373}%
  \BibitemOpen
  \bibfield  {author} {\bibinfo {author} {\bibfnamefont {H.}~\bibnamefont
  {Xie}}, \bibinfo {author} {\bibfnamefont {T.}~\bibnamefont {Zhang}}, \bibinfo
  {author} {\bibfnamefont {R.}~\bibnamefont {Xie}}, \bibinfo {author}
  {\bibfnamefont {Z.}~\bibnamefont {Hou}}, \bibinfo {author} {\bibfnamefont
  {X.}~\bibnamefont {Ji}}, \bibinfo {author} {\bibfnamefont {Y.}~\bibnamefont
  {Pang}}, \bibinfo {author} {\bibfnamefont {S.}~\bibnamefont {Chen}}, \bibinfo
  {author} {\bibfnamefont {M.-M.}\ \bibnamefont {Titirici}}, \bibinfo {author}
  {\bibfnamefont {H.}~\bibnamefont {Weng}},\ and\ \bibinfo {author}
  {\bibfnamefont {G.}~\bibnamefont {Chai}},\ }\bibfield  {title} {\bibinfo
  {title} {{Facet Engineering to Regulate Surface States of Topological
  Crystalline Insulator Bismuth Rhombic Dodecahedrons for Highly Energy
  Efficient Electrochemical CO2 Reduction}},\ }\href
  {https://doi.org/https://doi.org/10.1002/adma.202008373} {\bibfield
  {journal} {\bibinfo  {journal} {Adv. Mater.}\ }\textbf {\bibinfo {volume}
  {33}},\ \bibinfo {pages} {2008373} (\bibinfo {year} {2021})}\BibitemShut
  {NoStop}%
\bibitem [{\citenamefont {Malkova}\ and\ \citenamefont
  {Bryant}(2010)}]{PhysRevB.82.155314}%
  \BibitemOpen
  \bibfield  {author} {\bibinfo {author} {\bibfnamefont {N.}~\bibnamefont
  {Malkova}}\ and\ \bibinfo {author} {\bibfnamefont {G.~W.}\ \bibnamefont
  {Bryant}},\ }\bibfield  {title} {\bibinfo {title} {{Negative-band-gap quantum
  dots: Gap collapse, intrinsic surface states, excitonic response, and
  excitonic insulator phase}},\ }\href
  {https://doi.org/10.1103/PhysRevB.82.155314} {\bibfield  {journal} {\bibinfo
  {journal} {Phys. Rev. B}\ }\textbf {\bibinfo {volume} {82}},\ \bibinfo
  {pages} {155314} (\bibinfo {year} {2010})}\BibitemShut {NoStop}%
\bibitem [{\citenamefont {Imura}\ \emph {et~al.}(2012)\citenamefont {Imura},
  \citenamefont {Yoshimura}, \citenamefont {Takane},\ and\ \citenamefont
  {Fukui}}]{PhysRevB.86.235119}%
  \BibitemOpen
  \bibfield  {author} {\bibinfo {author} {\bibfnamefont {K.-I.}\ \bibnamefont
  {Imura}}, \bibinfo {author} {\bibfnamefont {Y.}~\bibnamefont {Yoshimura}},
  \bibinfo {author} {\bibfnamefont {Y.}~\bibnamefont {Takane}},\ and\ \bibinfo
  {author} {\bibfnamefont {T.}~\bibnamefont {Fukui}},\ }\bibfield  {title}
  {\bibinfo {title} {{Spherical topological insulator}},\ }\href
  {https://doi.org/10.1103/PhysRevB.86.235119} {\bibfield  {journal} {\bibinfo
  {journal} {Phys. Rev. B}\ }\textbf {\bibinfo {volume} {86}},\ \bibinfo
  {pages} {235119} (\bibinfo {year} {2012})}\BibitemShut {NoStop}%
\bibitem [{\citenamefont {Takane}\ and\ \citenamefont
  {Imura}(2013)}]{doi:10.7566/JPSJ.82.074712}%
  \BibitemOpen
  \bibfield  {author} {\bibinfo {author} {\bibfnamefont {Y.}~\bibnamefont
  {Takane}}\ and\ \bibinfo {author} {\bibfnamefont {K.-I.}\ \bibnamefont
  {Imura}},\ }\bibfield  {title} {\bibinfo {title} {{Unified Description of
  Dirac Electrons on a Curved Surface of Topological Insulators}},\ }\href
  {https://doi.org/10.7566/JPSJ.82.074712} {\bibfield  {journal} {\bibinfo
  {journal} {J. Phys. Soc. Jpn.}\ }\textbf {\bibinfo {volume} {82}},\ \bibinfo
  {pages} {074712} (\bibinfo {year} {2013})}\BibitemShut {NoStop}%
\bibitem [{\citenamefont {Lin}\ \emph {et~al.}(2015)\citenamefont {Lin},
  \citenamefont {Lin}, \citenamefont {Chi}, \citenamefont {Wu}, \citenamefont
  {Cheng}, \citenamefont {Tseng}, \citenamefont {He}, \citenamefont {Wu},
  \citenamefont {Lee},\ and\ \citenamefont {Lin}}]{lin2015using}%
  \BibitemOpen
  \bibfield  {author} {\bibinfo {author} {\bibfnamefont {Y.-H.}\ \bibnamefont
  {Lin}}, \bibinfo {author} {\bibfnamefont {S.-F.}\ \bibnamefont {Lin}},
  \bibinfo {author} {\bibfnamefont {Y.-C.}\ \bibnamefont {Chi}}, \bibinfo
  {author} {\bibfnamefont {C.-L.}\ \bibnamefont {Wu}}, \bibinfo {author}
  {\bibfnamefont {C.-H.}\ \bibnamefont {Cheng}}, \bibinfo {author}
  {\bibfnamefont {W.-H.}\ \bibnamefont {Tseng}}, \bibinfo {author}
  {\bibfnamefont {J.-H.}\ \bibnamefont {He}}, \bibinfo {author} {\bibfnamefont
  {C.-I.}\ \bibnamefont {Wu}}, \bibinfo {author} {\bibfnamefont {C.-K.}\
  \bibnamefont {Lee}},\ and\ \bibinfo {author} {\bibfnamefont {G.-R.}\
  \bibnamefont {Lin}},\ }\bibfield  {title} {\bibinfo {title} {{Using n-and
  p-type ${\mathrm{Bi}}_{2}{\mathrm{Te}}_{3}$ topological insulator
  nanoparticles to enable controlled femtosecond mode-locking of fiber
  lasers}},\ }\href
  {https://pubs.acs.org/doi/full/10.1021/acsphotonics.5b00031} {\bibfield
  {journal} {\bibinfo  {journal} {ACS Photonics}\ }\textbf {\bibinfo {volume}
  {2}},\ \bibinfo {pages} {481} (\bibinfo {year} {2015})}\BibitemShut {NoStop}%
\bibitem [{\citenamefont {Siroki}\ \emph {et~al.}(2016)\citenamefont {Siroki},
  \citenamefont {Lee}, \citenamefont {Haynes},\ and\ \citenamefont
  {Giannini}}]{Siroki:2016vz}%
  \BibitemOpen
  \bibfield  {author} {\bibinfo {author} {\bibfnamefont {G.}~\bibnamefont
  {Siroki}}, \bibinfo {author} {\bibfnamefont {D.~K.~K.}\ \bibnamefont {Lee}},
  \bibinfo {author} {\bibfnamefont {P.~D.}\ \bibnamefont {Haynes}},\ and\
  \bibinfo {author} {\bibfnamefont {V.}~\bibnamefont {Giannini}},\ }\bibfield
  {title} {\bibinfo {title} {Single-electron induced surface plasmons on a
  topological nanoparticle},\ }\href {https://doi.org/10.1038/ncomms12375}
  {\bibfield  {journal} {\bibinfo  {journal} {Nat. Commun.}\ }\textbf {\bibinfo
  {volume} {7}},\ \bibinfo {pages} {12375} (\bibinfo {year}
  {2016})}\BibitemShut {NoStop}%
\bibitem [{\citenamefont {Siroki}\ \emph {et~al.}(2017)\citenamefont {Siroki},
  \citenamefont {Haynes}, \citenamefont {Lee},\ and\ \citenamefont
  {Giannini}}]{PhysRevMaterials.1.024201}%
  \BibitemOpen
  \bibfield  {author} {\bibinfo {author} {\bibfnamefont {G.}~\bibnamefont
  {Siroki}}, \bibinfo {author} {\bibfnamefont {P.~D.}\ \bibnamefont {Haynes}},
  \bibinfo {author} {\bibfnamefont {D.~K.~K.}\ \bibnamefont {Lee}},\ and\
  \bibinfo {author} {\bibfnamefont {V.}~\bibnamefont {Giannini}},\ }\bibfield
  {title} {\bibinfo {title} {{Protection of surface states in topological
  nanoparticles}},\ }\href {https://doi.org/10.1103/PhysRevMaterials.1.024201}
  {\bibfield  {journal} {\bibinfo  {journal} {Phys. Rev. Mater.}\ }\textbf
  {\bibinfo {volume} {1}},\ \bibinfo {pages} {024201} (\bibinfo {year}
  {2017})}\BibitemShut {NoStop}%
\bibitem [{\citenamefont {Gioia}\ \emph {et~al.}(2019)\citenamefont {Gioia},
  \citenamefont {Christie}, \citenamefont {Z\"ulicke}, \citenamefont
  {Governale},\ and\ \citenamefont {Sneyd}}]{PhysRevB.100.205417}%
  \BibitemOpen
  \bibfield  {author} {\bibinfo {author} {\bibfnamefont {L.}~\bibnamefont
  {Gioia}}, \bibinfo {author} {\bibfnamefont {M.~G.}\ \bibnamefont {Christie}},
  \bibinfo {author} {\bibfnamefont {U.}~\bibnamefont {Z\"ulicke}}, \bibinfo
  {author} {\bibfnamefont {M.}~\bibnamefont {Governale}},\ and\ \bibinfo
  {author} {\bibfnamefont {A.~J.}\ \bibnamefont {Sneyd}},\ }\bibfield  {title}
  {\bibinfo {title} {{Spherical topological insulator nanoparticles: Quantum
  size effects and optical transitions}},\ }\href
  {https://doi.org/10.1103/PhysRevB.100.205417} {\bibfield  {journal} {\bibinfo
   {journal} {Phys. Rev. B}\ }\textbf {\bibinfo {volume} {100}},\ \bibinfo
  {pages} {205417} (\bibinfo {year} {2019})}\BibitemShut {NoStop}%
\bibitem [{\citenamefont {Chatzidakis}\ and\ \citenamefont
  {Yannopapas}(2020)}]{PhysRevB.101.165410}%
  \BibitemOpen
  \bibfield  {author} {\bibinfo {author} {\bibfnamefont {G.~D.}\ \bibnamefont
  {Chatzidakis}}\ and\ \bibinfo {author} {\bibfnamefont {V.}~\bibnamefont
  {Yannopapas}},\ }\bibfield  {title} {\bibinfo {title} {{Strong
  electromagnetic coupling in dimers of topological-insulator nanoparticles and
  quantum emitters}},\ }\href {https://doi.org/10.1103/PhysRevB.101.165410}
  {\bibfield  {journal} {\bibinfo  {journal} {Phys. Rev. B}\ }\textbf {\bibinfo
  {volume} {101}},\ \bibinfo {pages} {165410} (\bibinfo {year}
  {2020})}\BibitemShut {NoStop}%
\bibitem [{\citenamefont {Castro-Enriquez}\ \emph {et~al.}(2020)\citenamefont
  {Castro-Enriquez}, \citenamefont {Quezada},\ and\ \citenamefont
  {Mart\'{\i}n-Ruiz}}]{PhysRevA.102.013720}%
  \BibitemOpen
  \bibfield  {author} {\bibinfo {author} {\bibfnamefont {L.~A.}\ \bibnamefont
  {Castro-Enriquez}}, \bibinfo {author} {\bibfnamefont {L.~F.}\ \bibnamefont
  {Quezada}},\ and\ \bibinfo {author} {\bibfnamefont {A.}~\bibnamefont
  {Mart\'{\i}n-Ruiz}},\ }\bibfield  {title} {\bibinfo {title} {{Optical
  response of a topological-insulator--quantum-dot hybrid interacting with a
  probe electric field}},\ }\href {https://doi.org/10.1103/PhysRevA.102.013720}
  {\bibfield  {journal} {\bibinfo  {journal} {Phys. Rev. A}\ }\textbf {\bibinfo
  {volume} {102}},\ \bibinfo {pages} {013720} (\bibinfo {year}
  {2020})}\BibitemShut {NoStop}%
\bibitem [{\citenamefont {Castro-Enr{\'\i}quez}\ \emph
  {et~al.}(2022)\citenamefont {Castro-Enr{\'\i}quez}, \citenamefont
  {Mart{\'\i}n-Ruiz},\ and\ \citenamefont {Cambiaso}}]{Castro-Enriquez:2022ut}%
  \BibitemOpen
  \bibfield  {author} {\bibinfo {author} {\bibfnamefont {L.~A.}\ \bibnamefont
  {Castro-Enr{\'\i}quez}}, \bibinfo {author} {\bibfnamefont {A.}~\bibnamefont
  {Mart{\'\i}n-Ruiz}},\ and\ \bibinfo {author} {\bibfnamefont {M.}~\bibnamefont
  {Cambiaso}},\ }\bibfield  {title} {\bibinfo {title} {{Topological signatures
  in the entanglement of a topological insulator-quantum dot hybrid}},\ }\href
  {https://doi.org/10.1038/s41598-022-24939-3} {\bibfield  {journal} {\bibinfo
  {journal} {Sci. Rep.}\ }\textbf {\bibinfo {volume} {12}},\ \bibinfo {pages}
  {20856} (\bibinfo {year} {2022})}\BibitemShut {NoStop}%
\bibitem [{\citenamefont {Hasan}\ and\ \citenamefont
  {Kane}(2010)}]{RevModPhys.82.3045}%
  \BibitemOpen
  \bibfield  {author} {\bibinfo {author} {\bibfnamefont {M.~Z.}\ \bibnamefont
  {Hasan}}\ and\ \bibinfo {author} {\bibfnamefont {C.~L.}\ \bibnamefont
  {Kane}},\ }\bibfield  {title} {\bibinfo {title} {{Colloquium: Topological
  insulators}},\ }\href {https://doi.org/10.1103/RevModPhys.82.3045} {\bibfield
   {journal} {\bibinfo  {journal} {Rev. Mod. Phys.}\ }\textbf {\bibinfo
  {volume} {82}},\ \bibinfo {pages} {3045} (\bibinfo {year}
  {2010})}\BibitemShut {NoStop}%
\bibitem [{\citenamefont {Qi}\ and\ \citenamefont
  {Zhang}(2011)}]{RevModPhys.83.1057}%
  \BibitemOpen
  \bibfield  {author} {\bibinfo {author} {\bibfnamefont {X.-L.}\ \bibnamefont
  {Qi}}\ and\ \bibinfo {author} {\bibfnamefont {S.-C.}\ \bibnamefont {Zhang}},\
  }\bibfield  {title} {\bibinfo {title} {{Topological insulators and
  superconductors}},\ }\href {https://doi.org/10.1103/RevModPhys.83.1057}
  {\bibfield  {journal} {\bibinfo  {journal} {Rev. Mod. Phys.}\ }\textbf
  {\bibinfo {volume} {83}},\ \bibinfo {pages} {1057} (\bibinfo {year}
  {2011})}\BibitemShut {NoStop}%
\bibitem [{\citenamefont {Fu}(2011)}]{PhysRevLett.106.106802}%
  \BibitemOpen
  \bibfield  {author} {\bibinfo {author} {\bibfnamefont {L.}~\bibnamefont
  {Fu}},\ }\bibfield  {title} {\bibinfo {title} {Topological crystalline
  insulators},\ }\href {https://doi.org/10.1103/PhysRevLett.106.106802}
  {\bibfield  {journal} {\bibinfo  {journal} {Phys. Rev. Lett.}\ }\textbf
  {\bibinfo {volume} {106}},\ \bibinfo {pages} {106802} (\bibinfo {year}
  {2011})}\BibitemShut {NoStop}%
\bibitem [{\citenamefont {Hsieh}\ \emph {et~al.}(2012)\citenamefont {Hsieh},
  \citenamefont {Lin}, \citenamefont {Liu}, \citenamefont {Duan}, \citenamefont
  {Bansil},\ and\ \citenamefont {Fu}}]{hsieh2012topological}%
  \BibitemOpen
  \bibfield  {author} {\bibinfo {author} {\bibfnamefont {T.~H.}\ \bibnamefont
  {Hsieh}}, \bibinfo {author} {\bibfnamefont {H.}~\bibnamefont {Lin}}, \bibinfo
  {author} {\bibfnamefont {J.}~\bibnamefont {Liu}}, \bibinfo {author}
  {\bibfnamefont {W.}~\bibnamefont {Duan}}, \bibinfo {author} {\bibfnamefont
  {A.}~\bibnamefont {Bansil}},\ and\ \bibinfo {author} {\bibfnamefont
  {L.}~\bibnamefont {Fu}},\ }\bibfield  {title} {\bibinfo {title} {{Topological
  crystalline insulators in the SnTe material class}},\ }\href
  {https://www.nature.com/articles/ncomms1969} {\bibfield  {journal} {\bibinfo
  {journal} {Nat. Commun.}\ }\textbf {\bibinfo {volume} {3}},\ \bibinfo {pages}
  {982} (\bibinfo {year} {2012})}\BibitemShut {NoStop}%
\bibitem [{\citenamefont {Slager}\ \emph {et~al.}(2013)\citenamefont {Slager},
  \citenamefont {Mesaros}, \citenamefont {Juri{\v c}i{\'c}},\ and\
  \citenamefont {Zaanen}}]{slager2013space}%
  \BibitemOpen
  \bibfield  {author} {\bibinfo {author} {\bibfnamefont {R.-J.}\ \bibnamefont
  {Slager}}, \bibinfo {author} {\bibfnamefont {A.}~\bibnamefont {Mesaros}},
  \bibinfo {author} {\bibfnamefont {V.}~\bibnamefont {Juri{\v c}i{\'c}}},\ and\
  \bibinfo {author} {\bibfnamefont {J.}~\bibnamefont {Zaanen}},\ }\bibfield
  {title} {\bibinfo {title} {The space group classification of topological
  band-insulators},\ }\href {https://doi.org/10.1038/nphys2513} {\bibfield
  {journal} {\bibinfo  {journal} {Nat. Phys.}\ }\textbf {\bibinfo {volume}
  {9}},\ \bibinfo {pages} {98} (\bibinfo {year} {2013})}\BibitemShut {NoStop}%
\bibitem [{\citenamefont {Kargarian}\ and\ \citenamefont
  {Fiete}(2013)}]{PhysRevLett.110.156403}%
  \BibitemOpen
  \bibfield  {author} {\bibinfo {author} {\bibfnamefont {M.}~\bibnamefont
  {Kargarian}}\ and\ \bibinfo {author} {\bibfnamefont {G.~A.}\ \bibnamefont
  {Fiete}},\ }\bibfield  {title} {\bibinfo {title} {{Topological Crystalline
  Insulators in Transition Metal Oxides}},\ }\href
  {https://doi.org/10.1103/PhysRevLett.110.156403} {\bibfield  {journal}
  {\bibinfo  {journal} {Phys. Rev. Lett.}\ }\textbf {\bibinfo {volume} {110}},\
  \bibinfo {pages} {156403} (\bibinfo {year} {2013})}\BibitemShut {NoStop}%
\bibitem [{\citenamefont {Liu}\ \emph {et~al.}(2014{\natexlab{b}})\citenamefont
  {Liu}, \citenamefont {Zhang},\ and\ \citenamefont
  {VanLeeuwen}}]{PhysRevB.90.085304}%
  \BibitemOpen
  \bibfield  {author} {\bibinfo {author} {\bibfnamefont {C.-X.}\ \bibnamefont
  {Liu}}, \bibinfo {author} {\bibfnamefont {R.-X.}\ \bibnamefont {Zhang}},\
  and\ \bibinfo {author} {\bibfnamefont {B.~K.}\ \bibnamefont {VanLeeuwen}},\
  }\bibfield  {title} {\bibinfo {title} {{Topological nonsymmorphic crystalline
  insulators}},\ }\href {https://doi.org/10.1103/PhysRevB.90.085304} {\bibfield
   {journal} {\bibinfo  {journal} {Phys. Rev. B}\ }\textbf {\bibinfo {volume}
  {90}},\ \bibinfo {pages} {085304} (\bibinfo {year}
  {2014}{\natexlab{b}})}\BibitemShut {NoStop}%
\bibitem [{\citenamefont {Shiozaki}\ and\ \citenamefont
  {Sato}(2014)}]{PhysRevB.90.165114}%
  \BibitemOpen
  \bibfield  {author} {\bibinfo {author} {\bibfnamefont {K.}~\bibnamefont
  {Shiozaki}}\ and\ \bibinfo {author} {\bibfnamefont {M.}~\bibnamefont
  {Sato}},\ }\bibfield  {title} {\bibinfo {title} {Topology of crystalline
  insulators and superconductors},\ }\href
  {https://doi.org/10.1103/PhysRevB.90.165114} {\bibfield  {journal} {\bibinfo
  {journal} {Phys. Rev. B}\ }\textbf {\bibinfo {volume} {90}},\ \bibinfo
  {pages} {165114} (\bibinfo {year} {2014})}\BibitemShut {NoStop}%
\bibitem [{\citenamefont {Shiozaki}\ \emph {et~al.}(2016)\citenamefont
  {Shiozaki}, \citenamefont {Sato},\ and\ \citenamefont
  {Gomi}}]{PhysRevB.93.195413}%
  \BibitemOpen
  \bibfield  {author} {\bibinfo {author} {\bibfnamefont {K.}~\bibnamefont
  {Shiozaki}}, \bibinfo {author} {\bibfnamefont {M.}~\bibnamefont {Sato}},\
  and\ \bibinfo {author} {\bibfnamefont {K.}~\bibnamefont {Gomi}},\ }\bibfield
  {title} {\bibinfo {title} {Topology of nonsymmorphic crystalline insulators
  and superconductors},\ }\href {https://doi.org/10.1103/PhysRevB.93.195413}
  {\bibfield  {journal} {\bibinfo  {journal} {Phys. Rev. B}\ }\textbf {\bibinfo
  {volume} {93}},\ \bibinfo {pages} {195413} (\bibinfo {year}
  {2016})}\BibitemShut {NoStop}%
\bibitem [{\citenamefont {Wang}\ \emph {et~al.}(2016)\citenamefont {Wang},
  \citenamefont {Alexandradinata}, \citenamefont {Cava},\ and\ \citenamefont
  {Bernevig}}]{wang2016hourglass}%
  \BibitemOpen
  \bibfield  {author} {\bibinfo {author} {\bibfnamefont {Z.}~\bibnamefont
  {Wang}}, \bibinfo {author} {\bibfnamefont {A.}~\bibnamefont
  {Alexandradinata}}, \bibinfo {author} {\bibfnamefont {R.~J.}\ \bibnamefont
  {Cava}},\ and\ \bibinfo {author} {\bibfnamefont {B.~A.}\ \bibnamefont
  {Bernevig}},\ }\bibfield  {title} {\bibinfo {title} {{Hourglass fermions}},\
  }\href {https://www.nature.com/articles/nature17410} {\bibfield  {journal}
  {\bibinfo  {journal} {Nature}\ }\textbf {\bibinfo {volume} {532}},\ \bibinfo
  {pages} {189} (\bibinfo {year} {2016})}\BibitemShut {NoStop}%
\bibitem [{\citenamefont {Lu}\ \emph {et~al.}(2016)\citenamefont {Lu},
  \citenamefont {Fang}, \citenamefont {Fu}, \citenamefont {Johnson},
  \citenamefont {Joannopoulos},\ and\ \citenamefont
  {Solja{\v{c}}i{\'c}}}]{lu2016symmetry}%
  \BibitemOpen
  \bibfield  {author} {\bibinfo {author} {\bibfnamefont {L.}~\bibnamefont
  {Lu}}, \bibinfo {author} {\bibfnamefont {C.}~\bibnamefont {Fang}}, \bibinfo
  {author} {\bibfnamefont {L.}~\bibnamefont {Fu}}, \bibinfo {author}
  {\bibfnamefont {S.~G.}\ \bibnamefont {Johnson}}, \bibinfo {author}
  {\bibfnamefont {J.~D.}\ \bibnamefont {Joannopoulos}},\ and\ \bibinfo {author}
  {\bibfnamefont {M.}~\bibnamefont {Solja{\v{c}}i{\'c}}},\ }\bibfield  {title}
  {\bibinfo {title} {{Symmetry-protected topological photonic crystal in three
  dimensions}},\ }\href {https://www.nature.com/articles/nphys3611} {\bibfield
  {journal} {\bibinfo  {journal} {Nat. Phys.}\ }\textbf {\bibinfo {volume}
  {12}},\ \bibinfo {pages} {337} (\bibinfo {year} {2016})}\BibitemShut
  {NoStop}%
\bibitem [{\citenamefont {Kim}\ and\ \citenamefont
  {Murakami}(2016)}]{PhysRevB.93.195138}%
  \BibitemOpen
  \bibfield  {author} {\bibinfo {author} {\bibfnamefont {H.}~\bibnamefont
  {Kim}}\ and\ \bibinfo {author} {\bibfnamefont {S.}~\bibnamefont {Murakami}},\
  }\bibfield  {title} {\bibinfo {title} {{Emergent spinless Weyl semimetals
  between the topological crystalline insulator and normal insulator phases
  with glide symmetry}},\ }\href {https://doi.org/10.1103/PhysRevB.93.195138}
  {\bibfield  {journal} {\bibinfo  {journal} {Phys. Rev. B}\ }\textbf {\bibinfo
  {volume} {93}},\ \bibinfo {pages} {195138} (\bibinfo {year}
  {2016})}\BibitemShut {NoStop}%
\bibitem [{\citenamefont {Chen}\ \emph {et~al.}(2017)\citenamefont {Chen},
  \citenamefont {Zhang}, \citenamefont {Yi}, \citenamefont {Song},
  \citenamefont {Zhang}, \citenamefont {Zhang}, \citenamefont {Shi},
  \citenamefont {Weng}, \citenamefont {Fang}, \citenamefont {Richard},\ and\
  \citenamefont {Ding}}]{PhysRevB.96.064102}%
  \BibitemOpen
  \bibfield  {author} {\bibinfo {author} {\bibfnamefont {D.}~\bibnamefont
  {Chen}}, \bibinfo {author} {\bibfnamefont {T.-T.}\ \bibnamefont {Zhang}},
  \bibinfo {author} {\bibfnamefont {C.-J.}\ \bibnamefont {Yi}}, \bibinfo
  {author} {\bibfnamefont {Z.-D.}\ \bibnamefont {Song}}, \bibinfo {author}
  {\bibfnamefont {W.-L.}\ \bibnamefont {Zhang}}, \bibinfo {author}
  {\bibfnamefont {T.}~\bibnamefont {Zhang}}, \bibinfo {author} {\bibfnamefont
  {Y.-G.}\ \bibnamefont {Shi}}, \bibinfo {author} {\bibfnamefont {H.-M.}\
  \bibnamefont {Weng}}, \bibinfo {author} {\bibfnamefont {Z.}~\bibnamefont
  {Fang}}, \bibinfo {author} {\bibfnamefont {P.}~\bibnamefont {Richard}},\ and\
  \bibinfo {author} {\bibfnamefont {H.}~\bibnamefont {Ding}},\ }\bibfield
  {title} {\bibinfo {title} {{Robustness of topological states with respect to
  lattice instability in the nonsymmorphic topological insulator KHgSb}},\
  }\href {https://doi.org/10.1103/PhysRevB.96.064102} {\bibfield  {journal}
  {\bibinfo  {journal} {Phys. Rev. B}\ }\textbf {\bibinfo {volume} {96}},\
  \bibinfo {pages} {064102} (\bibinfo {year} {2017})}\BibitemShut {NoStop}%
\bibitem [{\citenamefont {Wieder}\ \emph {et~al.}(2018)\citenamefont {Wieder},
  \citenamefont {Bradlyn}, \citenamefont {Wang}, \citenamefont {Cano},
  \citenamefont {Kim}, \citenamefont {Kim}, \citenamefont {Rappe},
  \citenamefont {Kane},\ and\ \citenamefont {Bernevig}}]{wieder2018wallpaper}%
  \BibitemOpen
  \bibfield  {author} {\bibinfo {author} {\bibfnamefont {B.~J.}\ \bibnamefont
  {Wieder}}, \bibinfo {author} {\bibfnamefont {B.}~\bibnamefont {Bradlyn}},
  \bibinfo {author} {\bibfnamefont {Z.}~\bibnamefont {Wang}}, \bibinfo {author}
  {\bibfnamefont {J.}~\bibnamefont {Cano}}, \bibinfo {author} {\bibfnamefont
  {Y.}~\bibnamefont {Kim}}, \bibinfo {author} {\bibfnamefont {H.-S.~D.}\
  \bibnamefont {Kim}}, \bibinfo {author} {\bibfnamefont {A.~M.}\ \bibnamefont
  {Rappe}}, \bibinfo {author} {\bibfnamefont {C.}~\bibnamefont {Kane}},\ and\
  \bibinfo {author} {\bibfnamefont {B.~A.}\ \bibnamefont {Bernevig}},\
  }\bibfield  {title} {\bibinfo {title} {{Wallpaper fermions and the
  nonsymmorphic Dirac insulator}},\ }\href
  {https://science.sciencemag.org/content/361/6399/246} {\bibfield  {journal}
  {\bibinfo  {journal} {Science}\ }\textbf {\bibinfo {volume} {361}},\ \bibinfo
  {pages} {246} (\bibinfo {year} {2018})}\BibitemShut {NoStop}%
\bibitem [{\citenamefont {Kim}\ \emph {et~al.}(2019)\citenamefont {Kim},
  \citenamefont {Shiozaki},\ and\ \citenamefont
  {Murakami}}]{PhysRevB.100.165202}%
  \BibitemOpen
  \bibfield  {author} {\bibinfo {author} {\bibfnamefont {H.}~\bibnamefont
  {Kim}}, \bibinfo {author} {\bibfnamefont {K.}~\bibnamefont {Shiozaki}},\ and\
  \bibinfo {author} {\bibfnamefont {S.}~\bibnamefont {Murakami}},\ }\bibfield
  {title} {\bibinfo {title} {{Glide-symmetric magnetic topological crystalline
  insulators with inversion symmetry}},\ }\href
  {https://doi.org/10.1103/PhysRevB.100.165202} {\bibfield  {journal} {\bibinfo
   {journal} {Phys. Rev. B}\ }\textbf {\bibinfo {volume} {100}},\ \bibinfo
  {pages} {165202} (\bibinfo {year} {2019})}\BibitemShut {NoStop}%
\bibitem [{\citenamefont {Kim}\ and\ \citenamefont
  {Murakami}(2020)}]{kim2020glide}%
  \BibitemOpen
  \bibfield  {author} {\bibinfo {author} {\bibfnamefont {H.}~\bibnamefont
  {Kim}}\ and\ \bibinfo {author} {\bibfnamefont {S.}~\bibnamefont {Murakami}},\
  }\bibfield  {title} {\bibinfo {title} {Glide-symmetric topological
  crystalline insulator phase in a nonprimitive lattice},\ }\href
  {https://doi.org/10.1103/PhysRevB.102.195202} {\bibfield  {journal} {\bibinfo
   {journal} {Phys. Rev. B}\ }\textbf {\bibinfo {volume} {102}},\ \bibinfo
  {pages} {195202} (\bibinfo {year} {2020})}\BibitemShut {NoStop}%
\bibitem [{\citenamefont {Sitte}\ \emph {et~al.}(2012)\citenamefont {Sitte},
  \citenamefont {Rosch}, \citenamefont {Altman},\ and\ \citenamefont
  {Fritz}}]{PhysRevLett.108.126807}%
  \BibitemOpen
  \bibfield  {author} {\bibinfo {author} {\bibfnamefont {M.}~\bibnamefont
  {Sitte}}, \bibinfo {author} {\bibfnamefont {A.}~\bibnamefont {Rosch}},
  \bibinfo {author} {\bibfnamefont {E.}~\bibnamefont {Altman}},\ and\ \bibinfo
  {author} {\bibfnamefont {L.}~\bibnamefont {Fritz}},\ }\bibfield  {title}
  {\bibinfo {title} {{Topological Insulators in Magnetic Fields: Quantum Hall
  Effect and Edge Channels with a Nonquantized $\ensuremath{\theta}$ Term}},\
  }\href {https://doi.org/10.1103/PhysRevLett.108.126807} {\bibfield  {journal}
  {\bibinfo  {journal} {Phys. Rev. Lett.}\ }\textbf {\bibinfo {volume} {108}},\
  \bibinfo {pages} {126807} (\bibinfo {year} {2012})}\BibitemShut {NoStop}%
\bibitem [{\citenamefont {Zhang}\ \emph {et~al.}(2013)\citenamefont {Zhang},
  \citenamefont {Kane},\ and\ \citenamefont {Mele}}]{PhysRevLett.110.046404}%
  \BibitemOpen
  \bibfield  {author} {\bibinfo {author} {\bibfnamefont {F.}~\bibnamefont
  {Zhang}}, \bibinfo {author} {\bibfnamefont {C.~L.}\ \bibnamefont {Kane}},\
  and\ \bibinfo {author} {\bibfnamefont {E.~J.}\ \bibnamefont {Mele}},\
  }\bibfield  {title} {\bibinfo {title} {{Surface State Magnetization and
  Chiral Edge States on Topological Insulators}},\ }\href
  {https://doi.org/10.1103/PhysRevLett.110.046404} {\bibfield  {journal}
  {\bibinfo  {journal} {Phys. Rev. Lett.}\ }\textbf {\bibinfo {volume} {110}},\
  \bibinfo {pages} {046404} (\bibinfo {year} {2013})}\BibitemShut {NoStop}%
\bibitem [{\citenamefont {Benalcazar}\ \emph
  {et~al.}(2017{\natexlab{a}})\citenamefont {Benalcazar}, \citenamefont
  {Bernevig},\ and\ \citenamefont {Hughes}}]{benalcazar2017quantized}%
  \BibitemOpen
  \bibfield  {author} {\bibinfo {author} {\bibfnamefont {W.~A.}\ \bibnamefont
  {Benalcazar}}, \bibinfo {author} {\bibfnamefont {B.~A.}\ \bibnamefont
  {Bernevig}},\ and\ \bibinfo {author} {\bibfnamefont {T.~L.}\ \bibnamefont
  {Hughes}},\ }\bibfield  {title} {\bibinfo {title} {{Quantized electric
  multipole insulators}},\ }\href {https://doi.org/10.1126/science.aah6442}
  {\bibfield  {journal} {\bibinfo  {journal} {Science}\ }\textbf {\bibinfo
  {volume} {357}},\ \bibinfo {pages} {61} (\bibinfo {year}
  {2017}{\natexlab{a}})}\BibitemShut {NoStop}%
\bibitem [{\citenamefont {Benalcazar}\ \emph
  {et~al.}(2017{\natexlab{b}})\citenamefont {Benalcazar}, \citenamefont
  {Bernevig},\ and\ \citenamefont {Hughes}}]{PhysRevB.96.245115}%
  \BibitemOpen
  \bibfield  {author} {\bibinfo {author} {\bibfnamefont {W.~A.}\ \bibnamefont
  {Benalcazar}}, \bibinfo {author} {\bibfnamefont {B.~A.}\ \bibnamefont
  {Bernevig}},\ and\ \bibinfo {author} {\bibfnamefont {T.~L.}\ \bibnamefont
  {Hughes}},\ }\bibfield  {title} {\bibinfo {title} {{Electric multipole
  moments, topological multipole moment pumping, and chiral hinge states in
  crystalline insulators}},\ }\href
  {https://doi.org/10.1103/PhysRevB.96.245115} {\bibfield  {journal} {\bibinfo
  {journal} {Phys. Rev. B}\ }\textbf {\bibinfo {volume} {96}},\ \bibinfo
  {pages} {245115} (\bibinfo {year} {2017}{\natexlab{b}})}\BibitemShut
  {NoStop}%
\bibitem [{\citenamefont {Langbehn}\ \emph {et~al.}(2017)\citenamefont
  {Langbehn}, \citenamefont {Peng}, \citenamefont {Trifunovic}, \citenamefont
  {von Oppen},\ and\ \citenamefont {Brouwer}}]{PhysRevLett.119.246401}%
  \BibitemOpen
  \bibfield  {author} {\bibinfo {author} {\bibfnamefont {J.}~\bibnamefont
  {Langbehn}}, \bibinfo {author} {\bibfnamefont {Y.}~\bibnamefont {Peng}},
  \bibinfo {author} {\bibfnamefont {L.}~\bibnamefont {Trifunovic}}, \bibinfo
  {author} {\bibfnamefont {F.}~\bibnamefont {von Oppen}},\ and\ \bibinfo
  {author} {\bibfnamefont {P.~W.}\ \bibnamefont {Brouwer}},\ }\bibfield
  {title} {\bibinfo {title} {{Reflection-Symmetric Second-Order Topological
  Insulators and Superconductors}},\ }\href
  {https://doi.org/10.1103/PhysRevLett.119.246401} {\bibfield  {journal}
  {\bibinfo  {journal} {Phys. Rev. Lett.}\ }\textbf {\bibinfo {volume} {119}},\
  \bibinfo {pages} {246401} (\bibinfo {year} {2017})}\BibitemShut {NoStop}%
\bibitem [{\citenamefont {Song}\ \emph {et~al.}(2017)\citenamefont {Song},
  \citenamefont {Fang},\ and\ \citenamefont {Fang}}]{PhysRevLett.119.246402}%
  \BibitemOpen
  \bibfield  {author} {\bibinfo {author} {\bibfnamefont {Z.}~\bibnamefont
  {Song}}, \bibinfo {author} {\bibfnamefont {Z.}~\bibnamefont {Fang}},\ and\
  \bibinfo {author} {\bibfnamefont {C.}~\bibnamefont {Fang}},\ }\bibfield
  {title} {\bibinfo {title} {{$(d\ensuremath{-}2)$-Dimensional Edge States of
  Rotation Symmetry Protected Topological States}},\ }\href
  {https://doi.org/10.1103/PhysRevLett.119.246402} {\bibfield  {journal}
  {\bibinfo  {journal} {Phys. Rev. Lett.}\ }\textbf {\bibinfo {volume} {119}},\
  \bibinfo {pages} {246402} (\bibinfo {year} {2017})}\BibitemShut {NoStop}%
\bibitem [{\citenamefont {Schindler}\ \emph
  {et~al.}(2018{\natexlab{a}})\citenamefont {Schindler}, \citenamefont {Cook},
  \citenamefont {Vergniory}, \citenamefont {Wang}, \citenamefont {Parkin},
  \citenamefont {Bernevig},\ and\ \citenamefont
  {Neupert}}]{schindler2018higher}%
  \BibitemOpen
  \bibfield  {author} {\bibinfo {author} {\bibfnamefont {F.}~\bibnamefont
  {Schindler}}, \bibinfo {author} {\bibfnamefont {A.~M.}\ \bibnamefont {Cook}},
  \bibinfo {author} {\bibfnamefont {M.~G.}\ \bibnamefont {Vergniory}}, \bibinfo
  {author} {\bibfnamefont {Z.}~\bibnamefont {Wang}}, \bibinfo {author}
  {\bibfnamefont {S.~S.}\ \bibnamefont {Parkin}}, \bibinfo {author}
  {\bibfnamefont {B.~A.}\ \bibnamefont {Bernevig}},\ and\ \bibinfo {author}
  {\bibfnamefont {T.}~\bibnamefont {Neupert}},\ }\bibfield  {title} {\bibinfo
  {title} {{Higher-order topological insulators}},\ }\href
  {https://doi.org/10.1126/sciadv.aat0346} {\bibfield  {journal} {\bibinfo
  {journal} {Sci. Adv.}\ }\textbf {\bibinfo {volume} {4}},\ \bibinfo {pages}
  {eaat0346} (\bibinfo {year} {2018}{\natexlab{a}})}\BibitemShut {NoStop}%
\bibitem [{\citenamefont {Fang}\ and\ \citenamefont
  {Fu}(2019)}]{fang2017rotation}%
  \BibitemOpen
  \bibfield  {author} {\bibinfo {author} {\bibfnamefont {C.}~\bibnamefont
  {Fang}}\ and\ \bibinfo {author} {\bibfnamefont {L.}~\bibnamefont {Fu}},\
  }\bibfield  {title} {\bibinfo {title} {{New classes of topological
  crystalline insulators having surface rotation anomaly}},\ }\href
  {https://advances.sciencemag.org/content/5/12/eaat2374} {\bibfield  {journal}
  {\bibinfo  {journal} {Sci. Adv.}\ }\textbf {\bibinfo {volume} {5}},\ \bibinfo
  {pages} {eaat2374} (\bibinfo {year} {2019})}\BibitemShut {NoStop}%
\bibitem [{\citenamefont {Geier}\ \emph {et~al.}(2018)\citenamefont {Geier},
  \citenamefont {Trifunovic}, \citenamefont {Hoskam},\ and\ \citenamefont
  {Brouwer}}]{PhysRevB.97.205135}%
  \BibitemOpen
  \bibfield  {author} {\bibinfo {author} {\bibfnamefont {M.}~\bibnamefont
  {Geier}}, \bibinfo {author} {\bibfnamefont {L.}~\bibnamefont {Trifunovic}},
  \bibinfo {author} {\bibfnamefont {M.}~\bibnamefont {Hoskam}},\ and\ \bibinfo
  {author} {\bibfnamefont {P.~W.}\ \bibnamefont {Brouwer}},\ }\bibfield
  {title} {\bibinfo {title} {{Second-order topological insulators and
  superconductors with an order-two crystalline symmetry}},\ }\href
  {https://doi.org/10.1103/PhysRevB.97.205135} {\bibfield  {journal} {\bibinfo
  {journal} {Phys. Rev. B}\ }\textbf {\bibinfo {volume} {97}},\ \bibinfo
  {pages} {205135} (\bibinfo {year} {2018})}\BibitemShut {NoStop}%
\bibitem [{\citenamefont {Kunst}\ \emph {et~al.}(2018)\citenamefont {Kunst},
  \citenamefont {van Miert},\ and\ \citenamefont
  {Bergholtz}}]{PhysRevB.97.241405}%
  \BibitemOpen
  \bibfield  {author} {\bibinfo {author} {\bibfnamefont {F.~K.}\ \bibnamefont
  {Kunst}}, \bibinfo {author} {\bibfnamefont {G.}~\bibnamefont {van Miert}},\
  and\ \bibinfo {author} {\bibfnamefont {E.~J.}\ \bibnamefont {Bergholtz}},\
  }\bibfield  {title} {\bibinfo {title} {{Lattice models with exactly solvable
  topological hinge and corner states}},\ }\href
  {https://doi.org/10.1103/PhysRevB.97.241405} {\bibfield  {journal} {\bibinfo
  {journal} {Phys. Rev. B}\ }\textbf {\bibinfo {volume} {97}},\ \bibinfo
  {pages} {241405(R)} (\bibinfo {year} {2018})}\BibitemShut {NoStop}%
\bibitem [{\citenamefont {Schindler}\ \emph
  {et~al.}(2018{\natexlab{b}})\citenamefont {Schindler}, \citenamefont {Wang},
  \citenamefont {Vergniory}, \citenamefont {Cook}, \citenamefont {Murani},
  \citenamefont {Sengupta}, \citenamefont {Kasumov}, \citenamefont {Deblock},
  \citenamefont {Jeon}, \citenamefont {Drozdov}, \citenamefont {Bouchiat},
  \citenamefont {Gu\'eron}, \citenamefont {Yazdani}, \citenamefont {Bernevig},\
  and\ \citenamefont {Neupert}}]{schindler2018higherbismuth}%
  \BibitemOpen
  \bibfield  {author} {\bibinfo {author} {\bibfnamefont {F.}~\bibnamefont
  {Schindler}}, \bibinfo {author} {\bibfnamefont {Z.}~\bibnamefont {Wang}},
  \bibinfo {author} {\bibfnamefont {M.~G.}\ \bibnamefont {Vergniory}}, \bibinfo
  {author} {\bibfnamefont {A.~M.}\ \bibnamefont {Cook}}, \bibinfo {author}
  {\bibfnamefont {A.}~\bibnamefont {Murani}}, \bibinfo {author} {\bibfnamefont
  {S.}~\bibnamefont {Sengupta}}, \bibinfo {author} {\bibfnamefont {A.~Y.}\
  \bibnamefont {Kasumov}}, \bibinfo {author} {\bibfnamefont {R.}~\bibnamefont
  {Deblock}}, \bibinfo {author} {\bibfnamefont {S.}~\bibnamefont {Jeon}},
  \bibinfo {author} {\bibfnamefont {I.}~\bibnamefont {Drozdov}}, \bibinfo
  {author} {\bibfnamefont {H.}~\bibnamefont {Bouchiat}}, \bibinfo {author}
  {\bibfnamefont {S.}~\bibnamefont {Gu\'eron}}, \bibinfo {author}
  {\bibfnamefont {A.}~\bibnamefont {Yazdani}}, \bibinfo {author} {\bibfnamefont
  {B.~A.}\ \bibnamefont {Bernevig}},\ and\ \bibinfo {author} {\bibfnamefont
  {T.}~\bibnamefont {Neupert}},\ }\bibfield  {title} {\bibinfo {title}
  {{Higher-order topology in bismuth}},\ }\href
  {https://doi.org/10.1038/s41567-018-0224-7} {\bibfield  {journal} {\bibinfo
  {journal} {Nat. Phys.}\ }\textbf {\bibinfo {volume} {14}},\ \bibinfo {pages}
  {918} (\bibinfo {year} {2018}{\natexlab{b}})}\BibitemShut {NoStop}%
\bibitem [{\citenamefont {Xie}\ \emph {et~al.}(2018)\citenamefont {Xie},
  \citenamefont {Wang}, \citenamefont {Wang}, \citenamefont {Zhu},
  \citenamefont {Jiang}, \citenamefont {Lu},\ and\ \citenamefont
  {Chen}}]{PhysRevB.98.205147}%
  \BibitemOpen
  \bibfield  {author} {\bibinfo {author} {\bibfnamefont {B.-Y.}\ \bibnamefont
  {Xie}}, \bibinfo {author} {\bibfnamefont {H.-F.}\ \bibnamefont {Wang}},
  \bibinfo {author} {\bibfnamefont {H.-X.}\ \bibnamefont {Wang}}, \bibinfo
  {author} {\bibfnamefont {X.-Y.}\ \bibnamefont {Zhu}}, \bibinfo {author}
  {\bibfnamefont {J.-H.}\ \bibnamefont {Jiang}}, \bibinfo {author}
  {\bibfnamefont {M.-H.}\ \bibnamefont {Lu}},\ and\ \bibinfo {author}
  {\bibfnamefont {Y.-F.}\ \bibnamefont {Chen}},\ }\bibfield  {title} {\bibinfo
  {title} {{Second-order photonic topological insulator with corner states}},\
  }\href {https://doi.org/10.1103/PhysRevB.98.205147} {\bibfield  {journal}
  {\bibinfo  {journal} {Phys. Rev. B}\ }\textbf {\bibinfo {volume} {98}},\
  \bibinfo {pages} {205147} (\bibinfo {year} {2018})}\BibitemShut {NoStop}%
\bibitem [{\citenamefont {Serra-Garcia}\ \emph {et~al.}(2018)\citenamefont
  {Serra-Garcia}, \citenamefont {Peri}, \citenamefont {S{\"u}sstrunk},
  \citenamefont {Bilal}, \citenamefont {Larsen}, \citenamefont {Villanueva},\
  and\ \citenamefont {Huber}}]{serra2018observationnature555}%
  \BibitemOpen
  \bibfield  {author} {\bibinfo {author} {\bibfnamefont {M.}~\bibnamefont
  {Serra-Garcia}}, \bibinfo {author} {\bibfnamefont {V.}~\bibnamefont {Peri}},
  \bibinfo {author} {\bibfnamefont {R.}~\bibnamefont {S{\"u}sstrunk}}, \bibinfo
  {author} {\bibfnamefont {O.~R.}\ \bibnamefont {Bilal}}, \bibinfo {author}
  {\bibfnamefont {T.}~\bibnamefont {Larsen}}, \bibinfo {author} {\bibfnamefont
  {L.~G.}\ \bibnamefont {Villanueva}},\ and\ \bibinfo {author} {\bibfnamefont
  {S.~D.}\ \bibnamefont {Huber}},\ }\bibfield  {title} {\bibinfo {title}
  {{Observation of a phononic quadrupole topological insulator}},\ }\href
  {https://doi.org/10.1038/nature25156} {\bibfield  {journal} {\bibinfo
  {journal} {Nature}\ }\textbf {\bibinfo {volume} {555}},\ \bibinfo {pages}
  {342} (\bibinfo {year} {2018})}\BibitemShut {NoStop}%
\bibitem [{\citenamefont {Peterson}\ \emph {et~al.}(2018)\citenamefont
  {Peterson}, \citenamefont {Benalcazar}, \citenamefont {Hughes},\ and\
  \citenamefont {Bahl}}]{peterson2018quantizedNature7695}%
  \BibitemOpen
  \bibfield  {author} {\bibinfo {author} {\bibfnamefont {C.~W.}\ \bibnamefont
  {Peterson}}, \bibinfo {author} {\bibfnamefont {W.~A.}\ \bibnamefont
  {Benalcazar}}, \bibinfo {author} {\bibfnamefont {T.~L.}\ \bibnamefont
  {Hughes}},\ and\ \bibinfo {author} {\bibfnamefont {G.}~\bibnamefont {Bahl}},\
  }\bibfield  {title} {\bibinfo {title} {{A quantized microwave quadrupole
  insulator with topologically protected corner states}},\ }\href
  {https://doi.org/10.1038/nature25777} {\bibfield  {journal} {\bibinfo
  {journal} {Nature}\ }\textbf {\bibinfo {volume} {555}},\ \bibinfo {pages}
  {346} (\bibinfo {year} {2018})}\BibitemShut {NoStop}%
\bibitem [{\citenamefont {Imhof}\ \emph {et~al.}(2018)\citenamefont {Imhof},
  \citenamefont {Berger}, \citenamefont {Bayer}, \citenamefont {Brehm},
  \citenamefont {Molenkamp}, \citenamefont {Kiessling}, \citenamefont
  {Schindler}, \citenamefont {Lee}, \citenamefont {Greiter}, \citenamefont
  {Neupert},\ and\ \citenamefont {Thomale}}]{imhof2018topolectricalnatphys}%
  \BibitemOpen
  \bibfield  {author} {\bibinfo {author} {\bibfnamefont {S.}~\bibnamefont
  {Imhof}}, \bibinfo {author} {\bibfnamefont {C.}~\bibnamefont {Berger}},
  \bibinfo {author} {\bibfnamefont {F.}~\bibnamefont {Bayer}}, \bibinfo
  {author} {\bibfnamefont {J.}~\bibnamefont {Brehm}}, \bibinfo {author}
  {\bibfnamefont {L.~W.}\ \bibnamefont {Molenkamp}}, \bibinfo {author}
  {\bibfnamefont {T.}~\bibnamefont {Kiessling}}, \bibinfo {author}
  {\bibfnamefont {F.}~\bibnamefont {Schindler}}, \bibinfo {author}
  {\bibfnamefont {C.~H.}\ \bibnamefont {Lee}}, \bibinfo {author} {\bibfnamefont
  {M.}~\bibnamefont {Greiter}}, \bibinfo {author} {\bibfnamefont
  {T.}~\bibnamefont {Neupert}},\ and\ \bibinfo {author} {\bibfnamefont
  {R.}~\bibnamefont {Thomale}},\ }\bibfield  {title} {\bibinfo {title}
  {{Topolectrical-circuit realization of topological corner modes}},\ }\href
  {https://doi.org/10.1038/s41567-018-0246-1} {\bibfield  {journal} {\bibinfo
  {journal} {Nat. Phys.}\ }\textbf {\bibinfo {volume} {14}},\ \bibinfo {pages}
  {925} (\bibinfo {year} {2018})}\BibitemShut {NoStop}%
\bibitem [{\citenamefont {Peng}\ and\ \citenamefont
  {Refael}(2019)}]{PhysRevLett.123.016806}%
  \BibitemOpen
  \bibfield  {author} {\bibinfo {author} {\bibfnamefont {Y.}~\bibnamefont
  {Peng}}\ and\ \bibinfo {author} {\bibfnamefont {G.}~\bibnamefont {Refael}},\
  }\bibfield  {title} {\bibinfo {title} {Floquet second-order topological
  insulators from nonsymmorphic space-time symmetries},\ }\href
  {https://doi.org/10.1103/PhysRevLett.123.016806} {\bibfield  {journal}
  {\bibinfo  {journal} {Phys. Rev. Lett.}\ }\textbf {\bibinfo {volume} {123}},\
  \bibinfo {pages} {016806} (\bibinfo {year} {2019})}\BibitemShut {NoStop}%
\bibitem [{\citenamefont {Wang}\ \emph
  {et~al.}(2019{\natexlab{b}})\citenamefont {Wang}, \citenamefont {Wieder},
  \citenamefont {Li}, \citenamefont {Yan},\ and\ \citenamefont
  {Bernevig}}]{wang2018higher}%
  \BibitemOpen
  \bibfield  {author} {\bibinfo {author} {\bibfnamefont {Z.}~\bibnamefont
  {Wang}}, \bibinfo {author} {\bibfnamefont {B.~J.}\ \bibnamefont {Wieder}},
  \bibinfo {author} {\bibfnamefont {J.}~\bibnamefont {Li}}, \bibinfo {author}
  {\bibfnamefont {B.}~\bibnamefont {Yan}},\ and\ \bibinfo {author}
  {\bibfnamefont {B.~A.}\ \bibnamefont {Bernevig}},\ }\bibfield  {title}
  {\bibinfo {title} {{Higher-Order Topology, Monopole Nodal Lines, and the
  Origin of Large Fermi Arcs in Transition Metal Dichalcogenides
  $X{\mathrm{Te}}_{2}$ ($X=\mathrm{Mo},\mathrm{W}$)}},\ }\href
  {https://doi.org/10.1103/PhysRevLett.123.186401} {\bibfield  {journal}
  {\bibinfo  {journal} {Phys. Rev. Lett.}\ }\textbf {\bibinfo {volume} {123}},\
  \bibinfo {pages} {186401} (\bibinfo {year} {2019}{\natexlab{b}})}\BibitemShut
  {NoStop}%
\bibitem [{\citenamefont {Sheng}\ \emph {et~al.}(2019)\citenamefont {Sheng},
  \citenamefont {Chen}, \citenamefont {Liu}, \citenamefont {Chen},
  \citenamefont {Yu}, \citenamefont {Zhao},\ and\ \citenamefont
  {Yang}}]{sheng2019two}%
  \BibitemOpen
  \bibfield  {author} {\bibinfo {author} {\bibfnamefont {X.-L.}\ \bibnamefont
  {Sheng}}, \bibinfo {author} {\bibfnamefont {C.}~\bibnamefont {Chen}},
  \bibinfo {author} {\bibfnamefont {H.}~\bibnamefont {Liu}}, \bibinfo {author}
  {\bibfnamefont {Z.}~\bibnamefont {Chen}}, \bibinfo {author} {\bibfnamefont
  {Z.-M.}\ \bibnamefont {Yu}}, \bibinfo {author} {\bibfnamefont {Y.~X.}\
  \bibnamefont {Zhao}},\ and\ \bibinfo {author} {\bibfnamefont {S.~A.}\
  \bibnamefont {Yang}},\ }\bibfield  {title} {\bibinfo {title} {Two-dimensional
  second-order topological insulator in graphdiyne},\ }\href
  {https://doi.org/10.1103/PhysRevLett.123.256402} {\bibfield  {journal}
  {\bibinfo  {journal} {Phys. Rev. Lett.}\ }\textbf {\bibinfo {volume} {123}},\
  \bibinfo {pages} {256402} (\bibinfo {year} {2019})}\BibitemShut {NoStop}%
\bibitem [{\citenamefont {Fukui}\ and\ \citenamefont
  {Hatsugai}(2018)}]{PhysRevB.98.035147}%
  \BibitemOpen
  \bibfield  {author} {\bibinfo {author} {\bibfnamefont {T.}~\bibnamefont
  {Fukui}}\ and\ \bibinfo {author} {\bibfnamefont {Y.}~\bibnamefont
  {Hatsugai}},\ }\bibfield  {title} {\bibinfo {title} {Entanglement
  polarization for the topological quadrupole phase},\ }\href
  {https://doi.org/10.1103/PhysRevB.98.035147} {\bibfield  {journal} {\bibinfo
  {journal} {Phys. Rev. B}\ }\textbf {\bibinfo {volume} {98}},\ \bibinfo
  {pages} {035147} (\bibinfo {year} {2018})}\BibitemShut {NoStop}%
\bibitem [{\citenamefont {Okugawa}\ \emph {et~al.}(2019)\citenamefont
  {Okugawa}, \citenamefont {Hayashi},\ and\ \citenamefont
  {Nakanishi}}]{okugawa2019second}%
  \BibitemOpen
  \bibfield  {author} {\bibinfo {author} {\bibfnamefont {R.}~\bibnamefont
  {Okugawa}}, \bibinfo {author} {\bibfnamefont {S.}~\bibnamefont {Hayashi}},\
  and\ \bibinfo {author} {\bibfnamefont {T.}~\bibnamefont {Nakanishi}},\
  }\bibfield  {title} {\bibinfo {title} {{Second-order topological phases
  protected by chiral symmetry}},\ }\href
  {https://doi.org/10.1103/PhysRevB.100.235302} {\bibfield  {journal} {\bibinfo
   {journal} {Phys. Rev. B}\ }\textbf {\bibinfo {volume} {100}},\ \bibinfo
  {pages} {235302} (\bibinfo {year} {2019})}\BibitemShut {NoStop}%
\bibitem [{\citenamefont {Ghosh}\ \emph {et~al.}(2020)\citenamefont {Ghosh},
  \citenamefont {Paul},\ and\ \citenamefont {Saha}}]{ghosh2019engineering}%
  \BibitemOpen
  \bibfield  {author} {\bibinfo {author} {\bibfnamefont {A.~K.}\ \bibnamefont
  {Ghosh}}, \bibinfo {author} {\bibfnamefont {G.~C.}\ \bibnamefont {Paul}},\
  and\ \bibinfo {author} {\bibfnamefont {A.}~\bibnamefont {Saha}},\ }\bibfield
  {title} {\bibinfo {title} {{Higher order topological insulator via periodic
  driving}},\ }\href {https://doi.org/10.1103/PhysRevB.101.235403} {\bibfield
  {journal} {\bibinfo  {journal} {Phys. Rev. B}\ }\textbf {\bibinfo {volume}
  {101}},\ \bibinfo {pages} {235403} (\bibinfo {year} {2020})}\BibitemShut
  {NoStop}%
\bibitem [{\citenamefont {Agarwala}\ \emph {et~al.}(2020)\citenamefont
  {Agarwala}, \citenamefont {Juri\ifmmode \check{c}\else
  \v{c}\fi{}i\ifmmode~\acute{c}\else \'{c}\fi{}},\ and\ \citenamefont
  {Roy}}]{agarwala2019higher}%
  \BibitemOpen
  \bibfield  {author} {\bibinfo {author} {\bibfnamefont {A.}~\bibnamefont
  {Agarwala}}, \bibinfo {author} {\bibfnamefont {V.}~\bibnamefont {Juri\ifmmode
  \check{c}\else \v{c}\fi{}i\ifmmode~\acute{c}\else \'{c}\fi{}}},\ and\
  \bibinfo {author} {\bibfnamefont {B.}~\bibnamefont {Roy}},\ }\bibfield
  {title} {\bibinfo {title} {Higher-order topological insulators in amorphous
  solids},\ }\href {https://doi.org/10.1103/PhysRevResearch.2.012067}
  {\bibfield  {journal} {\bibinfo  {journal} {Phys. Rev. Research}\ }\textbf
  {\bibinfo {volume} {2}},\ \bibinfo {pages} {012067(R)} (\bibinfo {year}
  {2020})}\BibitemShut {NoStop}%
\bibitem [{\citenamefont {Chen}\ \emph
  {et~al.}(2020{\natexlab{a}})\citenamefont {Chen}, \citenamefont {Chen},
  \citenamefont {Gao}, \citenamefont {Zhou},\ and\ \citenamefont
  {Xu}}]{chen2019higher}%
  \BibitemOpen
  \bibfield  {author} {\bibinfo {author} {\bibfnamefont {R.}~\bibnamefont
  {Chen}}, \bibinfo {author} {\bibfnamefont {C.-Z.}\ \bibnamefont {Chen}},
  \bibinfo {author} {\bibfnamefont {J.-H.}\ \bibnamefont {Gao}}, \bibinfo
  {author} {\bibfnamefont {B.}~\bibnamefont {Zhou}},\ and\ \bibinfo {author}
  {\bibfnamefont {D.-H.}\ \bibnamefont {Xu}},\ }\bibfield  {title} {\bibinfo
  {title} {Higher-order topological insulators in quasicrystals},\ }\href
  {https://doi.org/10.1103/PhysRevLett.124.036803} {\bibfield  {journal}
  {\bibinfo  {journal} {Phys. Rev. Lett.}\ }\textbf {\bibinfo {volume} {124}},\
  \bibinfo {pages} {036803} (\bibinfo {year} {2020}{\natexlab{a}})}\BibitemShut
  {NoStop}%
\bibitem [{\citenamefont {Kheirkhah}\ \emph {et~al.}(2020)\citenamefont
  {Kheirkhah}, \citenamefont {Yan}, \citenamefont {Nagai},\ and\ \citenamefont
  {Marsiglio}}]{PhysRevLett.125.017001}%
  \BibitemOpen
  \bibfield  {author} {\bibinfo {author} {\bibfnamefont {M.}~\bibnamefont
  {Kheirkhah}}, \bibinfo {author} {\bibfnamefont {Z.}~\bibnamefont {Yan}},
  \bibinfo {author} {\bibfnamefont {Y.}~\bibnamefont {Nagai}},\ and\ \bibinfo
  {author} {\bibfnamefont {F.}~\bibnamefont {Marsiglio}},\ }\bibfield  {title}
  {\bibinfo {title} {{First- and Second-Order Topological Superconductivity and
  Temperature-Driven Topological Phase Transitions in the Extended Hubbard
  Model with Spin-Orbit Coupling}},\ }\href
  {https://doi.org/10.1103/PhysRevLett.125.017001} {\bibfield  {journal}
  {\bibinfo  {journal} {Phys. Rev. Lett.}\ }\textbf {\bibinfo {volume} {125}},\
  \bibinfo {pages} {017001} (\bibinfo {year} {2020})}\BibitemShut {NoStop}%
\bibitem [{\citenamefont {Hirayama}\ \emph {et~al.}(2020)\citenamefont
  {Hirayama}, \citenamefont {Takahashi}, \citenamefont {Matsuishi},
  \citenamefont {Hosono},\ and\ \citenamefont {Murakami}}]{hirayama2020higher}%
  \BibitemOpen
  \bibfield  {author} {\bibinfo {author} {\bibfnamefont {M.}~\bibnamefont
  {Hirayama}}, \bibinfo {author} {\bibfnamefont {R.}~\bibnamefont {Takahashi}},
  \bibinfo {author} {\bibfnamefont {S.}~\bibnamefont {Matsuishi}}, \bibinfo
  {author} {\bibfnamefont {H.}~\bibnamefont {Hosono}},\ and\ \bibinfo {author}
  {\bibfnamefont {S.}~\bibnamefont {Murakami}},\ }\bibfield  {title} {\bibinfo
  {title} {{Higher-order topological crystalline insulating phase and quantized
  hinge charge in topological electride apatite}},\ }\href
  {https://doi.org/10.1103/PhysRevResearch.2.043131} {\bibfield  {journal}
  {\bibinfo  {journal} {Phys. Rev. Research}\ }\textbf {\bibinfo {volume}
  {2}},\ \bibinfo {pages} {043131} (\bibinfo {year} {2020})}\BibitemShut
  {NoStop}%
\bibitem [{\citenamefont {Chen}\ \emph
  {et~al.}(2020{\natexlab{b}})\citenamefont {Chen}, \citenamefont {Song},
  \citenamefont {Zhao}, \citenamefont {Chen}, \citenamefont {Yu}, \citenamefont
  {Sheng},\ and\ \citenamefont {Yang}}]{chen2020universal}%
  \BibitemOpen
  \bibfield  {author} {\bibinfo {author} {\bibfnamefont {C.}~\bibnamefont
  {Chen}}, \bibinfo {author} {\bibfnamefont {Z.}~\bibnamefont {Song}}, \bibinfo
  {author} {\bibfnamefont {J.-Z.}\ \bibnamefont {Zhao}}, \bibinfo {author}
  {\bibfnamefont {Z.}~\bibnamefont {Chen}}, \bibinfo {author} {\bibfnamefont
  {Z.-M.}\ \bibnamefont {Yu}}, \bibinfo {author} {\bibfnamefont {X.-L.}\
  \bibnamefont {Sheng}},\ and\ \bibinfo {author} {\bibfnamefont {S.~A.}\
  \bibnamefont {Yang}},\ }\bibfield  {title} {\bibinfo {title} {{Universal
  Approach to Magnetic Second-Order Topological Insulator}},\ }\href
  {https://doi.org/10.1103/PhysRevLett.125.056402} {\bibfield  {journal}
  {\bibinfo  {journal} {Phys. Rev. Lett.}\ }\textbf {\bibinfo {volume} {125}},\
  \bibinfo {pages} {056402} (\bibinfo {year} {2020}{\natexlab{b}})}\BibitemShut
  {NoStop}%
\bibitem [{\citenamefont {Arai}\ and\ \citenamefont
  {Murakami}(2021)}]{Arai:2021vc}%
  \BibitemOpen
  \bibfield  {author} {\bibinfo {author} {\bibfnamefont {N.}~\bibnamefont
  {Arai}}\ and\ \bibinfo {author} {\bibfnamefont {S.}~\bibnamefont
  {Murakami}},\ }\bibfield  {title} {\bibinfo {title} {{Anisotropic Penetration
  Depths of Corner States in a Higher-Order Topological Insulator}},\ }\href
  {https://doi.org/10.7566/JPSJ.90.074711} {\bibfield  {journal} {\bibinfo
  {journal} {J. Phys. Soc. Jpn.}\ }\textbf {\bibinfo {volume} {90}},\ \bibinfo
  {pages} {074711} (\bibinfo {year} {2021})}\BibitemShut {NoStop}%
\bibitem [{\citenamefont {Nagasato}\ \emph {et~al.}(2021)\citenamefont
  {Nagasato}, \citenamefont {Takane}, \citenamefont {Yoshimura}, \citenamefont
  {Hayashi},\ and\ \citenamefont {Nakanishi}}]{Nagasato:2021tg}%
  \BibitemOpen
  \bibfield  {author} {\bibinfo {author} {\bibfnamefont {Y.}~\bibnamefont
  {Nagasato}}, \bibinfo {author} {\bibfnamefont {Y.}~\bibnamefont {Takane}},
  \bibinfo {author} {\bibfnamefont {Y.}~\bibnamefont {Yoshimura}}, \bibinfo
  {author} {\bibfnamefont {S.}~\bibnamefont {Hayashi}},\ and\ \bibinfo {author}
  {\bibfnamefont {T.}~\bibnamefont {Nakanishi}},\ }\bibfield  {title} {\bibinfo
  {title} {{Gapless States Localized along a Staircase Edge in Second-Order
  Topological Insulators}},\ }\href {https://doi.org/10.7566/JPSJ.90.104703}
  {\bibfield  {journal} {\bibinfo  {journal} {J. Phys. Soc. Jpn.}\ }\textbf
  {\bibinfo {volume} {90}},\ \bibinfo {pages} {104703} (\bibinfo {year}
  {2021})}\BibitemShut {NoStop}%
\bibitem [{\citenamefont {Ko\ifmmode~\check{s}\else \v{s}\fi{}ata}\ and\
  \citenamefont {Zilberberg}(2021)}]{PhysRevResearch.3.L032029}%
  \BibitemOpen
  \bibfield  {author} {\bibinfo {author} {\bibfnamefont {J.}~\bibnamefont
  {Ko\ifmmode~\check{s}\else \v{s}\fi{}ata}}\ and\ \bibinfo {author}
  {\bibfnamefont {O.}~\bibnamefont {Zilberberg}},\ }\bibfield  {title}
  {\bibinfo {title} {{Second-order topological modes in two-dimensional
  continuous media}},\ }\href
  {https://doi.org/10.1103/PhysRevResearch.3.L032029} {\bibfield  {journal}
  {\bibinfo  {journal} {Phys. Rev. Res.}\ }\textbf {\bibinfo {volume} {3}},\
  \bibinfo {pages} {L032029} (\bibinfo {year} {2021})}\BibitemShut {NoStop}%
\bibitem [{\citenamefont {Fu}\ \emph {et~al.}(2021)\citenamefont {Fu},
  \citenamefont {Hu},\ and\ \citenamefont {Shen}}]{PhysRevResearch.3.033177}%
  \BibitemOpen
  \bibfield  {author} {\bibinfo {author} {\bibfnamefont {B.}~\bibnamefont
  {Fu}}, \bibinfo {author} {\bibfnamefont {Z.-A.}\ \bibnamefont {Hu}},\ and\
  \bibinfo {author} {\bibfnamefont {S.-Q.}\ \bibnamefont {Shen}},\ }\bibfield
  {title} {\bibinfo {title} {{Bulk-hinge correspondence and three-dimensional
  quantum anomalous Hall effect in second-order topological insulators}},\
  }\href {https://doi.org/10.1103/PhysRevResearch.3.033177} {\bibfield
  {journal} {\bibinfo  {journal} {Phys. Rev. Res.}\ }\textbf {\bibinfo {volume}
  {3}},\ \bibinfo {pages} {033177} (\bibinfo {year} {2021})}\BibitemShut
  {NoStop}%
\bibitem [{\citenamefont {Zhao}\ \emph {et~al.}(2021)\citenamefont {Zhao},
  \citenamefont {Qiang}, \citenamefont {Lu},\ and\ \citenamefont
  {Xie}}]{PhysRevLett.127.176601}%
  \BibitemOpen
  \bibfield  {author} {\bibinfo {author} {\bibfnamefont {P.-L.}\ \bibnamefont
  {Zhao}}, \bibinfo {author} {\bibfnamefont {X.-B.}\ \bibnamefont {Qiang}},
  \bibinfo {author} {\bibfnamefont {H.-Z.}\ \bibnamefont {Lu}},\ and\ \bibinfo
  {author} {\bibfnamefont {X.~C.}\ \bibnamefont {Xie}},\ }\bibfield  {title}
  {\bibinfo {title} {{Coulomb Instabilities of a Three-Dimensional Higher-Order
  Topological Insulator}},\ }\href
  {https://doi.org/10.1103/PhysRevLett.127.176601} {\bibfield  {journal}
  {\bibinfo  {journal} {Phys. Rev. Lett.}\ }\textbf {\bibinfo {volume} {127}},\
  \bibinfo {pages} {176601} (\bibinfo {year} {2021})}\BibitemShut {NoStop}%
\bibitem [{\citenamefont {Wang}\ and\ \citenamefont
  {Wang}(2021)}]{PhysRevB.103.115118}%
  \BibitemOpen
  \bibfield  {author} {\bibinfo {author} {\bibfnamefont {C.}~\bibnamefont
  {Wang}}\ and\ \bibinfo {author} {\bibfnamefont {X.~R.}\ \bibnamefont
  {Wang}},\ }\bibfield  {title} {\bibinfo {title} {Robustness of helical hinge
  states of weak second-order topological insulators},\ }\href
  {https://doi.org/10.1103/PhysRevB.103.115118} {\bibfield  {journal} {\bibinfo
   {journal} {Phys. Rev. B}\ }\textbf {\bibinfo {volume} {103}},\ \bibinfo
  {pages} {115118} (\bibinfo {year} {2021})}\BibitemShut {NoStop}%
\bibitem [{\citenamefont {Qian}\ \emph {et~al.}(2021)\citenamefont {Qian},
  \citenamefont {Liu},\ and\ \citenamefont {Yao}}]{PhysRevB.104.245427}%
  \BibitemOpen
  \bibfield  {author} {\bibinfo {author} {\bibfnamefont {S.}~\bibnamefont
  {Qian}}, \bibinfo {author} {\bibfnamefont {C.-C.}\ \bibnamefont {Liu}},\ and\
  \bibinfo {author} {\bibfnamefont {Y.}~\bibnamefont {Yao}},\ }\bibfield
  {title} {\bibinfo {title} {{Second-order topological insulator state in
  hexagonal lattices and its abundant material candidates}},\ }\href
  {https://doi.org/10.1103/PhysRevB.104.245427} {\bibfield  {journal} {\bibinfo
   {journal} {Phys. Rev. B}\ }\textbf {\bibinfo {volume} {104}},\ \bibinfo
  {pages} {245427} (\bibinfo {year} {2021})}\BibitemShut {NoStop}%
\bibitem [{\citenamefont {Bunney}\ \emph {et~al.}(2022)\citenamefont {Bunney},
  \citenamefont {Mizoguchi}, \citenamefont {Hatsugai},\ and\ \citenamefont
  {Rachel}}]{PhysRevB.105.045113}%
  \BibitemOpen
  \bibfield  {author} {\bibinfo {author} {\bibfnamefont {M.}~\bibnamefont
  {Bunney}}, \bibinfo {author} {\bibfnamefont {T.}~\bibnamefont {Mizoguchi}},
  \bibinfo {author} {\bibfnamefont {Y.}~\bibnamefont {Hatsugai}},\ and\
  \bibinfo {author} {\bibfnamefont {S.}~\bibnamefont {Rachel}},\ }\bibfield
  {title} {\bibinfo {title} {{Competition of first-order and second-order
  topology on the honeycomb lattice}},\ }\href
  {https://doi.org/10.1103/PhysRevB.105.045113} {\bibfield  {journal} {\bibinfo
   {journal} {Phys. Rev. B}\ }\textbf {\bibinfo {volume} {105}},\ \bibinfo
  {pages} {045113} (\bibinfo {year} {2022})}\BibitemShut {NoStop}%
\bibitem [{\citenamefont {Naito}\ \emph {et~al.}(2022)\citenamefont {Naito},
  \citenamefont {Takahashi}, \citenamefont {Watanabe},\ and\ \citenamefont
  {Murakami}}]{PhysRevB.105.045126}%
  \BibitemOpen
  \bibfield  {author} {\bibinfo {author} {\bibfnamefont {K.}~\bibnamefont
  {Naito}}, \bibinfo {author} {\bibfnamefont {R.}~\bibnamefont {Takahashi}},
  \bibinfo {author} {\bibfnamefont {H.}~\bibnamefont {Watanabe}},\ and\
  \bibinfo {author} {\bibfnamefont {S.}~\bibnamefont {Murakami}},\ }\bibfield
  {title} {\bibinfo {title} {{Fractional hinge and corner charges in various
  crystal shapes with cubic symmetry}},\ }\href
  {https://doi.org/10.1103/PhysRevB.105.045126} {\bibfield  {journal} {\bibinfo
   {journal} {Phys. Rev. B}\ }\textbf {\bibinfo {volume} {105}},\ \bibinfo
  {pages} {045126} (\bibinfo {year} {2022})}\BibitemShut {NoStop}%
\bibitem [{\citenamefont {Tanaka}\ \emph
  {et~al.}(2022{\natexlab{a}})\citenamefont {Tanaka}, \citenamefont
  {Takahashi}, \citenamefont {Okugawa},\ and\ \citenamefont
  {Murakami}}]{PhysRevB.105.115119}%
  \BibitemOpen
  \bibfield  {author} {\bibinfo {author} {\bibfnamefont {Y.}~\bibnamefont
  {Tanaka}}, \bibinfo {author} {\bibfnamefont {R.}~\bibnamefont {Takahashi}},
  \bibinfo {author} {\bibfnamefont {R.}~\bibnamefont {Okugawa}},\ and\ \bibinfo
  {author} {\bibfnamefont {S.}~\bibnamefont {Murakami}},\ }\bibfield  {title}
  {\bibinfo {title} {{Rotoinversion-symmetric bulk-hinge correspondence and its
  applications to higher-order Weyl semimetals}},\ }\href
  {https://doi.org/10.1103/PhysRevB.105.115119} {\bibfield  {journal} {\bibinfo
   {journal} {Phys. Rev. B}\ }\textbf {\bibinfo {volume} {105}},\ \bibinfo
  {pages} {115119} (\bibinfo {year} {2022}{\natexlab{a}})}\BibitemShut
  {NoStop}%
\bibitem [{\citenamefont {Scammell}\ \emph {et~al.}(2022)\citenamefont
  {Scammell}, \citenamefont {Ingham}, \citenamefont {Geier},\ and\
  \citenamefont {Li}}]{PhysRevB.105.195149}%
  \BibitemOpen
  \bibfield  {author} {\bibinfo {author} {\bibfnamefont {H.~D.}\ \bibnamefont
  {Scammell}}, \bibinfo {author} {\bibfnamefont {J.}~\bibnamefont {Ingham}},
  \bibinfo {author} {\bibfnamefont {M.}~\bibnamefont {Geier}},\ and\ \bibinfo
  {author} {\bibfnamefont {T.}~\bibnamefont {Li}},\ }\bibfield  {title}
  {\bibinfo {title} {{Intrinsic first- and higher-order topological
  superconductivity in a doped topological insulator}},\ }\href
  {https://doi.org/10.1103/PhysRevB.105.195149} {\bibfield  {journal} {\bibinfo
   {journal} {Phys. Rev. B}\ }\textbf {\bibinfo {volume} {105}},\ \bibinfo
  {pages} {195149} (\bibinfo {year} {2022})}\BibitemShut {NoStop}%
\bibitem [{\citenamefont {Wu}\ \emph {et~al.}(2022)\citenamefont {Wu},
  \citenamefont {Guo}, \citenamefont {Zhang},\ and\ \citenamefont
  {Jiang}}]{PhysRevB.106.165401}%
  \BibitemOpen
  \bibfield  {author} {\bibinfo {author} {\bibfnamefont {B.-L.}\ \bibnamefont
  {Wu}}, \bibinfo {author} {\bibfnamefont {A.-M.}\ \bibnamefont {Guo}},
  \bibinfo {author} {\bibfnamefont {Z.-Q.}\ \bibnamefont {Zhang}},\ and\
  \bibinfo {author} {\bibfnamefont {H.}~\bibnamefont {Jiang}},\ }\bibfield
  {title} {\bibinfo {title} {Quantized charge-pumping in higher-order
  topological insulators},\ }\href
  {https://doi.org/10.1103/PhysRevB.106.165401} {\bibfield  {journal} {\bibinfo
   {journal} {Phys. Rev. B}\ }\textbf {\bibinfo {volume} {106}},\ \bibinfo
  {pages} {165401} (\bibinfo {year} {2022})}\BibitemShut {NoStop}%
\bibitem [{\citenamefont {Miao}\ \emph {et~al.}(2022)\citenamefont {Miao},
  \citenamefont {Sun},\ and\ \citenamefont {Zhang}}]{PhysRevB.106.165422}%
  \BibitemOpen
  \bibfield  {author} {\bibinfo {author} {\bibfnamefont {C.-M.}\ \bibnamefont
  {Miao}}, \bibinfo {author} {\bibfnamefont {Q.-F.}\ \bibnamefont {Sun}},\ and\
  \bibinfo {author} {\bibfnamefont {Y.-T.}\ \bibnamefont {Zhang}},\ }\bibfield
  {title} {\bibinfo {title} {{Second-order topological corner states in zigzag
  graphene nanoflake with different types of edge magnetic configurations}},\
  }\href {https://doi.org/10.1103/PhysRevB.106.165422} {\bibfield  {journal}
  {\bibinfo  {journal} {Phys. Rev. B}\ }\textbf {\bibinfo {volume} {106}},\
  \bibinfo {pages} {165422} (\bibinfo {year} {2022})}\BibitemShut {NoStop}%
\bibitem [{\citenamefont {Mu}\ \emph {et~al.}(2022{\natexlab{a}})\citenamefont
  {Mu}, \citenamefont {Liu}, \citenamefont {Hu},\ and\ \citenamefont
  {Wang}}]{Mu:2022tierg}%
  \BibitemOpen
  \bibfield  {author} {\bibinfo {author} {\bibfnamefont {H.}~\bibnamefont
  {Mu}}, \bibinfo {author} {\bibfnamefont {B.}~\bibnamefont {Liu}}, \bibinfo
  {author} {\bibfnamefont {T.}~\bibnamefont {Hu}},\ and\ \bibinfo {author}
  {\bibfnamefont {Z.}~\bibnamefont {Wang}},\ }\bibfield  {title} {\bibinfo
  {title} {{Kekul{\'e}Lattice in Graphdiyne: Coexistence of Phononic and
  Electronic Second-Order Topological Insulator}},\ }\href
  {https://doi.org/10.1021/acs.nanolett.1c04239} {\bibfield  {journal}
  {\bibinfo  {journal} {Nano Lett.}\ }\textbf {\bibinfo {volume} {22}},\
  \bibinfo {pages} {1122} (\bibinfo {year} {2022}{\natexlab{a}})}\BibitemShut
  {NoStop}%
\bibitem [{\citenamefont {Mu}\ \emph {et~al.}(2022{\natexlab{b}})\citenamefont
  {Mu}, \citenamefont {Zhao}, \citenamefont {Zhang},\ and\ \citenamefont
  {Wang}}]{Mu:2022ta}%
  \BibitemOpen
  \bibfield  {author} {\bibinfo {author} {\bibfnamefont {H.}~\bibnamefont
  {Mu}}, \bibinfo {author} {\bibfnamefont {G.}~\bibnamefont {Zhao}}, \bibinfo
  {author} {\bibfnamefont {H.}~\bibnamefont {Zhang}},\ and\ \bibinfo {author}
  {\bibfnamefont {Z.}~\bibnamefont {Wang}},\ }\bibfield  {title} {\bibinfo
  {title} {{Antiferromagnetic second-order topological insulator with
  fractional mass-kink}},\ }\href {https://doi.org/10.1038/s41524-022-00761-3}
  {\bibfield  {journal} {\bibinfo  {journal} {npj Comput. Mater.}\ }\textbf
  {\bibinfo {volume} {8}},\ \bibinfo {pages} {82} (\bibinfo {year}
  {2022}{\natexlab{b}})}\BibitemShut {NoStop}%
\bibitem [{\citenamefont {Tanaka}\ \emph {et~al.}(2012)\citenamefont {Tanaka},
  \citenamefont {Ren}, \citenamefont {Sato}, \citenamefont {Nakayama},
  \citenamefont {Souma}, \citenamefont {Takahashi}, \citenamefont {Segawa},\
  and\ \citenamefont {Ando}}]{Tanaka:2012wj}%
  \BibitemOpen
  \bibfield  {author} {\bibinfo {author} {\bibfnamefont {Y.}~\bibnamefont
  {Tanaka}}, \bibinfo {author} {\bibfnamefont {Z.}~\bibnamefont {Ren}},
  \bibinfo {author} {\bibfnamefont {T.}~\bibnamefont {Sato}}, \bibinfo {author}
  {\bibfnamefont {K.}~\bibnamefont {Nakayama}}, \bibinfo {author}
  {\bibfnamefont {S.}~\bibnamefont {Souma}}, \bibinfo {author} {\bibfnamefont
  {T.}~\bibnamefont {Takahashi}}, \bibinfo {author} {\bibfnamefont
  {K.}~\bibnamefont {Segawa}},\ and\ \bibinfo {author} {\bibfnamefont
  {Y.}~\bibnamefont {Ando}},\ }\bibfield  {title} {\bibinfo {title}
  {{Experimental realization of a topological crystalline insulator in SnTe}},\
  }\href {https://doi.org/10.1038/nphys2442} {\bibfield  {journal} {\bibinfo
  {journal} {Nature Physics}\ }\textbf {\bibinfo {volume} {8}},\ \bibinfo
  {pages} {800} (\bibinfo {year} {2012})}\BibitemShut {NoStop}%
\bibitem [{\citenamefont {Wang}\ \emph {et~al.}(2013)\citenamefont {Wang},
  \citenamefont {Tsai}, \citenamefont {Lin}, \citenamefont {Xu}, \citenamefont
  {Neupane}, \citenamefont {Hasan},\ and\ \citenamefont
  {Bansil}}]{PhysRevB.87.235317}%
  \BibitemOpen
  \bibfield  {author} {\bibinfo {author} {\bibfnamefont {Y.~J.}\ \bibnamefont
  {Wang}}, \bibinfo {author} {\bibfnamefont {W.-F.}\ \bibnamefont {Tsai}},
  \bibinfo {author} {\bibfnamefont {H.}~\bibnamefont {Lin}}, \bibinfo {author}
  {\bibfnamefont {S.-Y.}\ \bibnamefont {Xu}}, \bibinfo {author} {\bibfnamefont
  {M.}~\bibnamefont {Neupane}}, \bibinfo {author} {\bibfnamefont {M.~Z.}\
  \bibnamefont {Hasan}},\ and\ \bibinfo {author} {\bibfnamefont
  {A.}~\bibnamefont {Bansil}},\ }\bibfield  {title} {\bibinfo {title}
  {{Nontrivial spin texture of the coaxial Dirac cones on the surface of
  topological crystalline insulator SnTe}},\ }\href
  {https://doi.org/10.1103/PhysRevB.87.235317} {\bibfield  {journal} {\bibinfo
  {journal} {Phys. Rev. B}\ }\textbf {\bibinfo {volume} {87}},\ \bibinfo
  {pages} {235317} (\bibinfo {year} {2013})}\BibitemShut {NoStop}%
\bibitem [{\citenamefont {Kim}\ \emph {et~al.}(2015)\citenamefont {Kim},
  \citenamefont {Kane}, \citenamefont {Mele},\ and\ \citenamefont
  {Rappe}}]{PhysRevLett.115.086802}%
  \BibitemOpen
  \bibfield  {author} {\bibinfo {author} {\bibfnamefont {Y.}~\bibnamefont
  {Kim}}, \bibinfo {author} {\bibfnamefont {C.~L.}\ \bibnamefont {Kane}},
  \bibinfo {author} {\bibfnamefont {E.~J.}\ \bibnamefont {Mele}},\ and\
  \bibinfo {author} {\bibfnamefont {A.~M.}\ \bibnamefont {Rappe}},\ }\bibfield
  {title} {\bibinfo {title} {{Layered Topological Crystalline Insulators}},\
  }\href {https://doi.org/10.1103/PhysRevLett.115.086802} {\bibfield  {journal}
  {\bibinfo  {journal} {Phys. Rev. Lett.}\ }\textbf {\bibinfo {volume} {115}},\
  \bibinfo {pages} {086802} (\bibinfo {year} {2015})}\BibitemShut {NoStop}%
\bibitem [{\citenamefont {Cao}\ \emph {et~al.}(2021)\citenamefont {Cao},
  \citenamefont {Tang}, \citenamefont {Wang},\ and\ \citenamefont
  {Wan}}]{Cao_2021}%
  \BibitemOpen
  \bibfield  {author} {\bibinfo {author} {\bibfnamefont {Z.}~\bibnamefont
  {Cao}}, \bibinfo {author} {\bibfnamefont {F.}~\bibnamefont {Tang}}, \bibinfo
  {author} {\bibfnamefont {D.}~\bibnamefont {Wang}},\ and\ \bibinfo {author}
  {\bibfnamefont {X.}~\bibnamefont {Wan}},\ }\bibfield  {title} {\bibinfo
  {title} {{Systematic identification of mirror-protected topological
  crystalline insulators by first-principles calculations}},\ }\href
  {https://doi.org/10.1088/1367-2630/ac2ce2} {\bibfield  {journal} {\bibinfo
  {journal} {New. J. Phys.}\ }\textbf {\bibinfo {volume} {23}},\ \bibinfo
  {pages} {103032} (\bibinfo {year} {2021})}\BibitemShut {NoStop}%
\bibitem [{\citenamefont {Khalaf}(2018)}]{PhysRevB.97.205136}%
  \BibitemOpen
  \bibfield  {author} {\bibinfo {author} {\bibfnamefont {E.}~\bibnamefont
  {Khalaf}},\ }\bibfield  {title} {\bibinfo {title} {{Higher-order topological
  insulators and superconductors protected by inversion symmetry}},\ }\href
  {https://doi.org/10.1103/PhysRevB.97.205136} {\bibfield  {journal} {\bibinfo
  {journal} {Phys. Rev. B}\ }\textbf {\bibinfo {volume} {97}},\ \bibinfo
  {pages} {205136} (\bibinfo {year} {2018})}\BibitemShut {NoStop}%
\bibitem [{\citenamefont {Matsugatani}\ and\ \citenamefont
  {Watanabe}(2018)}]{PhysRevB.98.205129}%
  \BibitemOpen
  \bibfield  {author} {\bibinfo {author} {\bibfnamefont {A.}~\bibnamefont
  {Matsugatani}}\ and\ \bibinfo {author} {\bibfnamefont {H.}~\bibnamefont
  {Watanabe}},\ }\bibfield  {title} {\bibinfo {title} {{Connecting higher-order
  topological insulators to lower-dimensional topological insulators}},\ }\href
  {https://doi.org/10.1103/PhysRevB.98.205129} {\bibfield  {journal} {\bibinfo
  {journal} {Phys. Rev. B}\ }\textbf {\bibinfo {volume} {98}},\ \bibinfo
  {pages} {205129} (\bibinfo {year} {2018})}\BibitemShut {NoStop}%
\bibitem [{\citenamefont {Xu}\ \emph {et~al.}(2019)\citenamefont {Xu},
  \citenamefont {Song}, \citenamefont {Wang}, \citenamefont {Weng},\ and\
  \citenamefont {Dai}}]{PhysRevLett.122.256402}%
  \BibitemOpen
  \bibfield  {author} {\bibinfo {author} {\bibfnamefont {Y.}~\bibnamefont
  {Xu}}, \bibinfo {author} {\bibfnamefont {Z.}~\bibnamefont {Song}}, \bibinfo
  {author} {\bibfnamefont {Z.}~\bibnamefont {Wang}}, \bibinfo {author}
  {\bibfnamefont {H.}~\bibnamefont {Weng}},\ and\ \bibinfo {author}
  {\bibfnamefont {X.}~\bibnamefont {Dai}},\ }\bibfield  {title} {\bibinfo
  {title} {{Higher-Order Topology of the Axion Insulator
  ${\mathrm{EuIn}}_{2}{\mathrm{As}}_{2}$}},\ }\href
  {https://doi.org/10.1103/PhysRevLett.122.256402} {\bibfield  {journal}
  {\bibinfo  {journal} {Phys. Rev. Lett.}\ }\textbf {\bibinfo {volume} {122}},\
  \bibinfo {pages} {256402} (\bibinfo {year} {2019})}\BibitemShut {NoStop}%
\bibitem [{\citenamefont {Tanaka}\ \emph {et~al.}(2020)\citenamefont {Tanaka},
  \citenamefont {Takahashi}, \citenamefont {Zhang},\ and\ \citenamefont
  {Murakami}}]{PhysRevResearch.2.043274}%
  \BibitemOpen
  \bibfield  {author} {\bibinfo {author} {\bibfnamefont {Y.}~\bibnamefont
  {Tanaka}}, \bibinfo {author} {\bibfnamefont {R.}~\bibnamefont {Takahashi}},
  \bibinfo {author} {\bibfnamefont {T.}~\bibnamefont {Zhang}},\ and\ \bibinfo
  {author} {\bibfnamefont {S.}~\bibnamefont {Murakami}},\ }\bibfield  {title}
  {\bibinfo {title} {{Theory of inversion-${\mathbb{Z}}_{4}$ protected
  topological chiral hinge states and its applications to layered
  antiferromagnets}},\ }\href
  {https://doi.org/10.1103/PhysRevResearch.2.043274} {\bibfield  {journal}
  {\bibinfo  {journal} {Phys. Rev. Research}\ }\textbf {\bibinfo {volume}
  {2}},\ \bibinfo {pages} {043274} (\bibinfo {year} {2020})}\BibitemShut
  {NoStop}%
\bibitem [{\citenamefont {Tanaka}\ \emph
  {et~al.}(2022{\natexlab{b}})\citenamefont {Tanaka}, \citenamefont {Zhang},
  \citenamefont {Uwaha},\ and\ \citenamefont
  {Murakami}}]{PhysRevLett.129.046802}%
  \BibitemOpen
  \bibfield  {author} {\bibinfo {author} {\bibfnamefont {Y.}~\bibnamefont
  {Tanaka}}, \bibinfo {author} {\bibfnamefont {T.}~\bibnamefont {Zhang}},
  \bibinfo {author} {\bibfnamefont {M.}~\bibnamefont {Uwaha}},\ and\ \bibinfo
  {author} {\bibfnamefont {S.}~\bibnamefont {Murakami}},\ }\bibfield  {title}
  {\bibinfo {title} {{Anomalous Crystal Shapes of Topological Crystalline
  Insulators}},\ }\href {https://doi.org/10.1103/PhysRevLett.129.046802}
  {\bibfield  {journal} {\bibinfo  {journal} {Phys. Rev. Lett.}\ }\textbf
  {\bibinfo {volume} {129}},\ \bibinfo {pages} {046802} (\bibinfo {year}
  {2022}{\natexlab{b}})}\BibitemShut {NoStop}%
\bibitem [{\citenamefont {Fu}\ and\ \citenamefont
  {Kane}(2007)}]{PhysRevB.76.045302}%
  \BibitemOpen
  \bibfield  {author} {\bibinfo {author} {\bibfnamefont {L.}~\bibnamefont
  {Fu}}\ and\ \bibinfo {author} {\bibfnamefont {C.~L.}\ \bibnamefont {Kane}},\
  }\bibfield  {title} {\bibinfo {title} {{Topological insulators with inversion
  symmetry}},\ }\href {https://doi.org/10.1103/PhysRevB.76.045302} {\bibfield
  {journal} {\bibinfo  {journal} {Phys. Rev. B}\ }\textbf {\bibinfo {volume}
  {76}},\ \bibinfo {pages} {045302} (\bibinfo {year} {2007})}\BibitemShut
  {NoStop}%
\bibitem [{\citenamefont {Bradlyn}\ \emph {et~al.}(2017)\citenamefont
  {Bradlyn}, \citenamefont {Elcoro}, \citenamefont {Cano}, \citenamefont
  {Vergniory}, \citenamefont {Wang}, \citenamefont {Felser}, \citenamefont
  {Aroyo},\ and\ \citenamefont {Bernevig}}]{bradlyn2017topological}%
  \BibitemOpen
  \bibfield  {author} {\bibinfo {author} {\bibfnamefont {B.}~\bibnamefont
  {Bradlyn}}, \bibinfo {author} {\bibfnamefont {L.}~\bibnamefont {Elcoro}},
  \bibinfo {author} {\bibfnamefont {J.}~\bibnamefont {Cano}}, \bibinfo {author}
  {\bibfnamefont {M.}~\bibnamefont {Vergniory}}, \bibinfo {author}
  {\bibfnamefont {Z.}~\bibnamefont {Wang}}, \bibinfo {author} {\bibfnamefont
  {C.}~\bibnamefont {Felser}}, \bibinfo {author} {\bibfnamefont {M.~I.}\
  \bibnamefont {Aroyo}},\ and\ \bibinfo {author} {\bibfnamefont {B.~A.}\
  \bibnamefont {Bernevig}},\ }\bibfield  {title} {\bibinfo {title} {Topological
  quantum chemistry},\ }\href {https://www.nature.com/articles/nature23268}
  {\bibfield  {journal} {\bibinfo  {journal} {Nature}\ }\textbf {\bibinfo
  {volume} {547}},\ \bibinfo {pages} {298} (\bibinfo {year}
  {2017})}\BibitemShut {NoStop}%
\bibitem [{\citenamefont {Kruthoff}\ \emph {et~al.}(2017)\citenamefont
  {Kruthoff}, \citenamefont {de~Boer}, \citenamefont {van Wezel}, \citenamefont
  {Kane},\ and\ \citenamefont {Slager}}]{PhysRevX.7.041069}%
  \BibitemOpen
  \bibfield  {author} {\bibinfo {author} {\bibfnamefont {J.}~\bibnamefont
  {Kruthoff}}, \bibinfo {author} {\bibfnamefont {J.}~\bibnamefont {de~Boer}},
  \bibinfo {author} {\bibfnamefont {J.}~\bibnamefont {van Wezel}}, \bibinfo
  {author} {\bibfnamefont {C.~L.}\ \bibnamefont {Kane}},\ and\ \bibinfo
  {author} {\bibfnamefont {R.-J.}\ \bibnamefont {Slager}},\ }\bibfield  {title}
  {\bibinfo {title} {{Topological Classification of Crystalline Insulators
  through Band Structure Combinatorics}},\ }\href
  {https://doi.org/10.1103/PhysRevX.7.041069} {\bibfield  {journal} {\bibinfo
  {journal} {Phys. Rev. X}\ }\textbf {\bibinfo {volume} {7}},\ \bibinfo {pages}
  {041069} (\bibinfo {year} {2017})}\BibitemShut {NoStop}%
\bibitem [{\citenamefont {Po}\ \emph {et~al.}(2017)\citenamefont {Po},
  \citenamefont {Vishwanath},\ and\ \citenamefont {Watanabe}}]{po2017symmetry}%
  \BibitemOpen
  \bibfield  {author} {\bibinfo {author} {\bibfnamefont {H.~C.}\ \bibnamefont
  {Po}}, \bibinfo {author} {\bibfnamefont {A.}~\bibnamefont {Vishwanath}},\
  and\ \bibinfo {author} {\bibfnamefont {H.}~\bibnamefont {Watanabe}},\
  }\bibfield  {title} {\bibinfo {title} {{Symmetry-based indicators of band
  topology in the 230 space groups}},\ }\href
  {https://doi.org/10.1038/s41467-017-00133-2} {\bibfield  {journal} {\bibinfo
  {journal} {Nat. Commun.}\ }\textbf {\bibinfo {volume} {8}},\ \bibinfo {pages}
  {50} (\bibinfo {year} {2017})}\BibitemShut {NoStop}%
\bibitem [{\citenamefont {Watanabe}\ \emph {et~al.}(2018)\citenamefont
  {Watanabe}, \citenamefont {Po},\ and\ \citenamefont
  {Vishwanath}}]{watanabe2018structure}%
  \BibitemOpen
  \bibfield  {author} {\bibinfo {author} {\bibfnamefont {H.}~\bibnamefont
  {Watanabe}}, \bibinfo {author} {\bibfnamefont {H.~C.}\ \bibnamefont {Po}},\
  and\ \bibinfo {author} {\bibfnamefont {A.}~\bibnamefont {Vishwanath}},\
  }\bibfield  {title} {\bibinfo {title} {Structure and topology of band
  structures in the 1651 magnetic space groups},\ }\href
  {https://advances.sciencemag.org/content/4/8/eaat8685} {\bibfield  {journal}
  {\bibinfo  {journal} {Sci. Adv.}\ }\textbf {\bibinfo {volume} {4}},\ \bibinfo
  {pages} {eaat8685} (\bibinfo {year} {2018})}\BibitemShut {NoStop}%
\bibitem [{\citenamefont {Ono}\ and\ \citenamefont
  {Watanabe}(2018)}]{PhysRevB.98.115150}%
  \BibitemOpen
  \bibfield  {author} {\bibinfo {author} {\bibfnamefont {S.}~\bibnamefont
  {Ono}}\ and\ \bibinfo {author} {\bibfnamefont {H.}~\bibnamefont {Watanabe}},\
  }\bibfield  {title} {\bibinfo {title} {{Unified understanding of symmetry
  indicators for all internal symmetry classes}},\ }\href
  {https://doi.org/10.1103/PhysRevB.98.115150} {\bibfield  {journal} {\bibinfo
  {journal} {Phys. Rev. B}\ }\textbf {\bibinfo {volume} {98}},\ \bibinfo
  {pages} {115150} (\bibinfo {year} {2018})}\BibitemShut {NoStop}%
\bibitem [{\citenamefont {Elcoro}\ \emph {et~al.}(2021)\citenamefont {Elcoro},
  \citenamefont {Wieder}, \citenamefont {Song}, \citenamefont {Xu},
  \citenamefont {Bradlyn},\ and\ \citenamefont
  {Bernevig}}]{elcoro2021magnetic}%
  \BibitemOpen
  \bibfield  {author} {\bibinfo {author} {\bibfnamefont {L.}~\bibnamefont
  {Elcoro}}, \bibinfo {author} {\bibfnamefont {B.~J.}\ \bibnamefont {Wieder}},
  \bibinfo {author} {\bibfnamefont {Z.}~\bibnamefont {Song}}, \bibinfo {author}
  {\bibfnamefont {Y.}~\bibnamefont {Xu}}, \bibinfo {author} {\bibfnamefont
  {B.}~\bibnamefont {Bradlyn}},\ and\ \bibinfo {author} {\bibfnamefont {B.~A.}\
  \bibnamefont {Bernevig}},\ }\bibfield  {title} {\bibinfo {title} {{Magnetic
  topological quantum chemistry}},\ }\href
  {https://www.nature.com/articles/s41467-021-26241-8} {\bibfield  {journal}
  {\bibinfo  {journal} {Nat. Commun.}\ }\textbf {\bibinfo {volume} {12}},\
  \bibinfo {pages} {5965} (\bibinfo {year} {2021})}\BibitemShut {NoStop}%
\bibitem [{\citenamefont {Peng}\ \emph {et~al.}(2022)\citenamefont {Peng},
  \citenamefont {Jiang}, \citenamefont {Fang}, \citenamefont {Weng},\ and\
  \citenamefont {Fang}}]{peng2021topological}%
  \BibitemOpen
  \bibfield  {author} {\bibinfo {author} {\bibfnamefont {B.}~\bibnamefont
  {Peng}}, \bibinfo {author} {\bibfnamefont {Y.}~\bibnamefont {Jiang}},
  \bibinfo {author} {\bibfnamefont {Z.}~\bibnamefont {Fang}}, \bibinfo {author}
  {\bibfnamefont {H.}~\bibnamefont {Weng}},\ and\ \bibinfo {author}
  {\bibfnamefont {C.}~\bibnamefont {Fang}},\ }\bibfield  {title} {\bibinfo
  {title} {{Topological classification and diagnosis in magnetically ordered
  electronic materials}},\ }\href {https://doi.org/10.1103/PhysRevB.105.235138}
  {\bibfield  {journal} {\bibinfo  {journal} {Phys. Rev. B}\ }\textbf {\bibinfo
  {volume} {105}},\ \bibinfo {pages} {235138} (\bibinfo {year}
  {2022})}\BibitemShut {NoStop}%
\bibitem [{\citenamefont {Shiozaki}\ \emph {et~al.}(2015)\citenamefont
  {Shiozaki}, \citenamefont {Sato},\ and\ \citenamefont
  {Gomi}}]{PhysRevB.91.155120}%
  \BibitemOpen
  \bibfield  {author} {\bibinfo {author} {\bibfnamefont {K.}~\bibnamefont
  {Shiozaki}}, \bibinfo {author} {\bibfnamefont {M.}~\bibnamefont {Sato}},\
  and\ \bibinfo {author} {\bibfnamefont {K.}~\bibnamefont {Gomi}},\ }\bibfield
  {title} {\bibinfo {title} {${Z}_{2}$ topology in nonsymmorphic crystalline
  insulators: M\"obius twist in surface states},\ }\href
  {https://doi.org/10.1103/PhysRevB.91.155120} {\bibfield  {journal} {\bibinfo
  {journal} {Phys. Rev. B}\ }\textbf {\bibinfo {volume} {91}},\ \bibinfo
  {pages} {155120} (\bibinfo {year} {2015})}\BibitemShut {NoStop}%
\bibitem [{\citenamefont {Fang}\ and\ \citenamefont
  {Fu}(2015)}]{PhysRevB.91.161105}%
  \BibitemOpen
  \bibfield  {author} {\bibinfo {author} {\bibfnamefont {C.}~\bibnamefont
  {Fang}}\ and\ \bibinfo {author} {\bibfnamefont {L.}~\bibnamefont {Fu}},\
  }\bibfield  {title} {\bibinfo {title} {New classes of three-dimensional
  topological crystalline insulators: Nonsymmorphic and magnetic},\ }\href
  {https://doi.org/10.1103/PhysRevB.91.161105} {\bibfield  {journal} {\bibinfo
  {journal} {Phys. Rev. B}\ }\textbf {\bibinfo {volume} {91}},\ \bibinfo
  {pages} {161105(R)} (\bibinfo {year} {2015})}\BibitemShut {NoStop}%
\bibitem [{\citenamefont {Rahm}\ and\ \citenamefont
  {Erhart}(2020)}]{WulffPack}%
  \BibitemOpen
  \bibfield  {author} {\bibinfo {author} {\bibfnamefont {J.~M.}\ \bibnamefont
  {Rahm}}\ and\ \bibinfo {author} {\bibfnamefont {P.}~\bibnamefont {Erhart}},\
  }\bibfield  {title} {\bibinfo {title} {{WulffPack: A Python package for Wulff
  constructions}},\ }\href {https://doi.org/10.21105/joss.01944} {\bibfield
  {journal} {\bibinfo  {journal} {J. Open Source Softw.}\ }\textbf {\bibinfo
  {volume} {5}},\ \bibinfo {pages} {1944} (\bibinfo {year} {2020})}\BibitemShut
  {NoStop}%
\bibitem [{\citenamefont {Chen}\ \emph {et~al.}(2010)\citenamefont {Chen},
  \citenamefont {Chu}, \citenamefont {Analytis}, \citenamefont {Liu},
  \citenamefont {Igarashi}, \citenamefont {Kuo}, \citenamefont {Qi},
  \citenamefont {Mo}, \citenamefont {Moore}, \citenamefont {Lu} \emph
  {et~al.}}]{chen2010massive}%
  \BibitemOpen
  \bibfield  {author} {\bibinfo {author} {\bibfnamefont {Y.}~\bibnamefont
  {Chen}}, \bibinfo {author} {\bibfnamefont {J.-H.}\ \bibnamefont {Chu}},
  \bibinfo {author} {\bibfnamefont {J.}~\bibnamefont {Analytis}}, \bibinfo
  {author} {\bibfnamefont {Z.}~\bibnamefont {Liu}}, \bibinfo {author}
  {\bibfnamefont {K.}~\bibnamefont {Igarashi}}, \bibinfo {author}
  {\bibfnamefont {H.-H.}\ \bibnamefont {Kuo}}, \bibinfo {author} {\bibfnamefont
  {X.}~\bibnamefont {Qi}}, \bibinfo {author} {\bibfnamefont {S.-K.}\
  \bibnamefont {Mo}}, \bibinfo {author} {\bibfnamefont {R.}~\bibnamefont
  {Moore}}, \bibinfo {author} {\bibfnamefont {D.}~\bibnamefont {Lu}}, \emph
  {et~al.},\ }\bibfield  {title} {\bibinfo {title} {{Massive Dirac fermion on
  the surface of a magnetically doped topological insulator}},\ }\href
  {https://www.science.org/doi/full/10.1126/science.1189924} {\bibfield
  {journal} {\bibinfo  {journal} {Science}\ }\textbf {\bibinfo {volume}
  {329}},\ \bibinfo {pages} {659} (\bibinfo {year} {2010})}\BibitemShut
  {NoStop}%
\bibitem [{\citenamefont {Lee}\ \emph {et~al.}(2015)\citenamefont {Lee},
  \citenamefont {Kim}, \citenamefont {Lee}, \citenamefont {Billinge},
  \citenamefont {Zhong}, \citenamefont {Schneeloch}, \citenamefont {Liu},
  \citenamefont {Valla}, \citenamefont {Tranquada}, \citenamefont {Gu} \emph
  {et~al.}}]{lee2015imaging}%
  \BibitemOpen
  \bibfield  {author} {\bibinfo {author} {\bibfnamefont {I.}~\bibnamefont
  {Lee}}, \bibinfo {author} {\bibfnamefont {C.~K.}\ \bibnamefont {Kim}},
  \bibinfo {author} {\bibfnamefont {J.}~\bibnamefont {Lee}}, \bibinfo {author}
  {\bibfnamefont {S.~J.}\ \bibnamefont {Billinge}}, \bibinfo {author}
  {\bibfnamefont {R.}~\bibnamefont {Zhong}}, \bibinfo {author} {\bibfnamefont
  {J.~A.}\ \bibnamefont {Schneeloch}}, \bibinfo {author} {\bibfnamefont
  {T.}~\bibnamefont {Liu}}, \bibinfo {author} {\bibfnamefont {T.}~\bibnamefont
  {Valla}}, \bibinfo {author} {\bibfnamefont {J.~M.}\ \bibnamefont
  {Tranquada}}, \bibinfo {author} {\bibfnamefont {G.}~\bibnamefont {Gu}}, \emph
  {et~al.},\ }\bibfield  {title} {\bibinfo {title} {{Imaging Dirac-mass
  disorder from magnetic dopant atoms in the ferromagnetic topological
  insulator ${\mathrm{Cr}}_{x} ({\mathrm{Bi}}_{0. 1}{\mathrm{Sb}}_{0.
  9})_{2-x}{\mathrm{Te}}_3$}},\ }\href
  {https://www.pnas.org/doi/abs/10.1073/pnas.1424322112} {\bibfield  {journal}
  {\bibinfo  {journal} {Proceedings of the National Academy of Sciences}\
  }\textbf {\bibinfo {volume} {112}},\ \bibinfo {pages} {1316} (\bibinfo {year}
  {2015})}\BibitemShut {NoStop}%
\bibitem [{\citenamefont {Li}\ \emph {et~al.}(2019)\citenamefont {Li},
  \citenamefont {Li}, \citenamefont {Du}, \citenamefont {Wang}, \citenamefont
  {Gu}, \citenamefont {Zhang}, \citenamefont {He}, \citenamefont {Duan},\ and\
  \citenamefont {Xu}}]{li2019intrinsic}%
  \BibitemOpen
  \bibfield  {author} {\bibinfo {author} {\bibfnamefont {J.}~\bibnamefont
  {Li}}, \bibinfo {author} {\bibfnamefont {Y.}~\bibnamefont {Li}}, \bibinfo
  {author} {\bibfnamefont {S.}~\bibnamefont {Du}}, \bibinfo {author}
  {\bibfnamefont {Z.}~\bibnamefont {Wang}}, \bibinfo {author} {\bibfnamefont
  {B.-L.}\ \bibnamefont {Gu}}, \bibinfo {author} {\bibfnamefont {S.-C.}\
  \bibnamefont {Zhang}}, \bibinfo {author} {\bibfnamefont {K.}~\bibnamefont
  {He}}, \bibinfo {author} {\bibfnamefont {W.}~\bibnamefont {Duan}},\ and\
  \bibinfo {author} {\bibfnamefont {Y.}~\bibnamefont {Xu}},\ }\bibfield
  {title} {\bibinfo {title} {{Intrinsic magnetic topological insulators in van
  der Waals layered ${\mathrm{MnBi}}_{2}{\mathrm{Te}}_{4}$-family materials}},\
  }\href {https://advances.sciencemag.org/content/5/6/eaaw5685} {\bibfield
  {journal} {\bibinfo  {journal} {Sci. Adv.}\ }\textbf {\bibinfo {volume}
  {5}},\ \bibinfo {pages} {eaaw5685} (\bibinfo {year} {2019})}\BibitemShut
  {NoStop}%
\bibitem [{\citenamefont {Fujita}\ \emph {et~al.}(1996)\citenamefont {Fujita},
  \citenamefont {Wakabayashi}, \citenamefont {Nakada},\ and\ \citenamefont
  {Kusakabe}}]{doi:10.1143/JPSJ.65.1920}%
  \BibitemOpen
  \bibfield  {author} {\bibinfo {author} {\bibfnamefont {M.}~\bibnamefont
  {Fujita}}, \bibinfo {author} {\bibfnamefont {K.}~\bibnamefont {Wakabayashi}},
  \bibinfo {author} {\bibfnamefont {K.}~\bibnamefont {Nakada}},\ and\ \bibinfo
  {author} {\bibfnamefont {K.}~\bibnamefont {Kusakabe}},\ }\bibfield  {title}
  {\bibinfo {title} {{Peculiar Localized State at Zigzag Graphite Edge}},\
  }\href {https://doi.org/10.1143/JPSJ.65.1920} {\bibfield  {journal} {\bibinfo
   {journal} {J. Phys. Soc. Jpn.}\ }\textbf {\bibinfo {volume} {65}},\ \bibinfo
  {pages} {1920} (\bibinfo {year} {1996})}\BibitemShut {NoStop}%
\bibitem [{\citenamefont {Gan}\ and\ \citenamefont
  {Srolovitz}(2010)}]{PhysRevB.81.125445}%
  \BibitemOpen
  \bibfield  {author} {\bibinfo {author} {\bibfnamefont {C.~K.}\ \bibnamefont
  {Gan}}\ and\ \bibinfo {author} {\bibfnamefont {D.~J.}\ \bibnamefont
  {Srolovitz}},\ }\bibfield  {title} {\bibinfo {title} {{First-principles study
  of graphene edge properties and flake shapes}},\ }\href
  {https://doi.org/10.1103/PhysRevB.81.125445} {\bibfield  {journal} {\bibinfo
  {journal} {Phys. Rev. B}\ }\textbf {\bibinfo {volume} {81}},\ \bibinfo
  {pages} {125445} (\bibinfo {year} {2010})}\BibitemShut {NoStop}%
\bibitem [{\citenamefont {Artyukhov}\ \emph {et~al.}(2012)\citenamefont
  {Artyukhov}, \citenamefont {Liu},\ and\ \citenamefont
  {Yakobson}}]{doi:10.1073/pnas.1207519109}%
  \BibitemOpen
  \bibfield  {author} {\bibinfo {author} {\bibfnamefont {V.~I.}\ \bibnamefont
  {Artyukhov}}, \bibinfo {author} {\bibfnamefont {Y.}~\bibnamefont {Liu}},\
  and\ \bibinfo {author} {\bibfnamefont {B.~I.}\ \bibnamefont {Yakobson}},\
  }\bibfield  {title} {\bibinfo {title} {{Equilibrium at the edge and atomistic
  mechanisms of graphene growth}},\ }\href
  {https://doi.org/10.1073/pnas.1207519109} {\bibfield  {journal} {\bibinfo
  {journal} {Proc. Natl. Acad. Sci. U.S.A.}\ }\textbf {\bibinfo {volume}
  {109}},\ \bibinfo {pages} {15136} (\bibinfo {year} {2012})}\BibitemShut
  {NoStop}%
\end{thebibliography}
%

\end{document}